\documentclass[useAMS,usenatbib,times]{mn2e}
\usepackage{graphicx}
\usepackage{subfigure}
\def\pageoffset#1#2{\hoffset=#1\relax\voffset=#2\relax} 
\pageoffset{0pc}{2.5pc} 
 at 10truept

\def\and  {\it {et al.} \rm}

\def\spose#1{\hbox to 0pt{#1\hss}}
\def\simlt{\mathrel{\spose{\lower 3pt\hbox{$\mathchar"218$}}
     \raise 2.0pt\hbox{$\mathchar"13C$}}}
\def\simgt{\mathrel{\spose{\lower 3pt\hbox{$\mathchar"218$}}
     \raise 2.0pt\hbox{$\mathchar"13E$}}}
\def\be{\begin{equation}}
\def\ee{\end{equation}}
\def\bce{\begin{center}}
\def\ece{\end{center}}
\def\bea{\begin{eqnarray}}
\def\eea{\end{eqnarray}}
\def\ben{\begin{enumerate}}
\def\een{\end{enumerate}}

\def\ni{\noindent}

\def\brr{\begin{array}}
\def\err{\end{array}}

\def\btau{\overline{\tau}}

\def\nh1{n_{\rm HI}}

\def \p1dk {P_{\rm 1D}(k)}
\def \simlt {\mathrel{\spose{\lower 3pt\hbox{$\mathchar"218$}}
     \raise 2.0pt\hbox{$\mathchar"13C$}}}
\def \simgt {\mathrel{\spose{\lower 3pt\hbox{$\mathchar"218$}}
     \raise 2.0pt\hbox{$\mathchar"13E$}}}

\def \P {{\cal P}}

\def \vH {{\cal H}}

\def \be {\begin{equation}}
\def \en {\end{equation}}
\def \bea {\begin{eqnarray}}
\def \ena {\end{eqnarray}}

\def \bi {\begin{itemize}}
\def \ei {\end{itemize}}

\usepackage[usenames]{color}
\definecolor{Blue}{rgb}{0,0.08,0.65}
\definecolor{Red}{rgb}{0.65,0.08,0.05}
\definecolor{Green}{rgb}{0.15,0.45,0.25}

\def\adj{^\dagger}
\def\inv{^{-1}}

\def\Rset{R}

\newcommand\spi[1]{^{(#1)}} 

\usepackage{amsmath}
\DeclareMathOperator{\E}{E} 
\DeclareMathOperator{\cov}{Cov} 
\DeclareMathOperator{\tra}{tr}

\providecommand{\pscal}[2]{\bigl<#1 \mid #2\bigr>} 

\newcommand\bivec[2]{\begin{bmatrix} #1 \\ #2 \end{bmatrix}}
\newcommand\bimat[4]{\begin{bmatrix} #1 & #2 \\ #3 & #4 \end{bmatrix}}

\begin{document}
\title[The  local theory of the  cosmic skeleton]{The  local theory of the  cosmic skeleton}


\author[Pogosyan,  Pichon, Gay, Prunet,      Cardoso, Sousbie, Colombi  ]
{\ni D. Pogosyan$^1$, C. Pichon$^{2,3}$, C.  Gay$^{2}$,     \newauthor
 S.  Prunet$^{2}$, J.F. Cardoso$^{4,2}$, T. Sousbie$^{2}$ \& S. Colombi$^{2}$    \\
$^1$ Department of Physics, University of Alberta, 11322-89 Avenue, Edmonton, Alberta, T6G 2G7, Canada  \\
$^2$ Institut d'Astrophysique de Paris \& UPMC,  98 bis boulevard Arago, 75014 Paris, France \\
$^3$ Service d'Astrophysique, IRFU,  CEA-CNRS, L'orme des merisiers, 91
470, Gif sur Yvette, France, \\
$^4$ Laboratoire de Traitement et Communication de l'Information, LTCI/CNRS 46, rue Barrault, 75013 Paris, France. \\
}
\onecolumn

\maketitle

\begin{abstract}

The local theory of the critical lines of 2D and 3D Gaussian fields
that underline the cosmic structures is presented.  In the context of
cosmological matter distribution the subset of critical lines of
the 3D density field serves to delineate the skeleton of the observed
filamentary structure at large scales.  A stiff approximation used to
quantitatively describe the filamentary skeleton shows that the flux of the 
skeleton lines is related to the average Gaussian curvature of the one D
sections of the field, much in the same way as the density of the peaks.
The distribution of the length of the critical lines with threshold
is analyzed in detail, while the extended descriptors of the skeleton -
its curvature and its singular points, are introduced and briefly described. 
Theoretical predictions are compared to measurements of the skeleton in
realizations of Gaussian random fields in 2D and 3D.  It is found that
the stiff approximation predicts accurately the shape of the differential
length, allows for analytical insight, and explicit closed form solutions.
Finally, it provides a simple classification of the singular points of
the critical lines: i) critical  points; ii) bifurcation points;
iii) slopping plateaux.
\end{abstract}

\section{Introduction}

The concept of random fields is central to cosmology.  Random fields both
provide initial conditions for the  evolution of the matter distribution in the
Universe, and represent how the observed signals manifest themselves in 3D,
(e.g., in the galaxy or matter density inhomogeneities that form
the Large Scale Structure (LSS)), or on the 2D sky (e.g. for the Cosmic
Microwave Background (CMB) temperature and polarization, the convergence
or shear in weak lensing maps).
In the modern cosmological theories where initial seeds
for inhomogeneities observed as cosmic structures have quantum origin, the
fields of initial density fluctuations (and velocities) are Gaussian.
Subsequent evolution retains Gaussianity for the observables that evolve
linearly (CMB, very Large Scale Structure) while developing non-Gaussian
signature if non-linear effects are involved ({\sl e.g. } lensing and LSS
at smaller scales).

While comparing the observational data to cosmological theory, in particular
in order to estimate parameters of cosmological models, the emphasis is
traditionally placed on the statistical descriptors of the random fields.
For  Gaussian fields the two-point correlation function or the power spectrum
provide full statistical information, while non-Gaussian properties may be
reflected in multi-point correlations.
The understanding and the description of the morphology of structures in our
Universe, on the other hand, calls for the studies of the {\it geometry}
and {\it topology} of random fields. This subject has an extensive history
from the early description of the one-dimensional radio signal time-streams
in 1940's, to  the study of the 2D ocean wave patterns in 1960's
\citep{1957RSPTA.249..321L} to 3D dimensional fields
\citep{1981grf..book.....A} that found the most fruitful application
in cosmology \citep{1981UsMN...36..244A,1986ApJ...304...15B}. 
The most prominent geometrical objects in a typical realization of a random
field are rear events - regions of unusually high or low values of the field.
The rare events are usually related to the most spectacular  observed objects
-- clusters of galaxies at low $z$,  large protogalaxies at high-$z$ or
extensive voids. They are associated with the neighbourhoods of
extrema -- maxima or minima -- making studies of  such {\it critical} points
the first step in understanding typical geometry of a field 
\citep{1984ApJ...284L...9K,1986ApJ...304...15B,1995MNRAS.272..447R,2006MNRAS.365..615S}.
The behaviour of the field in the neighbourhood of a rare peak is highly 
correlated with the peak properties, which allows to describe not only extrema
but the extended peak-patch region \citep{1996ApJS..103....1B}
as a point process that involves the field and its  successive derivatives.
Including the shear flow into consideration gives a compelling
application of the geometry of rare events
to the description of cluster formation through the peak-patch collapse
\citep{1996ApJS..103...63B}.

The rare events  reflect the organization of the field around them and by and
large determine the way the high (low) field regions are interconnected
by the bridges of enhanced field values.
In application to cosmology,  the  ``Cosmic Web'' picture emerges, which relates
the observed clusters of galaxies, and filaments that link them, 
to the geometrical
properties of the initial density field that are enhanced but not yet
destroyed by the still mildly non-linear evolution on supercluster scales 
\citep{1996Natur.380..603B}. The study of the connectivity of filamentary
structures reveal the role of the remaining type of critical points,
the saddle extrema,  in establishing, in particular,
the percolation properties of the Web \citep{2000PhRvL..85.5515C}.
The next step naturally involves describing the statistical properties of
these filamentary structures \citep{1998wfsc.conf...61P,1999ApJ...526..568S} 
and developing techniques for mapping the filaments in the simulation
and data.
\cite{NCD} presented a 2D algorithm to 
trace the filaments of a density field while introducing 
the {\it skeleton} as the set of locally defined critical lines
emanating from the critical points.
\cite{SPCNP} (hereafter SPCNP) extended the local theory and algorithm
to three dimensions and provided the foundation for this work 
while introducing the ``stiff'' approximation.
 Recently, \cite{sousb08}
presented  an algorithm to map out a fully connected version of the skeleton 
that is defined according to the global properties as the lines
of intersections of the patches  (see also \cite{Rien1}, \cite{Rien2} for alternative algorithms). 
This approach connects the study of the
filamentary structure to the geometrical and topological aspects of the
theory of gradient flows \citep{jost} and returns the focus to the notions of
peak and void patches.

This paper presents a consistent local theory for the cosmic skeleton,
while focusing on  the stiff approximation to compute the differential length
of the skeleton as a function of the  contrast and  modulus of the gradient of
the field.  It allows us to define precisely how the properties of the skeleton
depend  analytically on the underlying spectral parameters, and understand what
type of line prevails where. The crucial advantage of the local approach
to the critical lines is that it allows to cast the statistical treatment 
of the linear objects as a point process that involves the field and its
derivatives, which allows for analytical insight, and explicit closed form
solutions.  Our purpose is to construct the theory of critical lines
of a given field corresponding to an intermediate representation of the field,
which is more extended than the knowledge of the critical points. 
 
The organization of the paper is the following.
Section~\ref{sec:defstiff} classifies the various critical lines in 2D and 3D,
connects the average length in a unit volume to the flux of the 
skeleton lines and, within the stiff approximation, to
the average Gaussian curvature of the field in transverse sections. It also 
discusses the   meaning of this approximation. 
Section~\ref{sec:diff2D} calculates the differential length of all sets of
critical lines in 2D, while  Section~\ref{sec:diff3D} 
investigates  the corresponding 3D set of critical lines.
More generally, the  expression for the differential length of the
N dimensional skeleton is sketched in Appendix~\ref{sec:ND}.
Section~\ref{sec:other} introduces the extended descriptors of the skeleton,
Section~\ref{sec:bif} describes their singular points, 
while Section~\ref{sec:conclusion} provides
the discussion and the summary.
Appendix~\ref{sec:cardoso} gives the general method for obtaining in
close form the joint distribution of the field and any combination of
its derivative tensors in arbitrary dimensions.  In particular, it
exhibits all the statistical invariants and their dependence on the
spectral parameters.

\section{The critical lines and the skeleton of a 3D random field} \label{sec:defstiff}

\subsection{Local definition and classification}
The subject of our investigation is a random field, $\rho(\mathbf r)$, that in 
a cosmological setting describes, for example, the density of the matter in the
Universe, or the projected distribution of Cosmic Microwave light on the
celestial sphere. Our focus is on the geometrical properties of the critical
lines, that connect extrema of the field mapping out the filamentary ridges
and valleys of the field.  SPCNP have introduced the definition of the
{\it local critical lines} as the set of points  where the
gradient of the density, $\nabla \rho$, is an eigenvector of 
its Hessian matrix, $\vH$, $ \vH \cdot \nabla \rho=\lambda  \nabla \rho$
i.e, the gradient and one of the principal curvature axes are collinear. 
Formally, this can be specified by a set of equations 
\begin{equation}
\mathbf{S}\equiv\, 
\left( \mathbf{\nabla}\rho  \cdot \vH \right) \cdot{\boldsymbol \epsilon} \cdot 
(\mathbf{\nabla} \rho )=\mathbf{0}~,
\label{eq:defS2}
\end{equation}
where $\boldsymbol{\epsilon}$ is the fully antisymmetric 
(Levi-Civita) tensor of rank $N$.  In general $\mathbf S$ is 
an antisymmetric $N-2$ tensor.

In 3D, the function $\mathbf S$ is vector-valued, 
$S^i=\sum_{klm}\epsilon^{ikl}(\nabla_m \rho) {\vH ^m}_k (\nabla_l \rho)$. 
However, zeroes of $\mathbf{S}$ determine a set of lines rather than isolated
points. Let us consider the behaviour of $\mathbf{S}$ function in the
neighbourhood of a point $\mathbf{r}=\mathbf{0}$ that satisfies criticality
condition $\mathbf{S(0)}=\mathbf{0}$:
\begin{equation}
\mathbf{S}(\delta \mathbf{r}) \approx  \mathbf 0 +
\sum_k \left(\nabla_k \mathbf S \right)
\delta r^k~. \label{eq:sss}
\end{equation}
In our case, under the condition $S^i=0$,
the matrix $\nabla_k S^i$ by definition possesses the {\it left} null-vector, 
furnished by the density gradient,
$\sum_i \nabla_i \rho \left( \nabla_k S^i \right) = 0$; hence 
the gradients $\nabla S^i$ are not linearly independent.
Consequently, there is a non-trivial solution for the {\it right} null-vector
$\sum_k \left(\nabla_k S^i \right) \delta r^k=\mathbf{0}$,  which determines
the local direction of the line along which the criticality condition
is maintained, $\mathbf{S}(\delta \mathbf{r})=\mathbf 0$.
The critical lines intersect where $\nabla_k S^i$ admits more than one
independent {\it right} null-vector.

When we take the eigenvalues of the Hessian to be sorted, 
$\lambda_1 \ge \lambda_2 \ge \lambda_3 $, the gradient of the field at
the critical line may be found aligned with the first, second, or third
eigenvector.
This gives rise to the classification of the critical lines based on
the choice of the eigenvector aligned with the gradient, that becomes more fine
grained when the magnitudes of the eigenvalues are taken into account.
Namely, we distinguish {\it primary} critical lines, which correspond to 
$\nabla \rho$  being aligned with the direction in which the field is
the least curved, i.e where the eigenvalue is the smallest in magnitude,
and {\it secondary} critical lines at which $\nabla \rho$ is aligned with
the eigenvalues of larger magnitude.  The primary type consists of
\begin{enumerate}
\item The {\it skeleton}, that has the gradient in the $\lambda_1$ direction
and is limited to the region
$|\lambda_1| \le |\lambda_2|$, which translates to the condition 
$\lambda_1 + \lambda_2 \le 0$.
The skeleton has always eigenvalues 
in the directions transverse to $\nabla \rho$ negative, 
$\lambda_3 \le \lambda_2 \le 0$ and corresponds to the filamentary ridges
spreading from the maxima in the direction of the slowest descent.
\item The {\it anti-skeleton}, that has the gradient in the $\lambda_3$
direction and is restricted to the region $|\lambda_3| \le |\lambda_2|$, i.e 
$\lambda_3+\lambda_2 \ge 0 $.
In the directions transverse to $\nabla \rho$ the anti-skeleton has always 
positive curvature $\lambda_1 \ge \lambda_2 \ge 0$.  
It corresponds to the filamentary valleys
spreading from the minima in the direction of the slowest ascent.
Anti-skeleton can be viewed as a skeleton of the $-\rho$ field.
\item The {\it intermediate skeleton} along which
the gradient is aligned with the middle 
eigen-direction of the Hessian where this direction
is the shallowest 
$|\lambda_2| < |\lambda_1|,|\lambda_3|$, i.e
$-\lambda_1 < \lambda_2  < -\lambda_3 $. This conditions is 
only possible in saddle-like regions where $\lambda_1 > 0 $ and $\lambda_3 < 0$.
\end{enumerate}

The formal classification of the critical lines is summarized in Table~\ref{tbl:types}.
\begin{table}
\begin{tabular}{llcl}
\multicolumn{2}{c}{Type} & Alignment & \multicolumn{1}{c}{Condition} \\
\hline 
Primary& Skeleton: & $  \vH\cdot \,\nabla\rho=\lambda_1 \nabla\rho $ & 
$ \lambda_1+\lambda_2 \le  0 $ \\
& Inter-skeleton: & $  \vH\cdot \,\nabla\rho=\lambda_2 \nabla\rho $ & 
$ \lambda_1+\lambda_2 >  0 $ and $ \lambda_3+\lambda_2 <  0 $  \\
& Anti-skeleton: & $  \vH\cdot \,\nabla\rho=\lambda_3 \nabla\rho $  &  
$ \lambda_3+\lambda_2 \ge  0 $   \\
\hline
Secondary && $ \vH\cdot \,\nabla\rho=\lambda_2 \nabla\rho $ &
$ \lambda_1+\lambda_2 \le  0 $ \\
&& $ \vH\cdot \,\nabla\rho=\lambda_3 \nabla\rho $ &
$ \lambda_1+\lambda_2 \le  0 $  \\
&& $ \vH\cdot \,\nabla\rho=\lambda_1 \nabla\rho $ &
$ \lambda_1+\lambda_2 >  0 $ and $ \lambda_3+\lambda_2 <  0 $  \\
&& $ \vH\cdot \,\nabla\rho=\lambda_3 \nabla\rho $ &
$ \lambda_1+\lambda_2 >  0 $ and $ \lambda_3+\lambda_2 <  0 $ \\
&& $ \vH\cdot \,\nabla\rho=\lambda_1 \nabla\rho $ &
$ \lambda_3+\lambda_2 \ge  0 $  \\
&& $ \vH\cdot \,\nabla\rho=\lambda_2 \nabla\rho $ &
$ \lambda_3+\lambda_2 \ge  0 $  \\
\hline
\end{tabular}
\caption{the classification of the critical lines in 3D.}
\label{tbl:types}
\end{table}

\subsection{The average flux (length per unit volume) of the critical lines}
\label{sec:defsR}
As   the average number density is the fundamental quantity that describes point
events, e.g. extrema of a field, conversely the most important characterization of the
critical lines or skeleton  is their flux, {\it i.e.} the  number of critical lines intersecting a given oriented surface\footnote{
For M dimensional objects in N dimensional space, in general,
one counts the average number of intersections between objects M and N-M
dimensional surfaces, per unit N-M volume.
From a statistical point of view this constitutes a {\it point} process that can
be evaluated knowing the distribution of the field and some of its
derivatives at one arbitrary point only.}.
This flux is equivalent to the length of the lines per unit volume.
Following NCD and SPCNP we shall preferentially use the latter terminology as
it highlights that we deal with the first geometrical parameter, the length,
of the lines.  The subsequent parameters of these linear objects are the
curvature, and, in 3D, the torsion.

In this paper we consider $\rho$ to be a homogeneous and
isotropic Gaussian random field of zero mean, described by the power
spectrum $P(k)$.
In the statistical description of the skeleton of the 
field $\rho$,  several linear scales are involved
\begin{equation}
R_{0} = \frac{\sigma_{0}}{\sigma_{1}}, \quad R_{*} = \frac{\sigma_{1}}
{\sigma_{2}}, \quad \tilde R  = \frac{\sigma_{2}}{\sigma_{3}}\,,
 \quad \hat R  = \frac{\sigma_{3}}{\sigma_{4}}\,,
   \label{eq:defR0}
\end{equation}
\begin{equation}
{\rm where} \quad 
\sigma^2_0 = \langle\rho^2\rangle, \quad
\sigma^2_1  =  \langle\left(\nabla \rho\right)^2\rangle, \quad
\displaystyle \displaystyle \sigma^2_2  =   \langle\left(\Delta \rho\right)^2
\rangle, \quad
 \sigma^2_3  =  \langle\left(\nabla \Delta \rho\right)
^2\rangle\,, \quad \mbox{and generally}\quad 
  \sigma_p^2 = \frac{2 \pi^{D/2}}{\Gamma[D/2]}
  \int_0^\infty k^{2p} P(k) k^{D-1} dk \,.
\label{eq:sigmadef}
\end{equation}
These scales are ordered $R_0 \ge R_* \ge \tilde R \ge \ldots $.
The first two have well-known meanings of typical separation  
between zero-crossing of the field $R_{0}$ and  mean distance between  
extrema, $R_{*}$ (Bardeen \& al. 1986), and
the third one, $\tilde R$ is, by analogy, the typical distance  
between the inflection points.
These three are the only ones that are involved in determination of the length
of the critical lines.  The higher order scale $\hat R$  appear
only in computation of the curvature and the torsion
(see Section~\ref{sec:other}).

Let us define a set of spectral parameters that depend on the shape of
the underlying power spectrum.
Out of these five scales four dimensionless ratios may be   constructed  
that are intrinsic parameters of the theory
\begin{equation}
\displaystyle \gamma   \equiv \frac{R_{*}}{R_{0}}=\frac{\sigma_1^2}
{\sigma_0\sigma_2} , \quad
{\tilde \gamma}  \equiv \frac{\tilde R}{R_{*}}=\frac{\sigma^2_2}
{\sigma_3 \sigma_1} \,,
 \quad
{\hat \gamma}  \equiv \frac{\hat R}{\tilde R}=\frac{\sigma^2_3}
{\sigma_4 \sigma_2} \,,
 \quad  \mbox{and generally}\quad 
   \gamma_{p,q} = \frac { \sigma_{(p+q)/2}^2}{\sigma_p\sigma_q}\,.
\label{eq:gammadef}
\end{equation}
From the geometrical point of view $\gamma$ specifies
how frequently one encounters  a maximum between two zero crossings
of the  field, while $\tilde \gamma$ describes, on average, how many  
inflection points are between two extrema. From  a statistical perspective,
$\gamma$'s are the cross-correlation coefficients between the field and its 
derivatives at the same point (see Appendix~\ref{sec:cardoso}).
\begin{equation}
\gamma=\frac{\langle \rho \Delta \rho \rangle}{\sigma_0 \sigma_2} , \quad
{\tilde \gamma} = 
\frac{\langle \nabla \rho \cdot \nabla \Delta \rho \rangle}{\sigma_1 \sigma_3},
\quad \ldots \label{eq:defgam}
\end{equation}
For  Gaussian fields, these parameters can be easily calculated from  
the power spectrum.  All $\gamma$'s range from zero to one.
For reference, for the power-law spectra with
index $n > -3$,  smoothed at small scales with a Gaussian window,
$ \gamma = \sqrt{({n+3})/({n+5})},$ $ \tilde\gamma = \sqrt{({n+5})/({n+7})}$.
Note that  cosmologically relevant density power spectra have $n > -3 $
and, thus, while $\gamma$ can attain low values, $\tilde\gamma$ are  
always close to unity\footnote{Cosmological density fields,  
therefore, have of order one inflection point per extremum, unlike,  
for, example, a mountain range, where one encounters many inflection  
points on a way from a mountain top to the bottom;
see also Section~\ref{sec:stiff}.} .

Let us introduce the dimensionless quantities for the field and its derivatives
as well as for the functions $S^i$ and their gradients $\nabla S^i$:
\begin{eqnarray}
\sigma_0 x \equiv \rho,\quad \sigma_1 x_{k}\equiv \nabla_k \rho,
\quad
\ \sigma_2 x_{kl}\equiv \nabla_k \nabla_l \rho,\quad 
\sigma_3 x_{klm}\equiv \nabla_m \nabla_l \nabla_k \rho, 
\quad
\sigma_2 \sigma_1^2 s^i \equiv {\cal S}^i,\quad \sigma_2^2 \sigma_1 \nabla s^i \equiv \nabla {\cal S}^i,
\quad \sigma_4 \sigma_1^2 \nabla\nabla s^i \equiv\nabla \nabla {\cal S}^i,
\label{eq:scalefr} \end{eqnarray}
giving
\begin{equation}
s^i= \sum_{klm} \epsilon^{imk} x_{ml} x_{l} x_{k}\,, \quad {\rm  and} \quad
\nabla_{m} s^i= 
\sum_{kln} \epsilon^{ink} \left( {\tilde \gamma}^{-1} x_{nlm} x_{l} x_{k}
+\left[ x_{nl} x_{lm} x_{k}+ x_{nl} x_{km} x_{l}
\right]\right)\,.
\label{eq:defsi}
\end{equation} 
Note the specific choice of scaling for $\nabla S$ which  is convenient in view
of the subsequent development of the so-called ``stiff'' approximation.
SPCNP has shown that in terms of these dimensionless quantities,
the {\it cumulative} length per
unit volume of the total set of critical lines 
below the threshold $\eta$ is given by
\begin{equation}
{\cal L}(\eta)=\left( \frac{1}{R_*}\right)^2 \int_{-\infty}^\eta dx
\int  d^{3}x_{k} d^{6} x_{kl}  d^{10} x_{klm} \
|\nabla s^i \times \nabla s^j| {P}(x,x_{k},x_{kl},x_{klm})
\delta_{\rm D}\left(s^i(x_{k},x_{kl},x_{klm})\right)
\delta_{\rm D}\left(s^j(x_{k},x_{kl},x_{klm})\right) \,,
 \label{eq:dLcum}
\end{equation}
where a pair $\nabla s^i, \nabla s^j$ can be chosen arbitrarily as long
as it is linearly independent. In this equation
$|\nabla s^i \times \nabla s^j|$ reflects the inverse
characteristic area orthogonal to a critical line per one such line
while the two $\delta_{\rm D}$-functions account for the critical line 
condition~(\ref{eq:defS2}). 
For the  complete set of critical lines, there are no restriction to the region
of integration. If one is interested in a particular type of the critical lines,
the integration should be restricted to the regions consistent with
Table~\ref{tbl:types}. The {\it differential} length (per unit volume) is simply given by the derivative of equation~(\ref{eq:dLcum})  with respect to $\eta$:
\begin{equation}
\frac{\partial{\cal L}}{\partial \eta}=\left( \frac{1}{R_*}\right)^2
\int  d^{3}x_{k} d^{6} x_{kl}  d^{10} x_{klm} \
|\nabla s^i \times \nabla s^j| {P}(\eta,x_{k},x_{kl},x_{klm})
\delta_{D}\left(s^i(x_{k},x_{kl},x_{klm})\right)
\delta_{D}\left(s^j(x_{k},x_{kl},x_{klm})\right) \,,
 \label{eq:myeq2}
\end{equation}
while the  total length of critical lines is
\begin{equation}
L \equiv {\cal L}(\infty) = \int_{-\infty}^{\infty} 
d \eta \frac{\partial{\cal L}}{\partial \eta} \quad .
\end{equation}

Since for Gaussian field,  the derivatives of even order are uncorrelated with
the odd orders, the joint distribution function $P(x,x_{k},x_{kl},x_{klm})$
entering   equation~(\ref{eq:dLcum}) is factorized  as
\begin{equation}
{P}(x,x_{k},x_{kl},x_{klm})={ P}_{0}(x,x_{kl}) {P}_{1}(x_{k},x_{klm}).
\end{equation}
In $P_0$, the only dependence on the power spectrum of the field 
is through the parameter $\gamma$ (c.f. equation~(\ref{eq:gammadef}))
that describes the correlation between the field and its second derivatives.
Similarly  ${P_{1}}(x_{k},x_{klm})$ only involves $\tilde \gamma$ which
describes the correlation between the gradient of the field and its
third derivatives.  Therefore, ${\partial{\cal L}}/{\partial \eta}$ depends
only on $\eta$, $\tilde R$ $\gamma$  and $\tilde \gamma$.  The integrated
length, $L$
may depend only on  $\tilde \gamma$ and $\tilde R$ since the marginalization
of $P_{0}(\eta,x_{kl})$ over $\eta$ eliminates the dependency over $\gamma$.

\subsection{The ``stiff'' filament approximation}

Let us look at the dependence of the length of the critical lines on 
characteristic scales of the field in more detail. The $R_*^{-2}$ factor
that appeared in equation~(\ref{eq:myeq2}) reflect our choice of dimensionless
variables~(\ref{eq:defsi}) and is suggestive but not yet conclusive since
$|\nabla s^i \times \nabla s^j|$ that includes third derivative
terms, depends also on the other scale, $\tilde R$.
Let us write formally 
\begin{equation}
\nabla s^{i}\times \nabla s^{j} = {\tilde \gamma}^{-2}
{\mathbf A}(x_{k},x_{kl},x_{klm})
+{\tilde \gamma}^{-1} {\mathbf B} (x_{k},x_{kl},x_{klm})
+ {\mathbf C} (x_{k,}x_{kl}) \,.\label{eq:simple}
\end{equation}
If the third derivatives are important and the first term dominates,
then the length scaling $L \propto {\tilde \gamma}^{-2} R_*^{-2} = 
{\tilde R}^{-2}$ would reflect the mean separation between the inflection
points, $\tilde R$.  Indeed, by definition the local skeleton is
almost straight within
a volume that has one inflection point $\sim \tilde R^{3}$.
A straight segment through such volume has length $\sim \tilde R$,
thus the expected length per unit volume is $\sim 1 /\tilde R^{2}$.
But if the last term dominates statistically, the length per unit volume of
the skeleton will scale as $L \propto R_*^{-2}$ that can be interpreted
that the critical lines are almost straight within a large volume
volume $\sim R_*^{3}$ containing one extremum. 
This is consistent with observation that the integral
term does not depend on the third derivatives, thus inflection points play no
role, and any dependence on $\tilde\gamma$ drops out. 

Which regime holds can be established by measuring the dependence of the
critical lines length in the simulations as a function of smoothing length
for different spectral indexes. For the power-law spectra with Gaussian 
smoothing at the radial scale $\sigma$, in 3D, 
$R_* = \sqrt{2} \sigma/\sqrt{n+5}$,
while $\tilde R = \sqrt{2} \sigma/\sqrt{n+7}$. The measurements in SPCNP
found that $L \propto (n+5.5) \sigma^{-2}$ over the range of
spectral indexes relevant to cosmology, which points at the subdominant 
nature played by the third derivatives. In the  ``stiff'' approximation
we omit the third derivative, 
effectively assuming that the Hessian can be treated
as constant during the evaluation of $\nabla s$.
This picture corresponds to a  skeleton connecting extrema with relatively straight segments. 
In the ``stiff'' approximation,  equation~(\ref{eq:myeq2}) becomes
\begin{equation}
\frac{\partial{\cal L}}{\partial \eta} \approx \left( \frac{1}{R_{*}}\right)^2 \int d^{3}x_{k} d^{6} x_{kl}\ | {\mathbf C}(x_{k,}x_{lm}) |
\ {P}_{0}(\eta,x_{kl}) \  P_{1}(x_{k})
\delta_{D}\left(s^i(x_{k},x_{kl})\right)
\delta_{D}\left(s^j(x_{k},x_{kl})\right) \,.
\end{equation}
The differential length is then  only the function of $\gamma$ times $L$.

\begin{figure*}
  \centering
  \includegraphics[angle=0,width=12cm]{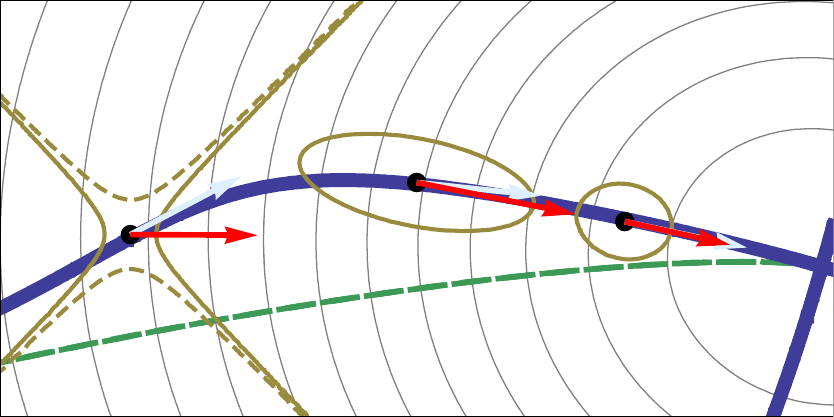}
  \caption{The neighbourhood of a local critical line (thick blue line).
  This is a zoomed section of the wide patch shown in Figure~\ref{fig:example}.
  Thin lines are isocontours of the field. Three sample points are
  investigated in detail. The signature, orientation
  and the magnitude of the local Hessian are represented
  by the golden shapes. Near the maximum on the right edge, the signature
  of the eigenvalues of the Hessian is (-1,-1), which is shown by ellipses
  oriented according to eigen-directions with longer semi-axis along the 
  direction of the least curvature. At the leftmost point the eigenvalue 
  signature is ``saddle-like'', (1,-1), which is represented by a pair of 
  hyperbolae, also oriented with respect to eigen-directions.
  By definition, on the critical line the gradient of the field $\nabla \rho$, 
  shown by red arrows, is aligned with one of the eigen-directions (i.e the
  axis of the ellipse or hyperbola in the graph).
  The light cyan arrows are the tangent vectors to the critical line
  $ \propto \boldsymbol{\epsilon} \cdot \nabla S$, while stiff approximation
  to them would be parallel to the gradient. The direction of the critical line
  is close to the gradient when it follows the ridge near the maximum,
  but slides at an angle in the ``saddle-like'' region, before joining the
  saddle extremal point beyond the left edge
  of the plot (see Figure~\ref{fig:example}). Note that the gradient line
  that takes us to the same saddle as a segment of the global skeleton
  (dashed green) does not follow the ridge too closely in this instance.
} 
\label{fig:example_detail}
\end{figure*}

\begin{figure*}
  \centering
  \includegraphics[angle=0,width=10cm]{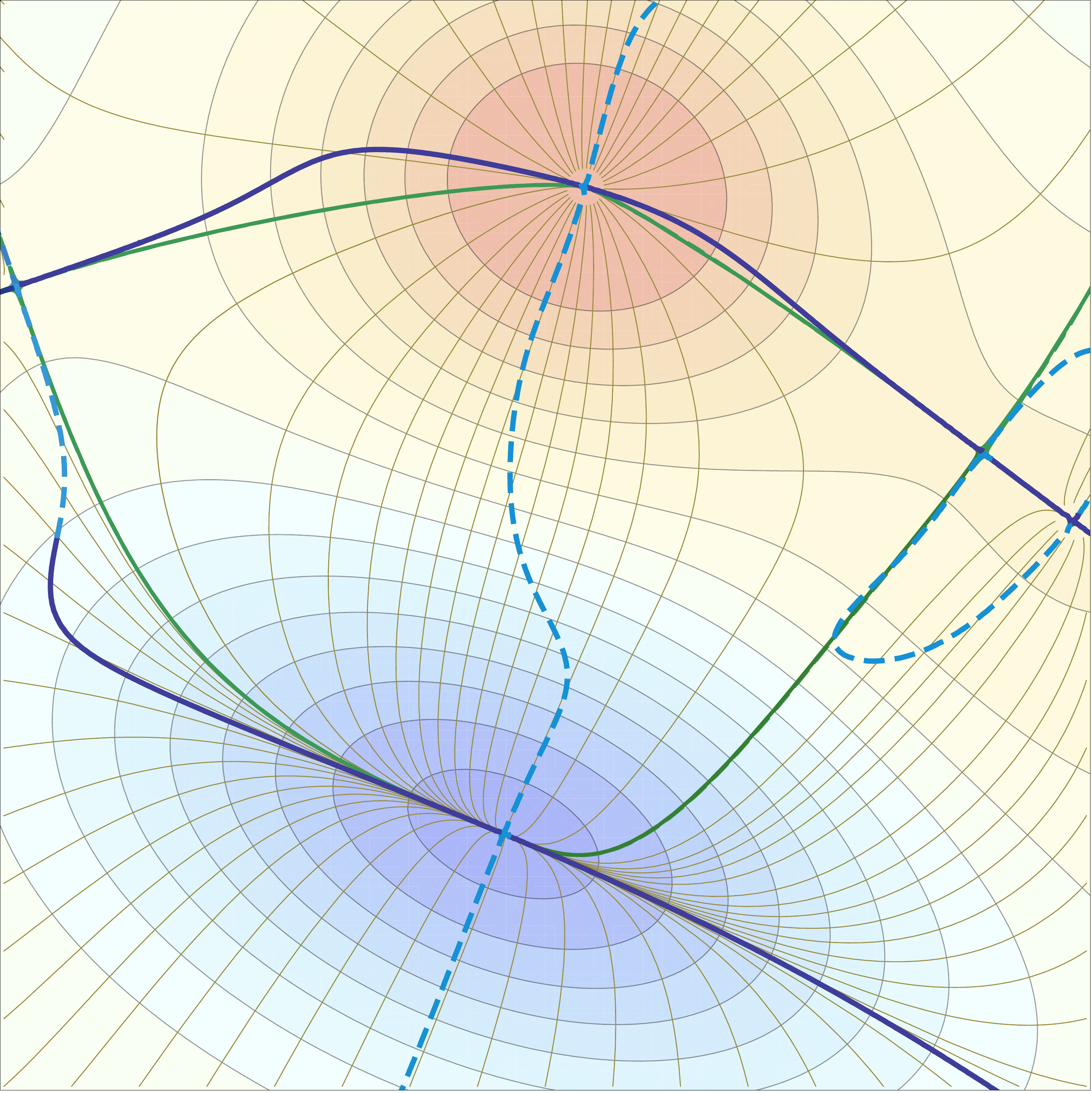}
  \caption{An example  of a generic patch of a 2D field.  
  The underlying isocontours correspond to the density field.
  The thin gold lines are the gradients lines of the field.
  The blue lines is the local set of  critical lines, given by
  the solution of ${\cal S}\equiv\, \left(\,\vH\cdot \nabla \rho \right)
  \times \nabla \rho =\mathbf{0}$. Primary lines are shown 
  in solid and the secondary lines are dashed.
  The green lines correspond to global critical lines: the skeleton and 
  the anti skeleton, which delineate a  special bundle of gradient
  lines \citep{jost} at the intersection of a peak patch and a void patch.
  The primary local lines follow fairly well the gradient lines,
  noticeably near the extrema, where the ``stiff'' approximation holds best.
  In contrast, the
  approximation worsens for the secondary critical lines.
  The main distinction between the global and local skeletons is that
  the global one follows the everywhere smooth gradient line that uniquely 
  connects a maximum to a saddle, at the cost of deviating from
  being exactly on the ridge (see how in the vicinity of the minimum at
  the bottom, the right green line does not follow the trough)
  The local skeleton tries to delineate the ridges as far from extrema
  as possible, but then the lines that follow this local
  procedure from different extrema do not meet and have to rather suddenly
  reconnect. 
  A particularly striking example of this is the loop on the right hand side.
  A zoomed view of the area left to the top maximum
  is shown in Figure~\ref{fig:example_detail}.
} 
\label{fig:example}
\end{figure*}

The ``stiff'' approximation can be looked at from another perspective.
By definition at a point on a local critical line, two of the characteristic
directions defined for the field, namely, the direction of the gradient,
$\nabla\rho$,
and one chosen eigen direction of the Hessian, ${\cal H}$, must coincide.
But the direction of the critical line itself, given by 
$ \nabla S^{i}\times \nabla S^{j} $,  is not, in general, 
aligned with the gradient of the field. Local critical lines are not the
gradient lines, and in this sense they differ from the skeleton lines defined
globally as void-patch intersections \citep{sousb08}.
In the ``stiff'' approximation, however, 
$\left(\nabla_m s^i \right)_\mathrm{stiff} \approx
\sum_{kln} \epsilon^{ink} 
\left[ x_{nl} x_{lm} x_{k}+ x_{nl} x_{km} x_{l}
\right]
$
and $ \left[ \left(\nabla s^i \right)_\mathrm{stiff} 
\times \left(\nabla s^j \right)_\mathrm{stiff} \right] 
\times \nabla \rho = 0 $,  {\it i.e. } it is parallel to the gradient.
Figure~\ref{fig:example_detail} shows the details of the calculations
for the high-resolution segment of the 2D field.
Thus, the essence of the stiff approximation lies in the assumption that the
mismatch between the critical lines and gradient directions is statistically
small.
As Figure~\ref{fig:example}, which contains an extended view of
the same field, illustrates, this assumption holds
particularly well for the primary critical lines which more closely correspond
to the intuitive picture of sharp ridges and deep valleys. Indeed, at a primary
line the gradient points to the least curved direction, i.e, in some sense in
the direction in which the changes of the field properties are the slowest.
Therefore one can expect that this is the direction in which  the condition
of criticality will be maintained, i.e which the critical line itself will
follow.  Figure~\ref{fig:example} shows
that the primary lines start to deviate from the gradient
flow mostly towards their end points when the curvature of the
field along the line becomes comparable in  magnitude to the transverse one.
Secondary critical lines are much less certain to follow the gradient,
sometimes exhibiting a ``sliding'' behaviour, on occasion almost orthogonal
to the gradient, as a loop-like secondary line near the right saddle
in Figure~\ref{fig:example} exhibits. So the stiff approximation
for the secondary lines should be taken with more caution,
although we include them for completeness.

The stiff approximation provides a framework to compute the total differential
length of the critical lines and the local skeleton almost completely
analytically. In the next two sections we will carry this calculation in two
and three dimensions and argue that it can straightforwardly be extended in
N dimensions (see Appendix~\ref{sec:ND}).  
In what follows we shall omit in the derivation for brevity the $1/R_*$ (in 2D)
and $1/R_*^2$ (in 3D) factors, but keep in mind that all the length quantities
below scale accordingly.  In section~\ref{sec:stiff}, after the computational
machinery is developed, we return to the role the third derivative may play
in description of the critical lines.

\section{Critical lines of 2D fields}\label{sec:diff2D}
Even though the large scale structures of the universe are three dimensional, 
other important observed data sets involve 2D maps such as the  
cosmic microwave background or  lensing convergence maps. 
Hence analyzing the local  statistical properties of  filaments in
two dimensions is astrophysically well motivated. The 2D case is
also a convenient starting point  to introduce the details of the calculations
that can be generalized to 3D and higher dimensions.

The 2D case affords several simplifications over the 3D case.
In 2D, $S$ is a (pseudo) scalar function and its zero level, orthogonal
to $\nabla S$, determines the critical lines. The expression for the
differential length simplifies to
\begin{equation}
\frac{\partial{\cal L}}{\partial \eta}=\frac{1}{R_*}
\int  d^{2}x_{k} d^{3} x_{kl}  d^{4} x_{klm} \
|\nabla s| {P}(\eta,x_{k},x_{kl},x_{klm})
\delta_{D}\left(s(x_{k},x_{kl})\right) \,.
 \label{eq:dL2D_gen}
\end{equation}
There are just four types of critical lines: two primary,  the
skeleton and the anti-skeleton, and two corresponding secondary ones.
The classification of the 2D critical lines is summarized in Table~\ref{tbl:types2D}.
We shall focus on the most interesting primary lines in the main text, 
leaving the secondaries to the Appendix~\ref{app:2Dcrit}.
In Figure~\ref{fig:example} the critical lines of
different types are shown for an example generic patch of a 2D field.

\begin{table}
\begin{tabular}{llcl}
\multicolumn{2}{c}{Type} & Alignment & \multicolumn{1}{c}{Condition} \\
\hline 
Primary& Skeleton: & $  \vH\cdot \,\nabla\rho=\lambda_1 \nabla\rho $ & 
$ \lambda_1+\lambda_2 \le  0 $ \\
& Anti-skeleton: & $  \vH\cdot \,\nabla\rho=\lambda_2 \nabla\rho $ & 
$ \lambda_1+\lambda_2 >  0 $  \\
\hline
Secondary && $ \vH\cdot \,\nabla\rho=\lambda_2 \nabla\rho $ &
$ \lambda_1+\lambda_2 \le  0 $ \\
&& $ \vH\cdot \,\nabla\rho=\lambda_1 \nabla\rho $ &
$ \lambda_1+\lambda_2 >  0 $  \\
\hline
\end{tabular}
\caption{the classification of the critical lines in 2D.}
\label{tbl:types2D}
\end{table}

\subsection{The differential length of the critical lines of 2D fields} \label{sec:crit2D}
For 2D Gaussian fields, the calculation
of the length of the critical lines can be carried almost completely
analytically in the stiff approximation.
\subsubsection{Direct derivation in the field's frame} 
Let us first proceed in the original coordinate frame.
 Defining
\begin{equation}
s = x_1 x_2 \left( x_{11}-x_{22} \right) + x_{12} \left( x_2^2-x_1^2 \right) 
= 2 w x_1 x_2 + x_{12} \left( x_2^2-x_1^2 \right) \, ,
\label{eq:fullS}
\end{equation}
the stiff approximation to $\nabla s$ involves only up to second
derivatives of the field
\begin{equation}
\left|\nabla s \right|^2 = 
\left(x_1^2+x_2^2 \right) \left(w^2+x_{12}^2\right) \left(u^2+4 \left(w^2+x_{12}^2\right)
-4 \frac{2 x_1 x_2 x_{12} +w \left(x_1^2-x_2^2\right)}{x_1^2+x_2^2} u \right)\,,
\label{eq:fullgradS}
\end{equation}
and 
equation~(\ref{eq:dL2D_gen}) becomes explicitly
\begin{equation}
\frac{\partial {\cal L}}{\partial \eta} = 
\frac{16}{(2 \pi)^{3} \sqrt{1-\gamma^2}} \exp \left[-\frac{\eta^2}{2} \right]
\int\!\!\!\int\!\!\!\int\!\!\!\int\!\!\!\int
du dw dx_{12} dx_1 dx_2 \left| \nabla s \right| \delta_{\rm D}(s)
 \exp \left[-\frac{(u-\gamma \eta)^2}{2 (1-\gamma^2)} - 
 4 w^2 - 4 x_{12}^2 - x_1^2 - x_2^2 \right] ~ ,
\label{eq:2D}
\end{equation}
where the second derivatives are described using $u=-(x_{11}+x_{22})$, 
$w=(x_{11}-x_{22})/2$ and $x_{12}$.
Let us integrate over $x_{12}$ using the $\delta_{\rm D}$-function, which leads to a substitution
$ x_{12} \to({2 x_1 x_2 w})/({x_1^2-x_2^2}) $ with the Jacobian $|1/(x_1^2-x_2^2)|$.
Then equation~(\ref{eq:2D}) becomes
\begin{equation}
\frac{\partial {\cal L}}{\partial \eta} = 
\frac{16}{(2 \pi)^{3} \sqrt{1-\gamma^2}} \exp \left[-\frac{\eta^2}{2} \right]
\int\!\!\!\int\!\!\!\int\!\!\!\int
du dw  dx_1 dx_2 \frac{\left| \nabla s \right|}{\left|x_1^2-x_2^2\right|}
 \exp \left[-\frac{(u-\gamma \eta)^2}{2 (1-\gamma^2)} - 
 4 w^2 \frac{(x_1^2+x_2^2)^2}{(x_1^2-x_2^2)^2}  - x_1^2 - x_2^2\right] \quad ,
\end{equation}
where
\begin{equation}
\left|\nabla s \right|^2 =w^2 \frac{(x_1^2+x_2^2)^3}{(x_1^2-x_2^2)^2}
\left(u + 2 w \frac{(x_1^2+x_2^2)}{(x_1^2-x_2^2)}\right)^2 \quad .
\end{equation}
Let us now substitute\footnote{here we made a choice of sign. Now in the coordinate frame that has  the first direction aligned with the gradient of the field, i.e. $x_2=0$, $\tilde w = (x_{11}-x_{22})/2$, while in the frame that
has gradient aligned with the second direction, $x_1=0$, $\tilde w = (x_{22}-x_{11})/2$}
\begin{equation}
\tilde w = w \frac{(x_1^2+x_2^2)}{(x_1^2-x_2^2)}\,, \quad   {\rm noting~that} \quad \tilde w^2 =( w^2 +x_{12}^2)\,,
\end{equation}
to obtain
\begin{equation}
\frac{\partial {\cal L}}{\partial \eta} = 
\frac{16}{(2 \pi)^{3} \sqrt{1-\gamma^2}} \exp \left[-\frac{\eta^2}{2} \right]
\int\!\!\!\! \int \!\!\!\!\int \!\!\!\!\int
du d\tilde w  dx_1 dx_2 \frac{\left| \tilde w (u + 2 \tilde w) \right|}{\sqrt{x_1^2+x_2^2}} 
 \exp \left[-\frac{(u-\gamma \eta)^2}{2 (1-\gamma^2)} - 
 4 \tilde w^2   - x_1^2 - x_2^2 \right]\,.
\end{equation}
The integration over the first derivatives is now easily performed in the polar coordinates of the  $x_{1},x_{2}$ plane
to give
\begin{equation}
\frac{\partial {\cal L}}{\partial \eta} = 
\frac{2}{\pi^{3/2} \sqrt{1-\gamma^2}} \exp \left[-\frac{\eta^2}{2} \right]
\int_{-\infty}^\infty \!\int_{-\infty}^\infty \! du d\tilde w
\left| \tilde w (u + 2 \tilde w) \right|
 \exp \left[-\frac{(u-\gamma \eta)^2}{2 (1-\gamma^2)} - 
 4 \tilde w^2   \right]\,. \label{eq:2Dfinal}
\end{equation}
This is the final integral form which can be easily investigated in the  $u,\tilde w$ plane. 

\subsubsection{Derivation in the Hessian eigenframe}
To generalize the derivation to higher dimensions we note 
 that one can just perform all the calculations in the Hessian eigenframe. 
We shall denote all quantities {\it evaluated} in the eigenframe with tilde, e.g., $\tilde x_{11} (=\lambda_1),
\tilde x_{22} (=\lambda_2), \tilde u, \tilde w, \tilde x_1, \tilde x_2$.
What must be taken into account is that, in general, these quantities are not Gaussian random variables (while the corresponding ones in the fixed frame are),
since the transformation from the fixed to eigenframe is non-linear. The
Gaussian nature is only preserved for  $\tilde u, \tilde x_1, \tilde x_2$. 
In the Hessian eigenframe $\tilde x_{12}=0$. From
equations~(\ref{eq:fullS}-\ref{eq:fullgradS})
\begin{equation}
\tilde s = \tilde x_1 \tilde x_2 ( \tilde x_{11} - \tilde x_{22}) =
\tilde x_1 \tilde x_2 (\lambda_1-\lambda_2) = 2 \tilde x_1 \tilde x_2 \tilde w \,,\quad
\left|{\tilde \nabla s} \right| = | \tilde w |
\sqrt{(\tilde u + 2 \tilde w)^2 \tilde x _1^2+(\tilde u - 2 \tilde w)^2 \tilde x_2^2 } \quad .
\end{equation}
In equation~(\ref{eq:dL2D_gen}) the averaging is now carried
over the distribution of the eigenvalues with the measure 
$\pi (\lambda_1 - \lambda_2 )$ \citep{1970Ap......6..320D}
that accounts for eigenvalues being sorted, $\lambda_1 \ge \lambda_2$:
\begin{equation}
\frac{\partial {\cal L}}{\partial \eta} = 
\frac{8\cdot 2\cdot \pi}{(2 \pi)^{3} \sqrt{1-\gamma^2}} \exp \left[-\eta^2/2 \right] 
\int\!\!\!\! \int \!\!\!\!\int \!\!\!\!\int (\lambda_1-\lambda_2) d \lambda_1 d \lambda_2 d \tilde x_1 d \tilde x_2 
 \left| \tilde{\nabla s} \right| \delta_{\rm D}(\tilde s)
 \exp \left[-\frac{(\tilde u-\gamma \eta)^2}{2 (1-\gamma^2)} - 
 4 \tilde w^2  - \tilde x_1^2- \tilde x_2^2 \right]\,,
\end{equation}
or in terms of $\tilde u, \tilde w$
\begin{equation}
\frac{\partial {\cal L}}{\partial \eta} = 
\frac{16}{(2 \pi)^{2} \sqrt{1-\gamma^2}} \exp \left[-\eta^2/2 \right] 
\int\!\!\!\! \int \!\!\!\!\int \!\!\!\!\int |\tilde w| d \tilde u d \tilde w d \tilde x_1 d \tilde x_2 
 \left| \tilde{\nabla s} \right| \delta_{\rm D}(\tilde s)
 \exp \left[-\frac{(\tilde u-\gamma \eta)^2}{2 (1-\gamma^2)} - 
 4 \tilde w^2  - \tilde x_1^2 - \tilde x_2^2\right]\,. \label{eq:toto}
\end{equation}

In the argument of the delta-function in equation~(\ref{eq:toto})
$\tilde w$ can be zero only at special field
points
, not at a generic point on a skeleton.
So vanishing $\tilde s$ requires either $\tilde x_1=0$ or
$\tilde x_2=0$ which describes, as expected, that in the Hessian eigenframe
one of the component of the gradient vanishes on the critical line.
Since we have already chosen the coordinates so that the direction ``1'' is 
aligned with the largest eigenvalue and the critical lines can go in both 
eigen-directions, these two possibilities add up:\footnote{If we do not sort the eigenvalues and, thus, do 
not restrict the  $\tilde w$ to be non-negative, then the notions of first and second direction
are undefined, and we could choose now that the skeleton goes in, say,
the first direction and $\tilde x_2=0$. We will loose here factor of two which
is recovered by having to extend $\tilde w$ integration to negative values}
\begin{equation} 
\frac{\partial {\cal L}}{\partial \eta} = 
\frac{4 \sqrt{2}}{(2 \pi)^{3/2}\sqrt{1-\gamma^2}} \exp \left[-\eta^2/2 \right] 
\int_0^\infty \!\! d \tilde w \tilde w 
\int_{-\infty}^\infty \!\!\!\! d\tilde u
\left( \left| 2 \tilde w + \tilde u \right| + 
\left| 2 \tilde w - \tilde u \right|  \right)
\exp \left[-\frac{(\tilde u-\gamma \eta)^2}{2 (1-\gamma^2)} - 
 4 \tilde w^2 \right] \quad .
\label{eq:expl_hessian_deriv}
\end{equation}
Note that $|\tilde w - \tilde u/2 | = |\lambda _1| = |\tilde x_{11}|$
and $|\tilde u/2 + \tilde w| = |\lambda _2| = |\tilde x_{22}|$.
That is, the length
of the critical lines per unit volume is given by the average absolute value
of the
Gaussian curvature of the field in the space orthogonal to the skeleton,
given that in stiff approximation the direction of the skeleton is assumed to
coincide with the gradient of the field.
The reason for this is clear - the higher the curvature, the closer the next
neighbouring segment of the skeleton can be, thus increasing the flux {\sl i.e.} the length per unit volume.
If we replace $\tilde w \to -\tilde w $ in the second integral, we return
to the formula~(\ref{eq:2Dfinal}) 
with integration over both positive and negative $\tilde w$.
The integrated length of the critical lines 
is reduced to
\begin{equation}
L = \frac{2 \sqrt{2}}{\pi} \int_0^\infty \tilde w d \tilde w
\int_{-\infty}^\infty \!\!\! d\tilde u  \left( 
\left|2 \tilde w + \tilde u \right| + \left|2 \tilde w - \tilde u \right | \right)
 \exp \left[-\frac{\tilde u^2}{2} - 4 \tilde w^2   \right]\,.
 \label{eq:deriv_explicit}
\end{equation}
equations (\ref{eq:expl_hessian_deriv}) and (\ref{eq:deriv_explicit}) are the
results of the stiff approximation for the threshold dependent differential
and the integrated lengths  of the critical lines in 2D respectively.

\subsection{Primary critical lines in 2D: Skeleton and anti-Skeleton.}\label{sec:2D}
The local skeleton is the subset of all the critical lines, which includes the parts
that appear as the ridges in the field profile, rather than the
valleys. This subset is described by the constraints  that the skeleton lines should go
along the largest eigenvalue $\lambda_1 $ and, in addition, that
this direction has the smallest curvature, $|\lambda_1| \le |\lambda_2|$.
The anti-skeleton is a mirror structure describing the valley of the field and
in all the results can be obtained by replacing $\eta \to -\eta$ in the
formulae for the skeleton.

\begin{figure*}
  \centering 
  \includegraphics[angle=0,width=7cm]{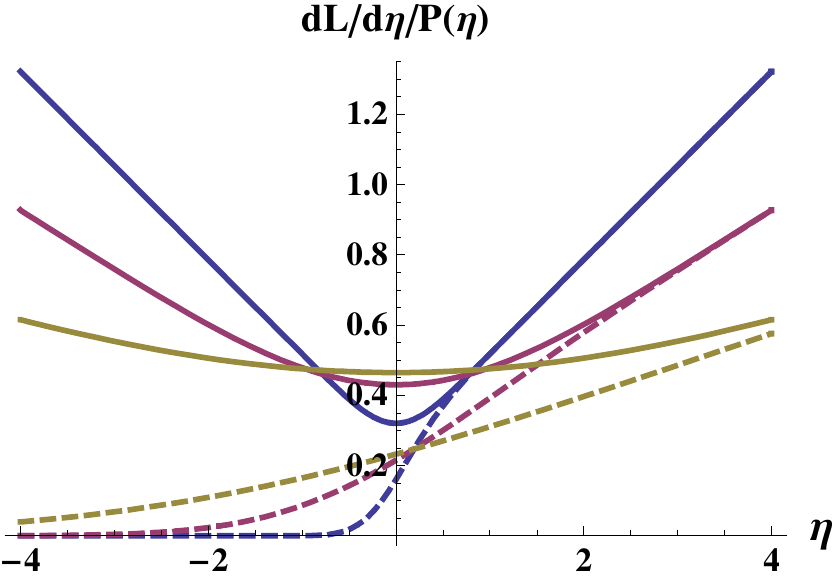}
  \hspace{1cm}
  \includegraphics[angle=0,width=7cm]{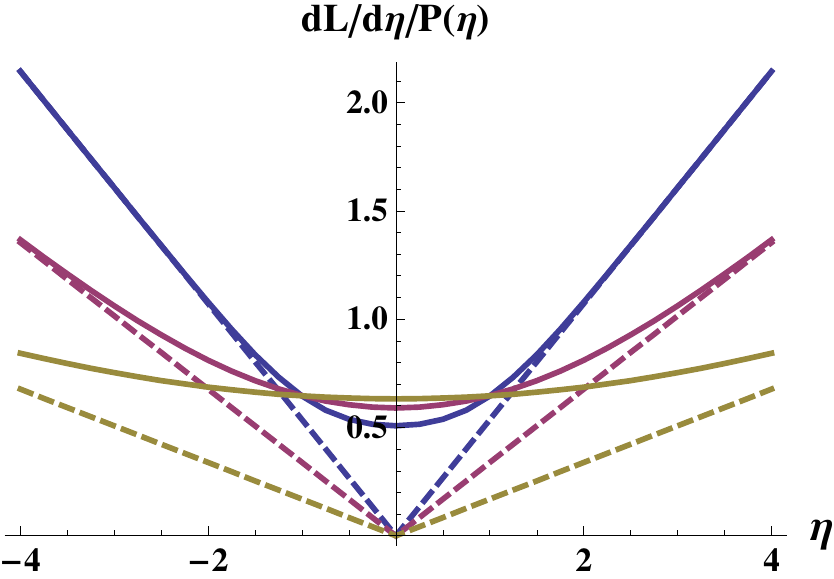}
\caption{{\sl left}:
${\partial {\cal L}^\mathrm{skel}}/{\partial \eta}/P(\eta)$ (dashed)
  and ${\partial {\cal L}^\mathrm{skel+antiskel}}/{\partial \eta}/P(\eta)$ 
  (solid) in 2D for the spectral parameter values $\gamma=0.3,0.6,0.95$.
  {\sl Right:} ${\partial {\cal L}}/{\partial \eta}/ {P(\eta)}$ (solid)
  and its asymptotic behaviour  
  (dashed) in 2D for the same spectral parameter values
  }
\label{fig:dLdtskel}
\label{fig:dLdt2Dall}
\end{figure*}

To derive the expression for the skeleton differential length 
let us return to equation~(\ref{eq:expl_hessian_deriv}). The critical lines
with $\nabla \rho$ aligned with the largest eigenvalue direction
have $\tilde x_2=0$. Thus, only one term is selected by the $\delta_{\rm D}$-function: it is
$\propto 2 |\lambda_2|=\left| 2 \tilde w + \tilde u \right| $. The magnitude
restrictions translates into $\tilde u \ge 0$, thus
\begin{equation} 
\frac{\partial {\cal L}^\mathrm{skel}}{\partial \eta} = 
\frac{4 \sqrt{2}}{(2 \pi)^{3/2}\sqrt{1-\gamma^2}} \exp \left[-\eta^2/2 \right] 
\int_0^\infty \!\! d \tilde w \tilde w 
\int_{0}^\infty \!\!\!\! d\tilde u
\left( 2 \tilde w + \tilde u \right)  
\exp \left[-\frac{(\tilde u-\gamma \eta)^2}{2 (1-\gamma^2)} - 
 4 \tilde w^2 \right] \quad .
\end{equation}
This result should not be confused with equation~(\ref{eq:deriv_explicit}),
where $\tilde w$ is integrated over full range of 
negative and positive values and
which is strictly equivalent to equation~(\ref{eq:expl_hessian_deriv}), 
counting critical lines aligned both with the lowest and
the largest eigen-directions.
Performing the last two integrals one obtains for the differential length in closed form
\begin{equation}
\frac{\partial {\cal L}^\mathrm{skel}}{\partial \eta} =
\frac{1}{\sqrt{2 \pi}} \exp\left[-\frac{\eta^2}{2}\right]
\left[\frac{1}{8} \left( 1+\frac{2}{\sqrt{\pi}} \gamma \eta \right) 
\left(1 + \mathrm{Erf}\left[\frac{\gamma \eta}{\sqrt{2} \sqrt{1-\gamma^2}}\right] \right)
+\frac{\sqrt{1-\gamma^2}}{2 \sqrt{2} \pi} 
\exp\left(-\frac{\gamma^2 \eta^2}{2 (1-\gamma^2)}\right)
\right]\,,
\label{eq:skeleton}
\end{equation}
and for the integrated skeleton length\footnote{In other words, one expect
to find one segment of skeleton per linear section of $\approx (4.2 R_*)$.}
\begin{equation} 
L^\mathrm{skel} = \frac{1}{8}+\frac{\sqrt{2}}{4 \pi}
 = 0.23754 ~ (\times R_*^{-1}) \quad .
\label{eq:skeleton_tot}
\end{equation} 
Note that modulo the stiff approximation, equation~(\ref{eq:skeleton_tot}) 
gives a universal, spectral parameter independent, scaling. 
Figure~\ref{fig:dLdtskel} demonstrates the threshold behaviour of the
differential lengths for several values of the spectral parameter  $\gamma$.

The most important and robust result of our theory is the behaviour of the 
differential length at high density thresholds
\begin{equation}
\frac{\partial {\cal L}^\mathrm{skel}}{\partial \eta} 
\stackrel{\gamma \eta \to \infty}{\sim}
\frac{1}{\sqrt{2 \pi}} \exp\left[-\frac{\eta^2}{2}\right]
\times \frac{1}{4} \left(
1+\frac{2}{\sqrt{\pi}} \gamma \eta \right)\,,
\label{eq:2Dhigheta}
\end{equation}
It represents a bias similar to 
the one found in \cite{1984ApJ...284L...9K} for the clustering of
high critical points - maxima.  According to the latter, 
the number density of peaks in  regions above high thresholds is higher than
on average.  Similarly, the length density of critical lines above
high threshold is enhanced relative to the mean.
From the point of view of measurements, perhaps a 
more interesting quantity than the differential length is 
the length per unit volume within the regions of high excursions of the field
${\cal L}(> \eta) $. 
In terms of the cumulative length given by equation~(\ref{eq:dLcum}),
${\cal L}(> \eta) = L-{\cal L}(\eta)$.
Its asymptotic behaviour at high $\eta$ for the skeleton 
is found by direct integration of equation~(\ref{eq:2Dhigheta})
\begin{equation}
{\cal L}^{\mathrm{skel}}(> \eta) \sim
\frac{1}{2}\mathrm{Erfc}\left(\frac{\eta}{\sqrt{2}}\right)
\times \frac{1}{4} \left(
1+\frac{2}{\sqrt{\pi}} \gamma \eta \right)\,.
\end{equation}
The first factor here is the fractional volume occupied by these
high excursions of the  field. 
{Note that, at large $\eta$ the differential length divided by the PDF scales like $\eta \gamma/R_*=\eta/R_0$ once the proper scaling with 
$1/R_*$
is introduced. Hence the differential length as a function of $\eta$ together with the total length give access to two characteristic scales $R_0$ and $R_*$. See Appendix~\ref{sec:ND} for a general proof of this result in N dimensions. }

The threshold dependence of  the statistics  of critical lines in the stiff approximation
is determined solely by the spectral parameter $\gamma$. In the limit
$\gamma=0$, 
when the distribution of the second derivatives of the field $\rho$ is
completely independent on the threshold, the length of the skeleton
per unit volume within the regions with $\rho/\sigma_0$ in the interval
$\eta,\eta+d\eta$ is just proportional to the fraction of the unit volume that
these regions occupy.  Completely generally, for any type of critical line,
\begin{equation}
\frac{\partial {\cal L}}{\partial \eta}(\gamma = 0) =
\frac{L}{\sqrt{2 \pi}} \exp\left[-\frac{\eta^2}{2}\right]\,.
\end{equation}
When $\gamma \to 1$ the trace of the Hessian $u$ becomes uniquely determined 
by the field level $\eta$ (recall equation (\ref{eq:defgam})).
For over-dense regions with positive $\eta$
equation~(\ref{eq:2Dhigheta}) is exact for $\gamma=1$, while no skeleton exists
in under-dense regions in this limit. 

Near zero (mean density) threshold the dependence of 
${\partial {\cal L}^\mathrm{skel}}/{\partial \eta} $ is 
\begin{equation}
\frac{\partial {\cal L}^\mathrm{skel}}{\partial \eta} 
\stackrel{\eta \to 0}{\sim}
\frac{1}{\sqrt{2 \pi}} \exp\left[-\frac{\eta^2}{2}\right] \times
\frac{1}{4}\left( 
\frac{1}{2}+ \frac{\sqrt{2}\sqrt{1-\gamma^2}}{\pi} + \frac{1}{\sqrt{\pi}}
\frac{1+\sqrt{1-\gamma^2}}{\sqrt{1-\gamma^2}} \gamma \eta \right)\,.
\label{eq:2Dloweta}
\end{equation}
Its details, in particular a step-like cutoff at negative 
$\eta$ when $\gamma \to 1$, are sensitive to the definition of the primary
lines.
In under-dense regions with large negative densities the skeleton
is exponentially suppressed.

Starting from equation~(\ref{eq:skeleton}) with $\eta \to -\eta$
for anti-skeleton, we obtain for the union of both primary critical lines 
\begin{equation}
\frac{\partial {\cal L}^\mathrm{skel+antiskel}}{\partial \eta} =
\frac{1}{\sqrt{2 \pi}} \exp\left[-\frac{\eta^2}{2}\right]
\left[\frac{1}{4} + \frac{1}{2 \sqrt{\pi}}  
\mathrm{Erf}\left(\frac{\gamma \eta}{\sqrt{2} \sqrt{1-\gamma^2}} \right)
\gamma \eta
+\frac{\sqrt{1-\gamma^2}}{\sqrt{2} \pi} 
\exp\left(-\frac{\gamma^2 \eta^2}{2 (1-\gamma^2)}\right)
\right]\,,
\label{eq:skel+anti}
\end{equation}
with twice the integrated length 
\begin{equation}
L^\mathrm{skel+antiskel} =\frac{1}{4}
+ \frac{1}{\sqrt{2} \pi} = 0.47508 ~ ( \times R_*^{-1}) \quad .
\label{eq:Ltot_skel+antiskel_2D}
\end{equation}
This function is now symmetric in $\eta$ with the skeleton providing
the dominant contribution described by equation~(\ref{eq:2Dhigheta})
in over-dense regions of space,  and the anti-skeleton dominating the 
under-dense regions. Near the mean, zero, threshold of the field, both
critical lines are present
\begin{equation}
\frac{\partial {\cal L}^\mathrm{skel+antiskel}}{\partial \eta} 
\stackrel{\eta \to 0}{\sim} 
\frac{1}{\sqrt{2 \pi}} \exp\left[-\frac{\eta^2}{2}\right]
\left(\frac{1}{4} + \frac{\sqrt{1-\gamma^2}}{\sqrt{2} \pi} 
+ \frac{\gamma^2 \eta^2}{2 \sqrt{2} \pi \sqrt{1-\gamma^2}}
\right)\,.
\end{equation}

\subsection{Secondary critical lines in 2D  }
Secondary critical lines do not allow for a full analytical treatment and
are investigated in Appendix~\ref{app:2Dcrit}. They are particularly important
near zero threshold, since at this transitional regime the exact behaviour of
primary or secondary lines depends significantly on our somewhat arbitrary
separation of the critical lines in types.  
In this paper we are tracking the skeleton --- density ridges --- as primary
lines emanating from the maxima, until the largest eigenvalue ceases to be
the shallowest. Alternative definition may, for example, somewhat extend
the skeleton at the expense of secondary lines at lower densities 
as long as all the eigenvalues transverse to
the gradient are negative, i.e until $\lambda_2$ becomes positive.
As an advantage, the differential length of the skeleton and the corresponding
secondary lines defined this way would not exhibit
inflections at low densities that can be seen in 
Figures~\ref{fig:dLdtskel} and \ref{fig:dLdtsec} for high $\gamma$'s.
But the downside is that then one looses the ability to describe
the primary lines analytically in a closed form. 
At the high density excursions the properties of the skeleton 
remain robust with respect to the variations in their exact definition.

However the important advantage of the definition of the primary
lines adopted in this paper lies deeper. The magnitude of the eigenvalue along
the direction transverse  to the gradient is connected  to the stability of
these trajectories near the critical lines and to their possible bifurcations.
This is discussed  in  part in Section~\ref{sec:bif}.

Let us summarize the results for the total set of critical lines,
primary and secondary combined, which are, of course, universal whatever the definition of 
 the separate types. Summing up the results of this Section with
the corresponding ones in Appendix~\ref{app:2Dcrit} 
\begin{eqnarray}
\label{eq:totalL_2D}
L & = &
\frac{\sqrt{2}+\mathrm{acot}(\sqrt{2})}{\pi} = 0.646071~(\times R_*^{-1})\,. \\
\frac{\partial {\cal L}}{\partial \eta} 
& \stackrel{\eta \to \infty}{\sim} &
\frac{1}{\sqrt{2 \pi}} \exp\left[-\frac{\eta^2}{2}\right]
\times \frac{\gamma \eta}{\sqrt{\pi}}  \,, 
\label{eq:2Dallhigheta} \\
\frac{\partial {\cal L}}{\partial \eta} 
& \stackrel{\eta \to 0}{\sim} &
\frac{1}{\sqrt{2 \pi}} \exp\left[-\frac{\eta^2}{2}\right] \times
\left(\frac{\sqrt{2(1-\gamma^2)} 
+ \mathrm{acot}\left(\sqrt{2(1-\gamma^2)}\right)}{\pi}
+ \frac{\sqrt{2(1-\gamma^2)}}{\pi (3-2\gamma^2)} (\gamma \eta)^2
\right) \,.
\end{eqnarray}
The full behaviour of the total differential length is presented in 
Figure~\ref{fig:dLdt2Dall}. One should note the linear asymptotic behaviour
at high density levels and  the regular quadratic behaviour near zero density
threshold\footnote{For all $\gamma$ but $\gamma=1$, for which 
\begin{equation}
\frac{\partial {\cal L}}{\partial \eta}(\gamma \to 1, \eta \to 0) \sim
\frac{1}{\sqrt{2 \pi}}  \exp \left[-\eta^2/2 \right] 
\left(\frac{1}{2} + \frac{1}{3 \sqrt{\pi}} |\eta|^3
- \frac{1}{10 \sqrt{\pi}} |\eta|^5 \ldots \right)\,.
\end{equation}
}.
Finally recall that in the section~\ref{sec:diff2D} we have omitted almost
everywhere a $1/R_*$ factor for the quoted lengths and differential lengths.

\section{ Critical lines of 3D fields}\label{sec:diff3D}
In three dimensions, we carry the computations directly in the eigenframe
of the Hessian, following closely the derivation of Sections~ \ref{sec:crit2D}  and \ref{sec:2D}. We present the formalism first for all the critical lines and
then narrow our focus to the primary ones.

\subsection{The length of the critical lines of 3D fields} \label{sec:crit3D}
In 3D, let us use the variables $\tilde u=-(\lambda_1+\lambda_2+\lambda_3),
\tilde w = (\lambda_1-\lambda_3)/2,
\tilde v = (2 \lambda_2 - \lambda_1 - \lambda_3)/2 $.
In the Hessian eigenframe
\begin{equation}
\tilde s^1= (\lambda_2-\lambda_3) \tilde x_2 \tilde x_3 
=(\tilde w + \tilde v)  \tilde x_2 \tilde x_3 \,,\quad
\tilde s^2 = (\lambda_3-\lambda_1) \tilde x_1 \tilde x_3 
=-2 \tilde w \tilde x_1 \tilde x_3 \,, \quad
\tilde s^3 = (\lambda_1-\lambda_2) \tilde x_1 \tilde x_2 = 
(\tilde w - \tilde v)  \tilde x_1 \tilde x_2 \,,
\label{eq:S_in3D}
\end{equation}
and 
\begin{eqnarray}
\widetilde{\nabla s}^1 &=& 
\left\{0,~\lambda_2 (\lambda_2-\lambda_3) \tilde x_3,~
\lambda_3 (\lambda_2-\lambda_3) \tilde x_2 \right\} = 
\left\{
0,~ -\frac{1}{3} (\tilde u - 2 \tilde v)(\tilde w + \tilde v)\tilde x_3,
~-\frac{1}{3}(\tilde u + \tilde v + 3 \tilde w)(\tilde w + \tilde v)\tilde x_2
\right\}\,, \nonumber \\
\widetilde{\nabla s}^2 &=&
\left\{\lambda_1 (\lambda_3-\lambda_1) \tilde x_3,~0,~
\lambda_3 (\lambda_3-\lambda_1) \tilde x_1 \right\} =
\left\{
\frac{2}{3} (\tilde u + \tilde v - 3 \tilde w) \tilde w  \tilde x_3, ~0,
~\frac{2}{3}(\tilde u + \tilde v + 3 \tilde w) \tilde w  \tilde x_1
\right\} \,,\nonumber \\
\widetilde{\nabla s}^3 &=&
\left\{\lambda_1 (\lambda_1-\lambda_2) \tilde x_2,~
\lambda_2 (\lambda_1-\lambda_2) \tilde x_1,~0\right\} =
\left\{
~-\frac{1}{3}(\tilde u + \tilde v - 3 \tilde w)(\tilde w - \tilde v)\tilde x_2,
~-\frac{1}{3} (\tilde u - 2 \tilde v)(\tilde w - \tilde v)\tilde x_3,~0
\right\}\,.
\label{eq:gradS_in3D}
\end{eqnarray}
In the eigenvalue space 
the measure is $ 2 \pi^2
|(\lambda_1-\lambda_2)(\lambda_2-\lambda_3)(\lambda_3-\lambda_1)| $
and the eigenvalues are considered sorted. 
For sorted eigenvalues
the choice of the directions has been fixed and the $(s^2,s^3),(s^1,s^3)$
and $(s^1,s^2)$ pairs of surfaces describe different possibilities for the
critical line. Those choices add together in the average integrated length.
Using the variable $\tilde w, \tilde v$ the condition of eigenvalues being sorted is
$\tilde w \ge 0, \;\; -\tilde w \le \tilde v \le \tilde w$.

\begin{figure*}
  \centering
  \subfigure{\includegraphics[angle=0,width=7cm]{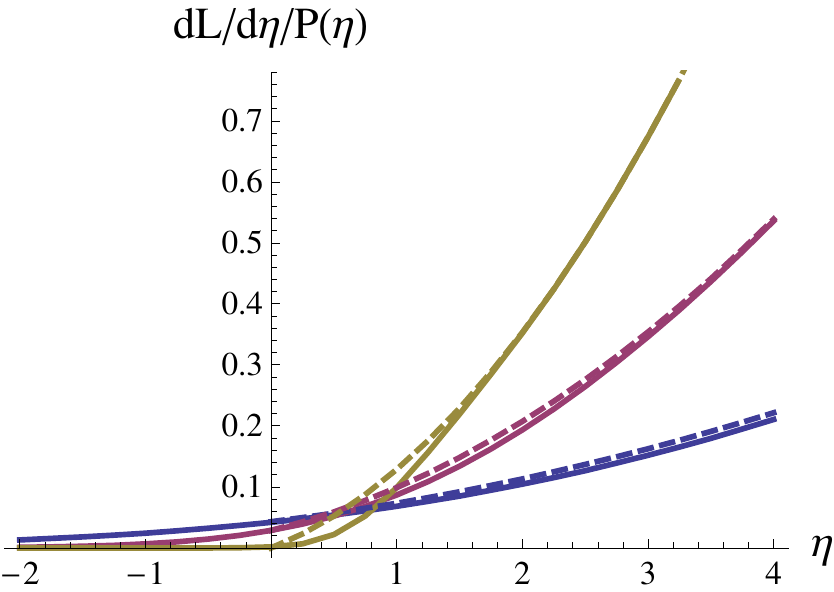}} 
  \hspace{1cm}
  \subfigure{\includegraphics[angle=0,width=7cm]{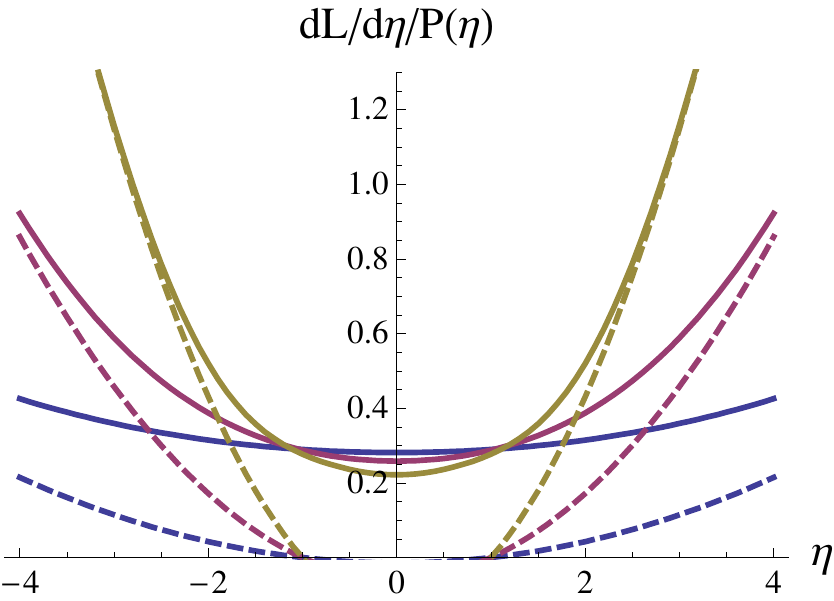}}
  \caption{{\sl Left:} The skeleton
   ${\partial {\cal L}^\mathrm{skel}}/{\partial \eta}/ {P(\eta)}$ (solid) and
   its asymptotic behaviour at high density thresholds (dashed) in 3D.
   The anti-skeleton is described by the curves symmetric with respect
  to a reflection of $\eta$.
  {\sl Right:}  ${\partial {\cal L}}/{\partial \eta}/ {P(\eta)}$
  its asymptotic behaviour for the total set of critical lines.
  The spectral parameter values are (from bottom to top at
  high $\eta$) $\gamma=0.3,0.6,0.95$.}
\label{fig:dLdt3Dskel}
\end{figure*}

Let us consider the critical lines that are the intersections of $(s^2,s^3)$.
Their differential length is given by
\begin{eqnarray}
\frac{\partial {\cal L}}{\partial \eta} = 2 \pi^2 \cdot \frac{3}{2} \cdot
\frac{3^{3/2} 15^2 5^{1/2}} {(2 \pi)^{5} \sqrt{1-\gamma^2}}
\exp \left[-\frac{1}{2} \eta^ 2\right] \!\! \!\!\!&&\!\!\!\!\!\!
 \int \!\left|(\lambda_1-\lambda_2)(\lambda_2-\lambda_3)(\lambda_3-\lambda_1)\right| 
 d\lambda_1 d \lambda_2 d \lambda_3 d \tilde x_1 d \tilde x_2 d \tilde x_3 
 \left| \widetilde{\nabla s}^2 \times \widetilde{\nabla s}^3 \right|
\delta_{\rm D}(\tilde s^2) \delta_{\rm D}(\tilde s^3)
\nonumber  \\
 &&\times \exp \left[-\frac{(\tilde u-\gamma \eta)^2}{2 (1-\gamma^2)} - 
 \frac{15}{2} \tilde w^2  - \frac{5}{2} \tilde v^2 - \frac{3}{2}\tilde x_1^2 - \frac{3}{2}\tilde x_2^2 - \frac{3}{2} \tilde x_3^2 \right]\,.
\label{eq:3DHess_start}
\end{eqnarray}
Integration over $\delta_{\rm D} (\tilde s_2)$ and $\delta_{\rm D} (\tilde s_3)$
leads to the only possibility $\tilde x_2 =0, \tilde x_3 = 0$. That
is, the choice of the surface $S_2$ and $S_3$ in the Hessian
eigenframe describes the skeleton along which the gradient is 
aligned with the direction $1$, correspondent to the largest eigenvalue,
while in the directions $2$ and $3$  the
components of the gradient of the field vanish.  With $\tilde x_2=
\tilde x_3 =0 $ we get a simple expression for
\begin{equation}
\left | \widetilde{\nabla s}_2 \times \widetilde{\nabla s}_3 \right| = 
\left| \lambda_2 \lambda_3 (\lambda_3-\lambda_1)(\lambda_1-\lambda_2) \right|
\tilde x_1^2= 
\frac{2}{9} \left| (\tilde u - 2 \tilde v) (\tilde u + \tilde v + 3 \tilde w)
(\tilde w - \tilde v) \tilde w \right | \tilde x_1^2\,,
\end{equation}  
while the subsequent integration over $\tilde x_2$ and $\tilde x_3$ using 
$\delta_{\rm D}$-functions and afterwards over $\tilde x_1$ gives
\begin{equation}
\frac{\partial {\cal L}^1}{\partial \eta} = 
\frac{3^4 5^{5/2}} {16 \pi^2 \sqrt{2 \pi(1-\gamma^2)}}
\exp \left[-\frac{1}{2} \eta^ 2\right] \!\!
\int
d \lambda_1 d \lambda_2 d \lambda_3  
 (\lambda_1-\lambda_2)(\lambda_2-\lambda_3)(\lambda_3-\lambda_1)
|\lambda_2 \lambda_3|  
\exp \left[-\frac{(\tilde u-\gamma \eta)^2}{2 (1-\gamma^2)} - 
 \frac{15}{2} \tilde w^2  - \frac{5}{2} \tilde v^2  \right] \; .
\label{eq:3DHess_next}
\end{equation}
Notice again that what the integrand involves the Gaussian curvature
in the direction orthogonal to the gradient, which in stiff approximation
is the direction of the filament itself.
The contributions of the critical lines directed along the second and third
eigen-direction is given by similar considerations and are added together
when all critical lines are considered. Changing variables one finally obtains
\begin{equation}
\frac{\partial {\cal L}}{\partial \eta} = 
\frac{3^3 5^{5/2}\exp \left[\eta^ 2/2\right] } {4 \pi^2 \sqrt{2 \pi(1-\gamma^2)}} 
\int  d \tilde u d \tilde w d \tilde v \;
\tilde w (\tilde w^2 - \tilde v^2) 
\left(|\lambda_2 \lambda_3|+|\lambda_1 \lambda_3|+|\lambda_1 \lambda_2|\right) 
 \exp \left[-\frac{(\tilde u-\gamma \eta)^2}{2 (1-\gamma^2)} - 
 \frac{15}{2} \tilde w^2  - \frac{5}{2} \tilde v^2  \right]\,, \label{eq:3Ddiff}
\end{equation}
while the integrated length is
\begin{equation}
L = \frac{3^3 5^{5/2}} {4 \pi^{2}} 
\int  d \tilde u d \tilde w d \tilde v \;
\tilde w (\tilde w^2 - \tilde v^2) 
\left(|\lambda_2 \lambda_3|+|\lambda_1 \lambda_3|+|\lambda_1 \lambda_2|\right) 
 \exp \left[-\frac{1}{2}\tilde u^2 - 
 \frac{15}{2} \tilde w^2  - \frac{5}{2} \tilde v^2  \right]\,,
 \label{eq:3Dlength}
\end{equation}
with
\footnote{ One should note the correspondence with the well-known result
for the number density of extrema of the field  \citep{1986ApJ...304...15B}
\begin{displaymath}
N_{\rm ext} \propto 
\int  d \tilde u d \tilde w d \tilde v \;
\tilde w (\tilde w^2 - \tilde v^2) 
|\lambda_1 \lambda_2 \lambda_3| 
 \exp \left[-\frac{1}{2}\tilde u^2 - 
 \frac{15}{2} \tilde w^2  - \frac{5}{2} \tilde v^2  \right] \quad .
\end{displaymath}
which is determined by the mean three-dimensional Gaussian curvature
 $|\lambda_1 \lambda_2 \lambda_3| $.
}
\begin{equation}
|\lambda_2 \lambda_3|+|\lambda_1 \lambda_3|+|\lambda_1 \lambda_2| =
\frac{1}{9}\left[ 
|( \tilde u-2 \tilde v)(\tilde u+\tilde v+3 \tilde w)|
+ |(\tilde u+ \tilde v+3 \tilde w)( \tilde u+ \tilde v-3 \tilde w)|
+ |( \tilde u-2 \tilde v)( \tilde u+ \tilde v-3 \tilde w)| \right] \quad .
\end{equation}
The equations~(\ref{eq:3Ddiff}) and (\ref{eq:3Dlength}) account for all the 
critical lines.  In Figure~\ref{fig:dLdt3Dskel} (right panel)
the results for 3D critical lines are plotted while the discussion of the
corresponding asymptotics is given in the Appendix~\ref{sec:3D_crit}.
We shall now turn our attention to the study of the primary lines and, in
particular, the 3D skeleton that delineates the over dense filamentary structure
and is of  more direct observational interest.

\subsection{Primary critical lines of 3D fields: Skeleton and Anti-Skeleton}

\begin{figure*}
  \centering
  \subfigure{\includegraphics[angle=0,width=5cm]{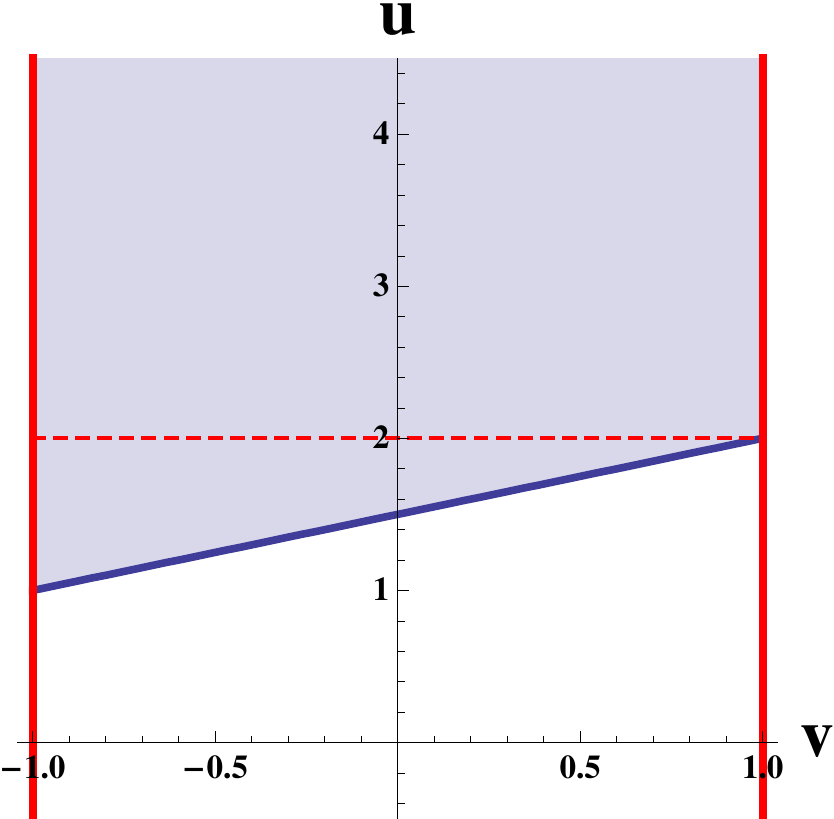}}
  \hspace{1cm}
  \subfigure{\includegraphics[angle=0,width=5cm]{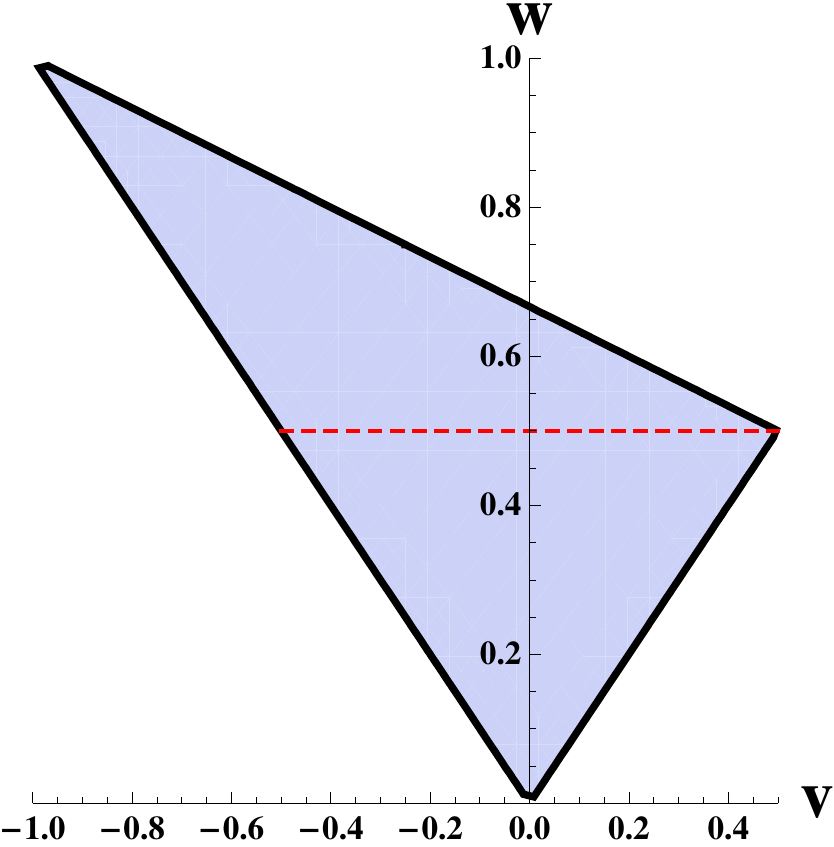}}
  \caption{{\sl Left}: Integration zones in  the $\tilde v, \tilde u$ plane
  for the  3D skeleton analysis.  Variables are given in units of $\tilde w$.
  $\tilde v$ varies from $-\tilde w$ to $+\tilde w$, while $\tilde u$ 
  must be greater than $\frac{1}{2}(\tilde v+3 \tilde w)$. 
  In the allowed shaded region $0 > \lambda_2 \ge \lambda_3$ everywhere.
  Horizontal lines mark the further subdivision of the integration space
  if the order of integration is
  changed according to equation~(\ref{eq:uwv_integral_skel}).
  {\sl Right}: Integration zones in the $(\tilde v , \tilde w )$ plane
  after $\tilde u$ has been mapped to the  $[0-\infty]$ interval.
  Variables are given in the units of $\tilde u$. 
  The lower triangular zone corresponds to the semi-open rectangular band
  above the  red dashed line in the left panel.
  In this region the integrand is given by the first term of
  equation~(\ref{eq:uwv_integral_skel}).
  It dominates the high $\eta$ asymptotics.
   }
\label{fig:uv_plane_skel}
\end{figure*}

The subset of critical lines identified with the skeleton
correspond to the lines with the gradient aligned with the largest eigenvalue
$\lambda_1$ while having  $\lambda_1 + \lambda_2 \le 0$.
In equation~(\ref{eq:3Ddiff}) such lines are described by 
the first term $ \sim |\lambda_2 \lambda_3|$.  
The differential length of the skeleton is then
\begin{eqnarray}
\frac{\partial {\cal L}^\mathrm{skel}}{\partial \eta} \!\!\! &=& \!\!\!\!\!
\frac{\displaystyle 3^3 5^{5/2} \exp \left[- \eta^ 2/2\right] } {4 \pi^2 \sqrt{2 \pi(1-\gamma^2)}} 
\int_0^\infty \!\!\!\!\!\!\ d \tilde w 
\int_{-\tilde w}^{\tilde w} \!\!\! d \tilde v
\int_{\frac{1}{2}(\tilde v+3 \tilde w)}^\infty \!\!\!\!\!\! d \tilde u \;
\tilde w (\tilde w^2 - \tilde v^2) 
\lambda_2 \lambda_3 
 \exp \left[-\frac{(\tilde u-\gamma \eta)^2}{2 (1-\gamma^2)} - 
 \frac{15}{2} \tilde w^2  - \frac{5}{2} \tilde v^2  \right]\,,
\label{eq:dLdt_3D_skel}\\
&=&\!\!\!\! \frac{\displaystyle 3 \cdot 5^{5/2} \exp \left[- \eta^ 2/2\right] } {4 \pi^2 \sqrt{2 \pi(1-\gamma^2)}} 
\int_0^\infty \!\!\!\!\!\!\ d \tilde w 
\int_{-\tilde w}^{\tilde w} \!\!\! d \tilde v
\int_{\frac{1}{2}(\tilde v+3 \tilde w)}^\infty \!\!\!\!\!\!\! d \tilde u \;
\tilde w (\tilde w^2 - \tilde v^2) 
(\tilde u - 2 \tilde v) (\tilde u + \tilde v + 3 \tilde  w)
 \exp \left[-\frac{(\tilde u-\gamma \eta)^2}{2 (1-\gamma^2)} - 
 \frac{15}{2} \tilde w^2  - \frac{5}{2} \tilde v^2  \right]\,.
\nonumber
\end{eqnarray}
The integration in $\tilde v$--$\tilde u $ plane is limited to the region
$\tilde u > \frac{1}{2} (\tilde v + 3 \tilde w)$, 
as shown in the left panel of Figure~\ref{fig:uv_plane_skel}.
The integrated length of the skeleton is 
\begin{equation}
L^{\mathrm{skel}} =0.046186 ~ ( \times R_*^{-2})\,,
\end{equation}
that is, one expect on average one skeleton line crossing a random
$\approx(5 R_*)^2$ surface element.
The results of integration of equation~(\ref{eq:dLdt_3D_skel}) are presented
in the left panel of Figure~\ref{fig:dLdt3Dskel}.
\subsubsection{Asymptotic behaviour at $\gamma\eta \to \infty$}
To study high $\eta$ asymptotes it is useful to change the order of
integration to have the $\tilde u$ integral as the outmost one. The inner
integration in $\tilde v$--$\tilde w$ plane is then carried out over the region
shown in the right panel of Figure~\ref{fig:uv_plane_skel}.
\begin{equation}
\int_0^\infty \!\!\!\!\!\!\ d \tilde w 
\int_{-\tilde w}^{\tilde w} \!\!\! d \tilde v
\int_{\frac{1}{2}(\tilde v+3 \tilde w)}^\infty \!\!\!\!\!\! d \tilde u \; \to
\int_0^\infty \!\! d\tilde u \int_0^{\tilde u/2} \!\!\!\! d\tilde w 
\int_{-\tilde w}^{\tilde w} \!\!\! d \tilde v  + 
\int_0^\infty \!\! d\tilde u \int_{\tilde u/2}^{\tilde u} \!\!\!\! d\tilde w 
\int_{-\tilde w}^{2 \tilde u - 3 \tilde w} \!\!\! d \tilde v
\label{eq:uwv_integral_skel} 
\end{equation}
The last term is exponentially suppressed as $\eta \to \infty$ while
the first one gives 
\begin{eqnarray}
\frac{\partial {\cal L}^{\mathrm{skel}}}{\partial \eta}
&\stackrel{\gamma \eta \to \infty}{\sim}&
\frac{3 \cdot 5^{5/2}\exp \left[-\eta^ 2/2\right] } {4 \pi^2 \sqrt{2 \pi (1-\gamma^2)}} 
\int_0^\infty \!\!\!\!\! d \tilde u \int_0^\infty \!\!\!\!\!\!\!\!\ d \tilde w \!
\int_{-\tilde w}^{\tilde w} \!\!\!\!\! d \tilde v \;
\tilde w (\tilde w^2 - \tilde v^2) 
(\tilde u - 2 \tilde v) (\tilde u + \tilde v + 3 \tilde  w)
 \exp \left[-\frac{(\tilde u-\gamma \eta)^2}{2 (1-\gamma^2)} - 
 \frac{15}{2} \tilde w^2  - \frac{5}{2} \tilde v^2  \right]
\nonumber \\
&\stackrel{\gamma \eta \to \infty}{\sim}&
\frac{1}{\sqrt{2 \pi}} \exp \left[-\frac{1}{2} \eta^ 2\right] 
\frac{(\gamma \eta)^2+{9}(\gamma \eta)/{\sqrt{10 \pi}} +(9/10-\gamma^2)}{6 \pi}\,.
\label{eq:3D_higheta}
\end{eqnarray}
The leading quadratic and the next linear terms 
can be recovered found by replacing $\tilde u \to \gamma \eta$ in the 
pre-exponential factor and treating the exponent as the 
$\delta_{\rm D}$-function. 
A more detailed asymptotic study of this Laplace-type integral is required to
recover the third-order constant term, that also contributes to the accuracy
of the expansion at the level demonstrated in Figure~\ref{fig:dLdt3Dskel}.

One finds that in the leading order in $\eta$ the skeleton
has the differential length growing as $(\gamma \eta)^2$ (see also
Appendix~\ref{sec:ND}) and involves, as expected, a third of all the critical
lines (compare with Appendix~\ref{sec:3D_crit}) in the regions of high 
excursions concentrated around the maxima of the field. However, at
intermediated thresholds, the skeleton constitutes more than a half of
all critical lines, highlighting enhanced importance of the filamentary
dense ridges among other critical lines. 
\footnote{Note  the appearance of the linear in $\gamma \eta$
term in the next to leading order for the skeleton, 
that canceled out for the critical lines.}
%
\subsubsection{Power series at $\eta \to 0$ and Hermite expansion}
Using two alternative series representations of the shifted Gaussian form that
encodes the dependence of the skeleton on the threshold $\eta$
\begin{eqnarray}
\frac{1}{\sqrt{2 \pi(1 - \gamma^2)}}
\exp\left[-\frac{(u - \gamma \eta)^2}{2 (1-\gamma^2)}\right] &=&
\frac{1}{\sqrt{2 \pi}} \exp\left[-\frac{u^2}{2}\right]
\sum_{k=0}^\infty \gamma^k H_k(\eta) H_k(u)\,, \\
&=&\frac{1}{\sqrt{2 \pi (1 - \gamma^2 )}} 
\exp\left[-\frac{u^2}{2(1-\gamma^2)}\right]
\sum_{k=0}^\infty \frac{1}{\sqrt{k!}}
\left(\frac{\gamma}{\sqrt{1-\gamma^2}} \eta\right)^k
H_k\left(\frac{u}{\sqrt{1-\gamma^2}}\right)\,,
\end{eqnarray}
we obtain either power series or Hermite\footnote{we use here the
{\it normalized} Hermite polynomials following probabilistic definition,
${1}/{\sqrt{2 \pi}}\int_{-\infty}^\infty d u \exp\left[-\frac{1}{2} u^2\right]
H_k(u) H_m(u) =\delta_{mk}$}  \citep{NCD} expansion 
of the differential length 
\begin{equation}
\frac{\partial {\cal L}^{\mathrm{skel}}}{\partial \eta} = 
\frac{1}{\sqrt{2 \pi}} \exp \left[-\eta^2/2 \right] \times
\left\{
\begin{array}{l}
\sum_{k=0}^\infty A_{k}(\gamma) (\gamma \eta)^{k}\,, \\ \\
\sum_{k=0}^\infty B_{k} \gamma^k H_k(\eta) \,,
\end{array}
\right.
\end{equation}
where
\begin{eqnarray}
\quad A_{k}(\gamma) &\equiv & \!\!\! 
\frac{3 \cdot 5^{5/2}}{4 \pi^2 \sqrt{k!} (1-\gamma^2)^{\frac{k+1}{2}}} 
\int_0^\infty \!\!\!\!\!\!\ d \tilde w 
\int_{-\tilde w}^{\tilde w} \!\!\! d \tilde v
\int_{\frac{1}{2}(\tilde v+3 \tilde w)}^\infty \!\!\!\!\!\! d \tilde u \;
\tilde w (\tilde w^2 - \tilde v^2) 
(\tilde u - 2 \tilde v) (\tilde u + \tilde v + 3 \tilde  w) \nonumber \\ &\times&
 \exp \left[-\frac{\tilde u^2}{2 (1-\gamma^2)} - 
 \frac{15}{2} \tilde w^2  - \frac{5}{2} \tilde v^2  \right]
H_{k}\left(\frac{\tilde u}{\sqrt{1-\gamma^2}} \right)
\quad ,
\label{eq:3Dskel_tto0_A}
\end{eqnarray}
and
\begin{equation}
\quad B_{k} \equiv 
\frac{3 \cdot 5^{5/2}}{4 \pi^2} 
\int_0^\infty \!\!\!\!\!\!\ d \tilde w 
\int_{-\tilde w}^{\tilde w} \!\!\! d \tilde v
\int_{\frac{1}{2}(\tilde v+3 \tilde w)}^\infty \!\!\!\!\!\! d \tilde u \;
\tilde w (\tilde w^2 - \tilde v^2) 
(\tilde u - 2 \tilde v) (\tilde u + \tilde v + 3 \tilde  w) 
 \exp \left[-\frac{\tilde u^2}{2} - 
 \frac{15}{2} \tilde w^2  - \frac{5}{2} \tilde v^2  \right]
H_{k}\left(\tilde u \right)
\quad .
\label{eq:3Dskel_tto0_B}
\end{equation}
These two expansions are similar but distinct. The power-law expansion is suitable
for an accurate analysis of the differential length near zero threshold for all
$\gamma < 1$.  On the other hand, the expansion in orthogonal Hermite
polynomials is useful as an approximation over an extended range of thresholds.
Both series are improper for $\gamma=1$.

\begin{figure*}
 \centering
 \subfigure{\includegraphics[angle=0,width=5cm]{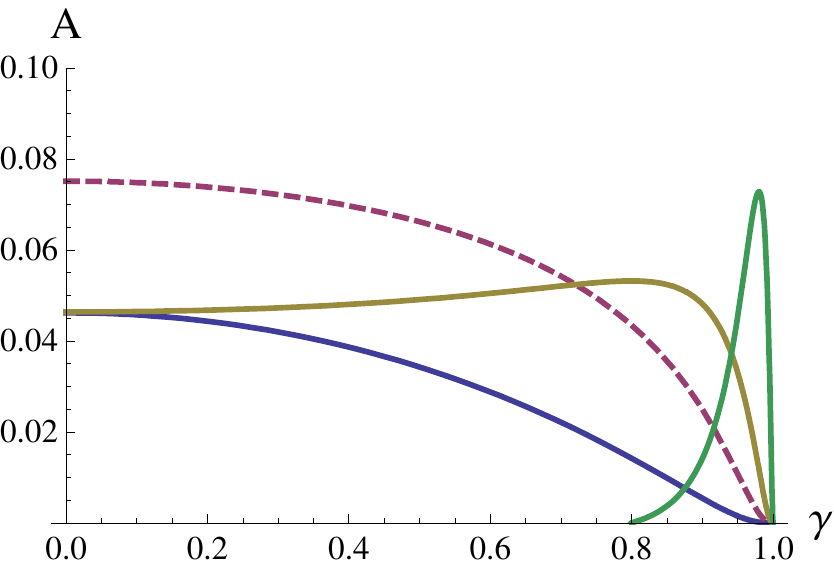}} \hspace{0.5cm}
 \subfigure{\includegraphics[angle=0,width=5cm]{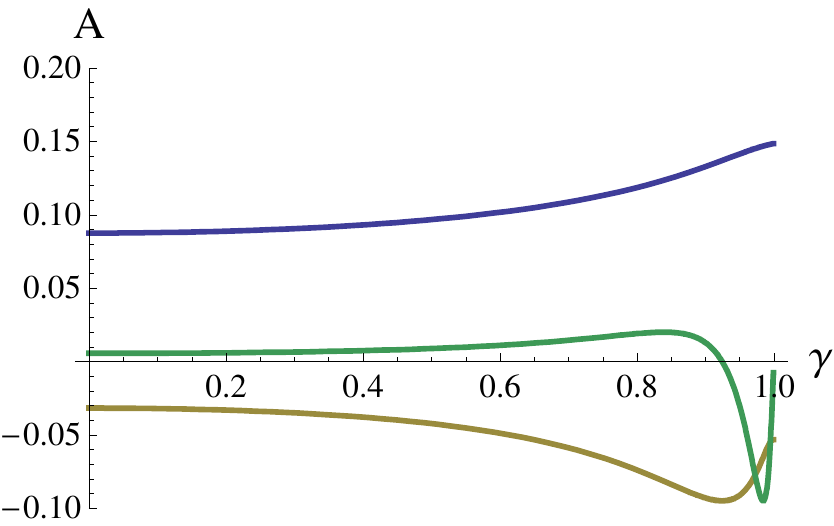}} \hspace{0.5cm}
 \subfigure{\includegraphics[angle=0,width=5cm]{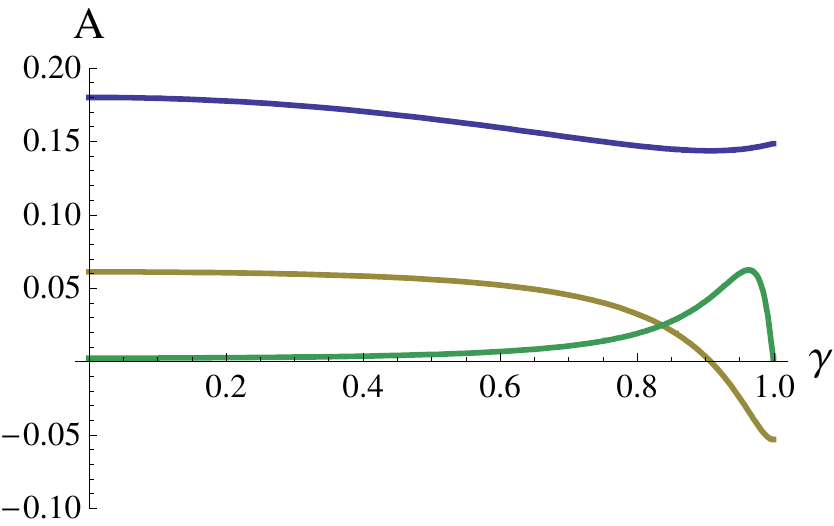}}
 \caption{
 First coefficients of low-$\eta$ power expansion
 ($A_0$ -- blue, $A_2$ -- yellow,
 $A_4$ -- green) of the differential lengths of the 3D skeleton ({\sl left}),
 inter-skeleton ({\sl middle}) and total set of primary critical lines
 ({\sl right}). The odd terms (e.g. $A_1$ -- dashed) are present only for
 the asymmetric case of the primary lines.
   }
\label{fig:Askel}
\end{figure*}

Although  these coefficients can be computed analytically, their expressions
are too cumbersome. Instead, we plot  several leading ones in
Figure~\ref{fig:Askel}. 
Remarkably, the power in Hermite expansion is concentrated in a  few low order 
terms, in particular, $k=0,1,2,3$ for the skeleton, with subsequent terms 
forming a slowly decaying oscillating series. This finding confirms in 3D
the conjecture of \cite{NCD}. The contribution of the
first three most dominant terms, $\sum_{k=0}^2 B_k H_k(\eta)=
0.0462+0.0751 \gamma\eta + 0.0464 \gamma^2(\eta^2-1)$ has the same structure
and remarkably similar coefficients as the high $\eta$
 asymptotics of equation~(\ref{eq:3D_higheta}) 
which evaluates to $0.0477+0.0852 \gamma\eta+0.0531\gamma^2(\eta^2-1) $.
 This explains why the high $\eta$ asymptotics provides 
a visually good fit through all thresholds when $\gamma$ is not too high.
At $\gamma \to 1$, the oscillatory tail of Hermite series provides 
the correction that reflects the irregular nature of the expansion in this 
limit. 

The power series expansion reflects the features of the Hermite expansion.
Starting, by definition, at
$A_0(0)=B_0={\cal L}_\mathrm{tot}^\mathrm{skel}=0.0462$, $A_0$
behaves as $A_0(\gamma) \approx {\cal L}_{\rm tot} (1-\gamma^2)$
over most of the $\gamma$ range.
Coupled with $A_2(\gamma)\approx {\rm const}= {\cal L}_{\rm tot}$
and $A_1(\gamma) \approx 0.0751 \sqrt{1-\gamma^2}$ we get
for the  first three orders 
$\sum_{k=0}^2 A_{k}(\gamma) (\gamma \eta)^{k} \approx
0.0462 + 0.0751 \sqrt{1-\gamma^2} \gamma \eta +
0.0462 \gamma^2 (\eta^2-1) $, close both to the  Hermite expansion
and to the high $\eta$ law for moderate $\gamma$. On the other hand,
the power series expansion explicitly demonstrates the increasing importance of 
higher-order terms for $\gamma > 0.8$. 

\subsection{Primary critical lines of 3D fields: Inter-Skeleton and the 
overall behaviour}
The intermediate primary critical lines are associated with saddle-like regions 
where the largest eigenvalues in magnitude are $\lambda_1 \ge 0 $ and 
$\lambda_3 \le 0$, and have opposite signs, and the shallowest direction  
aligned with the gradient is the second one with 
$  -\lambda_1 < \lambda_2 < -\lambda_3 $.
Their appearance reflects the  complexity of critical lines in 
space of more than two dimensions.

\begin{figure}
  \centering
 \subfigure{\includegraphics[angle=0,width=7cm]{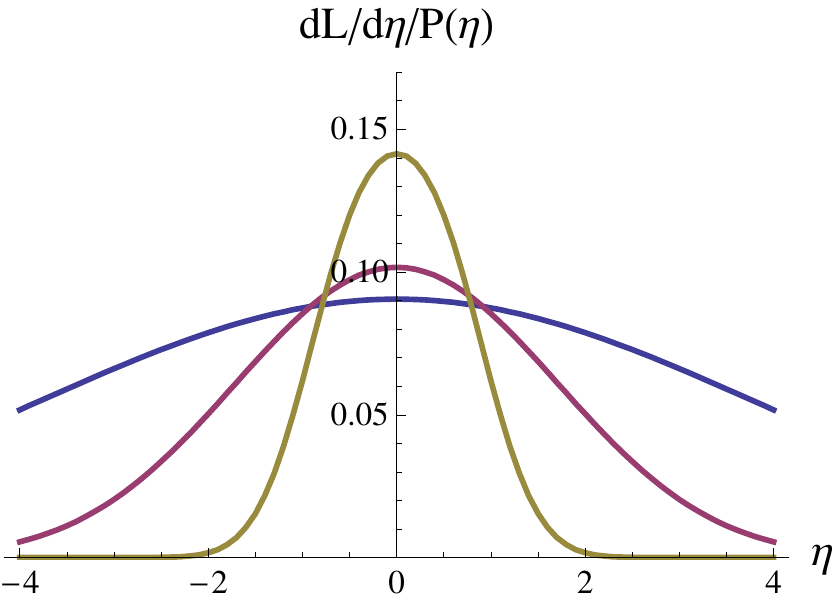}} 
 \hspace{1cm}
 \subfigure{\includegraphics[angle=0,width=7cm]{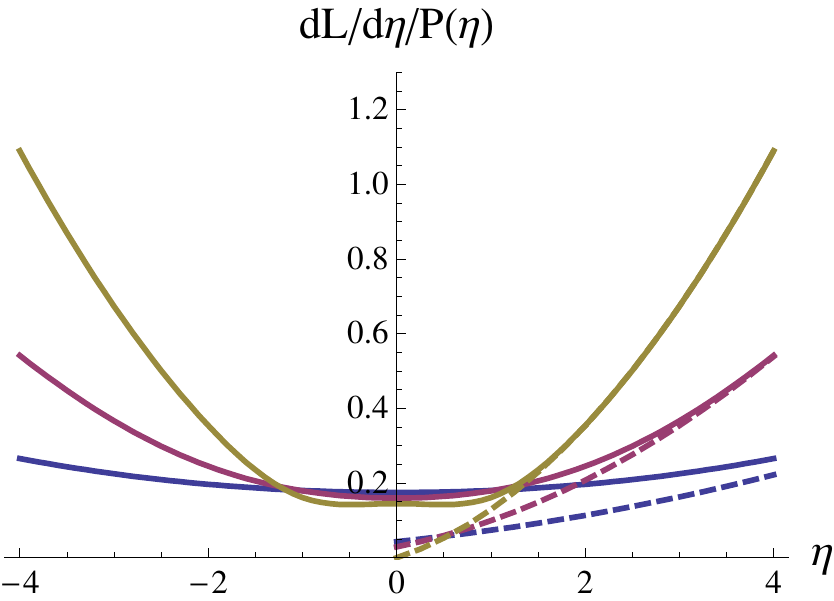}}
    \caption{Differential length ${\partial {\cal L}}/{\partial \eta}/P(\eta)$
    of the intermediate ({\sl left panel}) and combined primary
    lines ({\sl right panel}) as function of the threshold $\eta$ in 3D. 
    Different curves from blue to yellow correspond to
  the spectral parameter values $\gamma=0.3,0.6,0.95$. The dashed curves, drawn 
  only for positive $\eta$, correspond to high-$\eta$ asymptotic solutions.}
\label{fig:dLdt3Dprim}
\end{figure}

The differential length of the intra-skeleton computed in the stiff
approximation is presented in Figure~\ref{fig:dLdt3Dprim}.
The conditions for intermediate lines are prevalent for the
regions of the field of moderate values - within $2\sigma$ ($|\eta|<2$
of the zero mean for $\gamma > 0.6$. Although 
the occurrence of the intra-skeleton within these regions is never large 
($ {\partial {\cal L}}/{\partial \eta}/ {P(\eta)}$ is relatively small), 
the  regions  corresponding to a near mean density occupy large fractions of
the total volume, and as the result the total length of the intermediate
skeleton is almost twice that of the skeleton or the anti-skeleton:
\begin{equation}
L^\mathrm{inter} = 0.087533 ~ (\times R_*^{-2})\,.
\end{equation}
It constitutes nearly a half of the total length of the primary critical
lines \begin{equation}
L^\mathrm{prim} = L^\mathrm{skel}+L^\mathrm{antiskel}+L^\mathrm{inter} =
0.179905~ (\times R_*^{-2})\,.
\end{equation}
At high $|\eta|$ thresholds, in very dense regions near maxima or 
under-dense regions near  minima of the field, the intermediate skeleton
is rare. 

The total set of the primary critical line is even more than the skeleton
dominated by the low order terms in Hermite expansion. Indeed,
Figure~\ref{fig:Askel} demonstrates that just the first two terms
(odd orders are absent due to symmetry) in Hermite series are dominant,
$\sum_{k=0}^\infty B_{k} H_{k} (\eta) \approx 
{\cal L}_\mathrm{tot}^\mathrm{prim}\left(1 + 0.340 \gamma^2 (\eta^2-1)\right)$.

\subsection{Validity of the stiff approximation}\label{sec:stiff}
Let us consider the opposite to ``stiff'' regime, when the derivatives of the
Hessian dominate the $\nabla s$, 
\begin{equation}
(\nabla_{m} s^i)^{\rm lax} \approx {\tilde \gamma}^{-1}
\sum_{jkl} \epsilon^{ijk}  x_{jlm} x_{l} x_{k}\,.
\end{equation}
Although not natural for cosmology-inspired spectra, such a situation arises
when the power spectrum has an extended short wave tail with spectral index\footnote{In a cosmological framework this takes place when the density field with
$n < -1$ spectrum is smoothed with a top-hat window.}
$n$ between $-9$ and $-5$.  Such spectra have small $\tilde \gamma$,
$\tilde R \ll R_*$ and there are many inflection points of the field per
extremum. Interestingly, this regime also automatically means that 
the correlation between the gradient and third derivatives of the field
is small.  

Using the Hessian eigenframe formalism, we can obtain the important results
without explicit computation of the differential length. Let us focus on
the critical lines corresponding to the first eigenvalue.
Equations (\ref{eq:S_in3D}) for $\cal S$-surfaces  gives
rise to two $\delta_{\rm D}$-functions,
$\delta_{\rm D}(2 \tilde w \tilde x_1 \tilde x_3)
\delta_{\rm D}((\tilde w - \tilde v)  \tilde x_1 \tilde x_2) $ that after
integration over the transverse gradient components
$\tilde x_2$ and $\tilde x_3$ enforce $\tilde x_2=\tilde x_3=0$,
with the Jacobian factor $1/|2 \tilde w (\tilde w-\tilde v) \tilde x_1^2|$.
The length element in this frame obeys
\begin{equation}
|\nabla s^2 \times \nabla s^3|^{\rm lax} \approx
{\tilde \gamma}^{-2} x_1^4 \left|\sum_{ijmn}
\epsilon_{kmn} \epsilon^{2 i1} \epsilon^{3 j 1} x_{i1m} x_{j1n} \right|
\equiv {\tilde \gamma}^{-2} x_1^4 \psi(x_{klm})\,,
\end{equation}
where the last expression defines the $\psi(x_{klm})$ function.
The differential length is now given by 
\begin{eqnarray}
\frac{\partial {\cal L}}{\partial \eta}^{\rm lax} = &
\displaystyle \frac{1}{ {\tilde R}^2} 
\left\{  \displaystyle\frac{3^3 5^{5/2}} {8 \pi^3 \sqrt{1-\gamma^2}} 
\exp \left[-\frac{1}{2} \eta^ 2\right] 
\int  d \tilde u d \tilde w d \tilde v \; (\tilde w + \tilde v) 
 \exp \left[-\frac{(\tilde u-\gamma \eta)^2}{2 (1-\gamma^2)} - 
 \frac{15}{2} \tilde w^2  - \frac{5}{2} \tilde v^2  \right]  \right\} 
& \times \nonumber \\
&\displaystyle \times \left\{
\int x_1^2 d x_1 d^{10}x_{klm} \psi(x_{klm})
P_1(x_1,x_2=x_3=0,x_{klm}) \right\} &
\label{eq:nonstiff}
\end{eqnarray}
The last integral, with $P_1$ given
by equation~(\ref{eq:defp13D}), is a function of $\tilde \gamma$ only.
The first term  shows that, since the integrand prefactor is independent
on $\tilde u$,
the differential length does not depend on the threshold $\eta$ at large $\eta$
(it does at small $\eta$ only because of non-trivial integration boundaries
dependent on the exact type of critical lines). This is not surprising,
since in this limit, there is little link between the skeleton length and the
second derivatives, the only ones that are correlated with the field value.
Such threshold independent behaviour is not observed in simulations with
cosmological spectra, which argues once again
for the statistical validity of the ``stiff'' approximation.

\subsection{Measurements}\label{sec:measure}

In this section we compare the predictions of the local theory in stiff
approximation with the measurements of the statistical properties of the 
critical lines done on realizations of the Gaussian fields with different
power spectra. 

\begin{figure}
  \centering
  \includegraphics[angle=0,width=14cm]{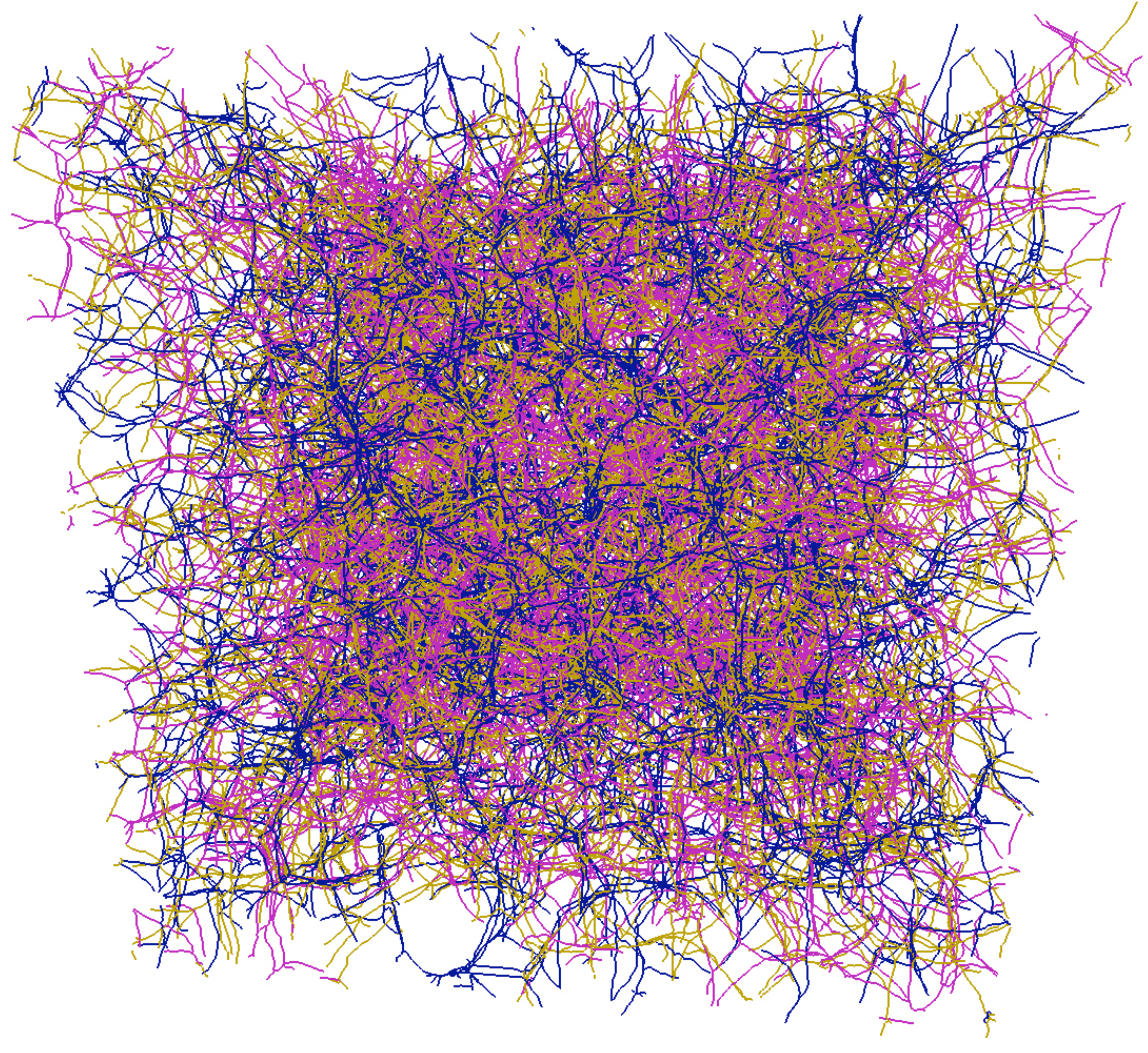}
  \caption{
  An example of set of primary critical lines (resp. skeleton in blue, 
  intermediate in magenta and anti skeleton in gold) for a scale invariant
  power spectrum with $\gamma=0.6$ in  a $256^3$ box smoothed over 5 pixels.
  }
\label{fig:allskl}
\end{figure}

We perform the measurements on critical lines found according
to the global definition.
The measurements are carried as follows: a set (typically $\sim 100$) of 
scale-invariant Gaussian random field of a N-$D$ maps (typically $1024^2$) or
cubes  (typically $256^3$) is generated with a given power index of
$n=0$, $-1$ or $-2$. The N-$D$ cube is then smoothed  via convolution with
a Gaussian kernel of width $6$ pixels.
The spectral parameters, $\gamma$, $\tilde \gamma$ {\it etc...}
are computed through the second moments of the derivative of the smoothed
field. The set of critical
lines is then extracted as the intersection of the peak patches and void
patches (see \cite{sousb08} for details). In Figure~\ref{fig:allskl}
an example realization of the primary critical lines in 3D cube is shown.
Since the algorithm 
produces a set of segments describing those critical lines tagged by the
underlying (smoothed) density field, it is straightforward to compute the
total and differential length per unit volume of the whole set.
The differential length per unit modulus gradient is extracted by tagging
the critical lines with this modulus (obtained via Fourier transform
differentiation) and proceeding as before. Finally, the curvature of the
skeleton is measured by computing the local curvature of
a set of adjacent segments via finite difference.

\begin{figure}
\centering
  \subfigure{\includegraphics[angle=0,width=7cm]{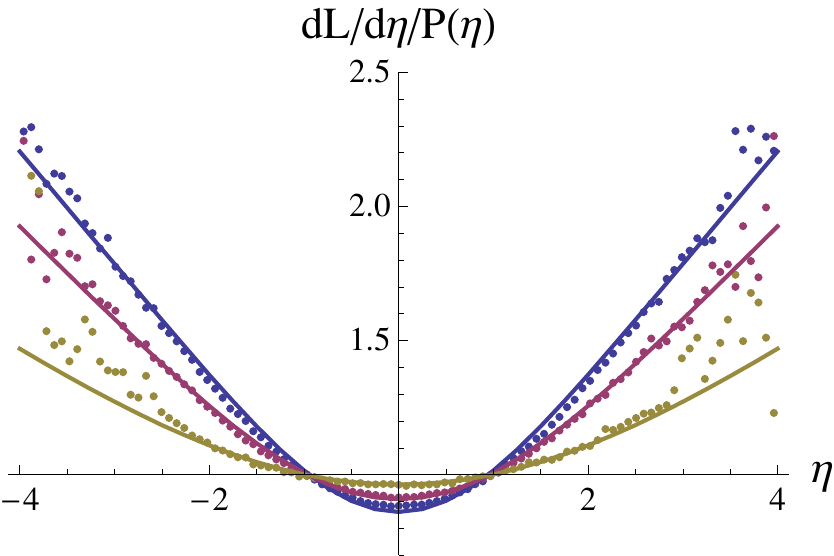}}
  \hspace{1cm}
  \subfigure{\includegraphics[angle=0,width=7cm]{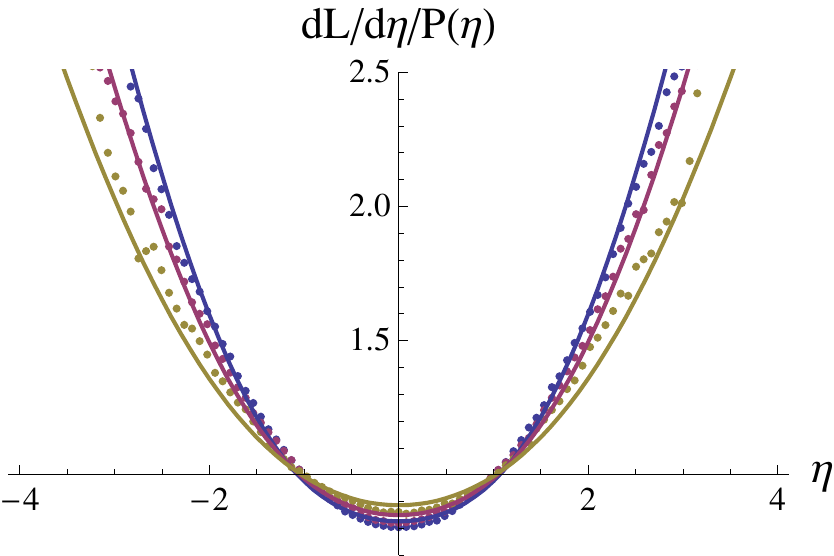}}
\caption{The relative differential length,
${\partial {\cal L}}/{\partial \eta}/\mathrm{PDF}$,
measured in simulation of 2D ({\sl left}) and 3D ({\sl right}) Gaussian
random fields with scale invariant power-law spectra versus predictions of
the local theory in stiff approximation (solid curves).
The spectral parameter $\gamma= 0.71, 0.59, 0.39$ for the 2D and
$\gamma=0.77,0.70,0.60$ for the 3D simulations.
} 
\label{fig:dLdtcomp}
\end{figure}

Let us emphasize that these measurements correspond to
properties of the global skeleton, whereas the theory developed in 
this paper is focused on the local skeleton.
Hence even more remarkable is the match between
the measured and the theoretical differential lengths for all values
of $\gamma$, that is exhibited in Figure~\ref{fig:dLdtcomp}.
This accuracy  should be considered as indicative of
the correspondence between the stiff approximation to the local theory
and the global set of critical lines.

\section{Other statistics and spectral parameters}\label{sec:other}

In the previous sections, the emphasis has been on the differential length
of the critical lines as a function of the excursion in density.
As argued in \cite{SPCNP} and demonstrated here,  it  provides means of 
constraining the shape parameter, $\gamma$. Let us now explore other
statistics which will allow us to 
constraint other shape parameters. In particular, let us demonstrate that 
the differential length as a function of the excursion in the modulus of the  
{\sl gradient} of the density  and the differential curvature depend on the
second shape parameter, $\tilde \gamma$.  Finally, we investigate the
number density of singular points on the critical lines.

\subsection{Differential length versus the gradient modulus}
The differential 
length of the skeleton with respect to the threshold $\eta$ carries information on the spectral parameter $\gamma$ thanks to the correlation between the
value $\eta$ and the Hessian of the field.  In the stiff approximation
the Hessian curvature completely determines the length
of the critical lines.  For the exact formulation, the 
length also depends on the third derivatives, that are correlated with the first
derivatives via the parameter $\tilde \gamma$. Thus, measuring  length
as a function of the modulus of the gradient should carry information
on $\tilde \gamma$ and provide an estimate of an impact the third derivatives
have on the length statistics of the critical lines.

To demonstrate the dependence of the skeleton length on the gradient of the
field in ``stiff'' approximation let us return to 
equation~(\ref{eq:3DHess_start}) which we take integrated over all density
thresholds. As before, we perform the integration over the $\delta$-functions
that enforces alignment of the gradient with the first eigen-direction,
$x_2=x_3=0$, however this time
we do not integrate over but rather take the differential of the result with
respect to $x_1$. Noting that $|x_1|=X \equiv \sqrt{x_1^2+x_2^2+x_3^2}$.
we obtain in place of equation~(\ref{eq:3DHess_next})
\begin{equation}
\frac{\partial {\cal L}}{\partial X} = \frac{3}{2} \cdot
\frac{3^{3/2} 15^2 5^{1/2}} {(2 \pi)^{5/2}}
\exp\left[-\frac{3}{2} X^2\right]  
\int\!
\left|(\lambda_1-\lambda_2)(\lambda_2-\lambda_3)(\lambda_3-\lambda_1)\right| 
 d\lambda_1 d \lambda_2 d \lambda_3  |\lambda_2 \lambda_3|
\exp \left[-\frac{1}{2}\tilde u^2 - 
 \frac{15}{2} \tilde w^2  - \frac{5}{2} \tilde v^2 \right]\,,
\end{equation}
where the last integral does not depend on $X$.
Dividing by the integrated length, 
$L=\int_0^\infty \partial {\cal L}/\partial X d X $,
and generalizing the result to fields in arbitrary $N$ dimensions  we
conclude that
\begin{equation}
\label{eq:dLdxstiff}
\left({\frac{1}{L }
\frac{\partial{\cal L}}{\partial X}}\right)^{\mathrm{stiff}}=
\sqrt{\frac{2N}{\pi}}
\exp\left[-\frac{N}{2} X^2 \right] \,.
\end{equation}

\begin{figure*}
\centering
{
\includegraphics[angle=0,width=7cm]{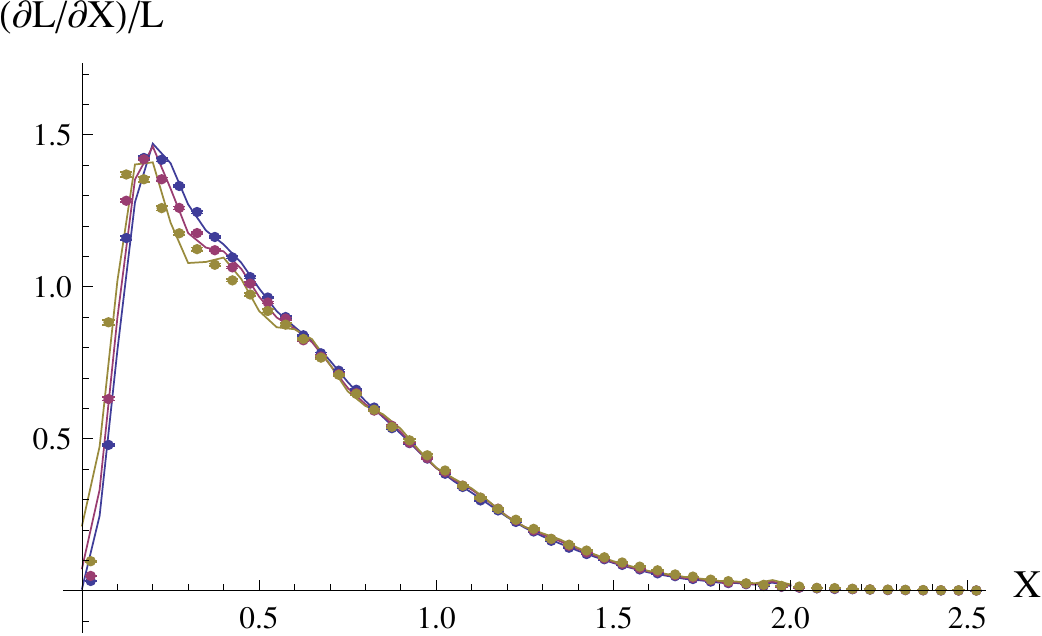}
\label{fig:pdfgrad2D}}\hspace{1cm}
{\includegraphics[angle=0,width=7cm]{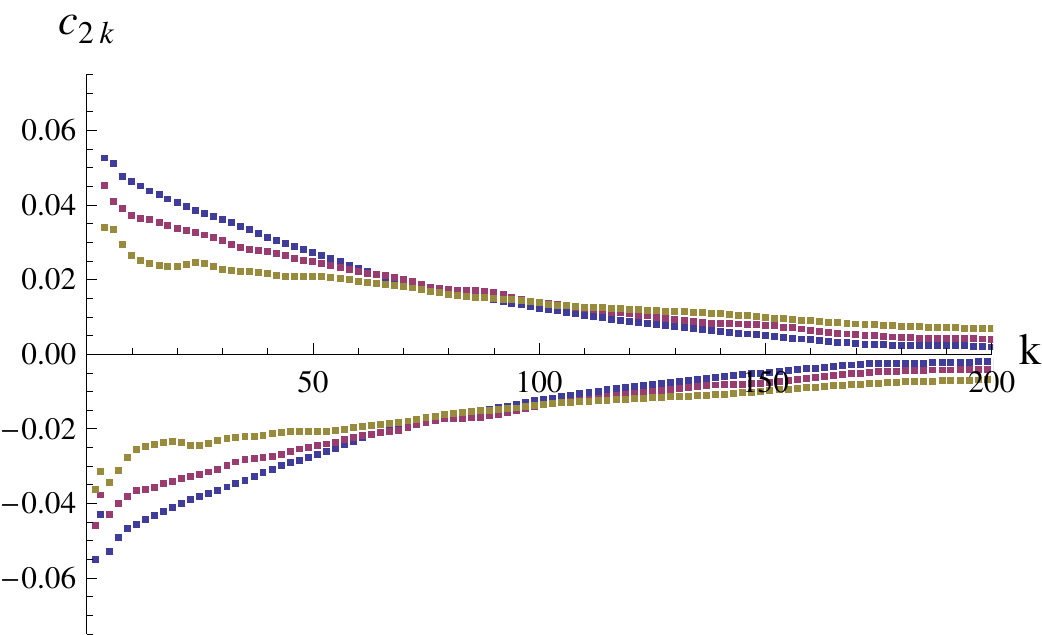}
\label{fig:pdfgradfit2D}}\\
  \caption{
  { \it left}: 
  measured $ \partial {\cal L}/ \partial X/L$ as a function of
  $X\equiv |\nabla \rho|$ for $\tilde\gamma= 0.71, 0.58$ and  $0.38$ using
  the set of 25 2D simulations of $1024^2$ Gaussian  random  fields 
  with scale invariant spectra smoothed over 7 pixels. 
 { \it Right}: value of the fit parameters $c_k$
 (see equation~(\ref{eq:dLdx})).  Note that $c_0=1$.
  }
\label{fig:gradrho2D}
\end{figure*}

The exact dependence of the differential lengths will deviate from this form
in a $\tilde \gamma$ -dependent way.  It is natural to parameterize such
deviation expanding the true statistics in Hermite series around the stiff
approximation
\begin{equation}
\frac{1}{ L }
\frac{\partial {\cal L}}{\partial X} = 
\sqrt{\frac{2 N}{\pi}} \exp\left[-\frac{N}{2} X^2 \right] 
\left(
\sum_{k=0}^{\infty} \frac{{c}_{2k}(\tilde \gamma)}{\sqrt{(2k)!}} 
H_{2k}( \sqrt{N}X ) \right)
 \,. \label{eq:dLdx}
\end{equation} 
This choice of expansion is dictated by the orthogonality
of the Hermite polynomials
with the weight $\propto \exp[-\sqrt{N} X^2/2]$ on the interval $X \ge 0$.
Thus, $c_0=1$. If the deviation from the stiff approximation is small,
one expects the expansion to be dominated by the $n=0$ term, while
the subsequent terms should quickly fall in a orderly fashion. 

To gain understanding on how the coefficients $c_{2k}(\tilde \gamma)$ behave
with $\tilde \gamma$, let us consider again the lax situation, opposite to the
stiff case, when the third derivatives of the field dominate the length
statistics.  Our starting point is equation~(\ref{eq:nonstiff})
which has the following structure when we consider the differential length with
respect to the $|x_1|=X$
\begin{equation}
\frac{\partial {\cal L}}{\partial X}^{\rm lax }\!\!\!\! \propto 
X^2 \exp\left[- \frac{3}{2} X^2 \right]
\int  d u_1 \left\{ \frac{\displaystyle
\exp\left[-\frac{3}{2}\frac{(u_1-\tilde\gamma X)^2}{1-{\tilde\gamma}^2} \right]+
\exp\left[-\frac{3}{2}\frac{(u_1+\tilde\gamma X)^2}{1-{\tilde\gamma}^2} \right]
}{\sqrt{2 \pi (1-{\tilde \gamma}^2)}}\right\}
\int  d^{9}x_{ijk} 
\psi(x_{ijk})
\bar P_1(x_2=x_3=0,x_{ijk}/u_1) 
\end{equation}
where $\bar P_1$ is given by equation~(\ref{eq:defp13D}) with the dependence on
$u_1$ factored out. The difference with the stiff 
approximation is large even for $\tilde \gamma=0$ as the gradient's dependence
becomes $\propto X^2 \exp[-3/2 X^2]$ in place of the stiff  scaling
$\propto \exp[-3/2 X^2]$. 
Using now this factor as the weight, for $\tilde \gamma \ne 0$
we expand the expression in the brackets  in generalized Laguerre
polynomials. The expansion coefficients are
of the form ${\tilde \gamma}^{2 k} \exp[-3 u_1^2/2] 
\sum_{m=0}^k d_m {\tilde \gamma}^{2m} H_{2m}(\sqrt{3} u_1) $;
denoting the result of  the integration of the expansion coefficients and all of
the residual factors over the third derivatives by $\Psi_k(\tilde \gamma)$
we obtain
\begin{equation}
\frac{1}{L} \frac{\partial {\cal L}}{\partial X}^{\rm lax} = 
3 \sqrt{\frac{6}{\pi}} X^2 \exp\left[- \frac{3}{2} X^2 \right]
\sum_{k=0}^\infty \frac{2^k k!}{(2k+1)!!}
{\tilde \gamma}^{2 k} L_k^{(1/2)}\left(3 X^2/2 \right) 
\Psi_k({\tilde \gamma})\,,\label{eq:temp}
\end{equation}
where, again, $\Psi_0(\tilde \gamma)=1$.
With the help of the relation between the Laguerre and Hermite polynomials
\begin{displaymath}
3 X^2 k! L_k^{(1/2)}\left(3/2 X^2\right)=
(-1)^k 2^{-k} \left(H_{2k+2}(\sqrt{3}X) + (2k+1) H_{2k}(\sqrt{3} X) \right)\,,
\end{displaymath}
we can cast equation~(\ref{eq:temp}) in the form of equation~(\ref{eq:dLdx})
\begin{eqnarray}
\frac{1}{L} \frac{\partial {\cal L}}{\partial X}^{\rm lax} &=& 
\sqrt{\frac{6}{\pi}} \exp\left[- \frac{3}{2} X^2 \right]
\sum_{k=0}^\infty \frac{(-1)^k}{(2k+1)!!}
{\tilde \gamma}^{2 k} 
\left(H_{2k+2}(\sqrt{3}X) + (2k+1) H_{2k}(\sqrt{3} X) \right)
\Psi_k({\tilde \gamma})\,, \nonumber \\
& = & \sqrt{\frac{6}{\pi}} \exp\left[- \frac{3}{2} X^2 \right]
\left[1+
\sum_{k=1}^\infty \frac{(-1)^{k-1}}{(2k-1)!!}
\left[ {\tilde \gamma}^{2 k-2} \Psi_{k-1}({\tilde \gamma})
-{\tilde \gamma}^{2 k} \Psi_k({\tilde \gamma}) \right]
H_{2k}(\sqrt{3} X)
\right]\,.
\end{eqnarray}
The coefficients $c_{2k}(\tilde \gamma) $ are
\begin{equation}
c_0=1~, \quad c_2 = \sqrt{2} \left(1-{\tilde \gamma}^2 \Psi_1(\tilde \gamma)\right)
~, \quad
c_{2k} = \left((-1)^{k-1} \sqrt{(2k)!}/(2k-1)!!\right) {\tilde \gamma}^{2k-2} 
\left(\Psi_{k-1}(\tilde\gamma)-{\tilde\gamma}^2 \Psi_k(\tilde\gamma)\right) ~.
\end{equation}
In particular, in the limit $\tilde \gamma \to 0$ 
the first two coefficients remain finite
and of equally significant magnitude $c_0=1, ~c_2=\sqrt{2}$, while all the
other ones vanish.

\begin{figure*}
\centering
{
\includegraphics[angle=0,width=7cm]{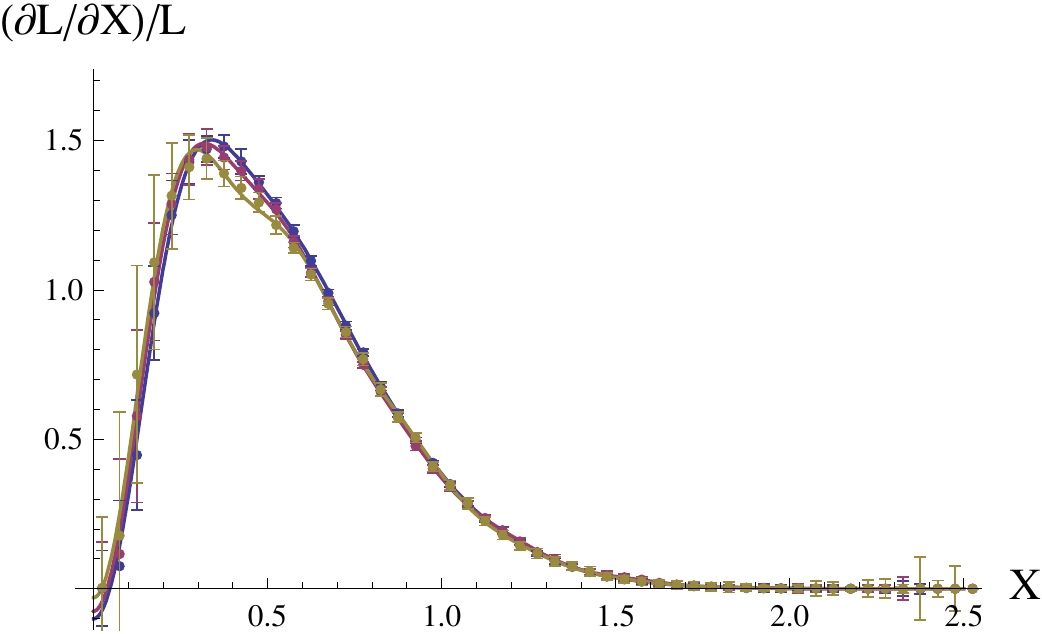}
\label{fig:pdfgrad}} \hspace{1cm}
{\includegraphics[angle=0,width=7cm]{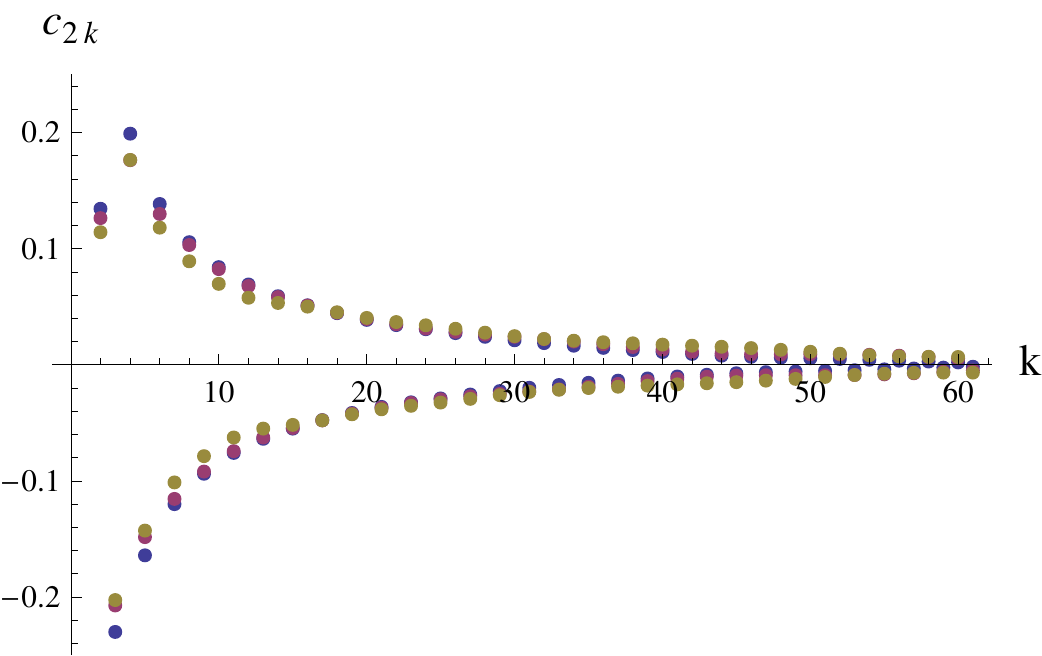}
\label{fig:pdfgradfit}}\\
  \caption{
  { \it left}: 
  measured $ \partial {\cal L}/ \partial X/L$ as a function of
  $X\equiv |\nabla \rho|$ for $\tilde\gamma= 0.86, 0.83$ and $0.79$
   using the set of 3D simulation of $128^3$ Gaussian  random  fields 
with scale invariant spectra smoothed over 7 pixels. 
 { \it Right}: value of the fit parameters $c_k$,
 Note its faster convergence combined with a larger 
 amplitude relative to the 2D case.
     }
\label{fig:gradrho}
\end{figure*}

Figure~\ref{fig:gradrho2D} and Figure~\ref{fig:gradrho} present
the measurements of $ {\partial {\cal L}}/{\partial X}/L $ in 2 and 3D
respectively, together with the corresponding coefficients given by
equation~(\ref{eq:dLdx}).
It is found that these coefficients are significantly smaller in two
dimensions, a clear indication that the stiff approximation holds better in 2D.
 
\subsection{Statistics of the curvature of the critical lines} \label{sec:curv}
The local curvature, ${\kappa}$, at a point on a curve specified by
the tangent vector
${\mathbf u}=d {\mathbf r}/ dt$ is determined by the acceleration of the
tangent vector $ \dot {\mathbf u} \equiv d {\mathbf u} / dt = \mathbf{u} \cdot
\left( \partial \mathbf{u} /\partial \mathbf{r} \right) $
transverse to the curve direction:
\begin{equation}
{\kappa} = \frac{ | \mathbf{u} \times \dot{\mathbf{u}} |}{|\mathbf{u}|^3}
= \frac{ | \mathbf{u} \times \left(\left(\nabla \mathbf{u}\right) \cdot \mathbf{u}\right) | } { | \mathbf{u} |^3 }\,.
\label{eq:curvature}
\end{equation}
Importantly, the curvature does not depend on parameterization $t$, nor
on normalization of the tangent vector ${\bf u}$.
In the local theory, the tangent vector to a critical line
is orthogonal to $\nabla s^i(x_{k},x_{kl})$ and can be taken to be
\begin{equation}
\mathbf{u} =  \boldsymbol \epsilon\cdot
		  \nabla s \quad (2D) ~, \quad \quad
\mathbf{u} =  \nabla s^i \cdot \boldsymbol \epsilon \cdot   \nabla s^j = \nabla s^i \times \nabla s^j \quad (3D) ~;  
\end{equation}
so the curvature $\kappa$ is the random quantity which involves the 
derivatives of the field up to fourth order,
\begin{eqnarray}
(2D) \quad {\kappa}&=&
{| \nabla s\cdot(\nabla\nabla s) \cdot \nabla s|} \big/ {|\nabla s|^{3}} \quad ,
\label{eq:defk2D} \\
(3D) \quad {\kappa}&=&
 \left|\left(\nabla s^{i}\times \nabla s^{j} \right)\times \left[
\left(\nabla s^{i}\times \nabla s^{j} \right)\cdot \nabla\left(
\nabla s^{i}\times \nabla s^{j} \right)
\right] \right|\big/\left|\nabla s^{i}\times \nabla s^{j} \right|^{3}\,.
\label{eq:defk3D}
\end{eqnarray}
The curvature of the critical lines fundamentally reflects the derivatives of
the field higher than the second. If they are neglected, the curvature is
identically zero.  Explicitly, the contributions that do not
involve higher derivatives, in 2D
\begin{eqnarray}
(2D) \quad (\mathbf{u} \times \dot{\mathbf{u}})^2 &=&
4 x_1^2 x_2^2 \lambda _1^4 \left(\lambda _1-\lambda _2\right)^6 \lambda _2^4
+ \ldots  \,, \\
(3D) \quad (\mathbf{u} \times \dot{\mathbf{u}} )^2 &=&
\left(x_2^2 \lambda _3^2+x_3^2 \lambda _2^2\right) 
\left(\lambda _1-\lambda _2\right)^4 \left(\lambda _1-\lambda_3\right)^4
\lambda _1^2 \left(3 x_1^2 \lambda _2^2 \lambda _3^2 
+\left(x_2^2 \lambda _3^2+x_3^2 \lambda_2^2\right) \lambda _1^2\right)^2
+ \ldots \,,
\end{eqnarray}
vanish when the correspondent critical line conditions $x_2=0$ or $x_2=x_3=0$
are applied\footnote{Note that by construction,  the torsion:
$ \tau =
 {  | \mathbf{u}\cdot ( \dot{\mathbf{u}}\times \ddot{\mathbf{u}}) |   }/
 {           | \mathbf{u} \times \dot{\mathbf{u}}              |^2 }  
$
contains only the terms proportional to at least the  third derivatives of
the field.
}.

The integrated curvature over the length of the line, $C=\int \kappa dL$ is
a useful dimensionless characteristics of the overall extend a line is curved.
We have seen that  the critical line length in volume $dV$ is
$d L \propto |\nabla s|\delta_{\rm D}(s) dV $ and 
$d L \propto |\nabla s^i \times \nabla s^j|
\delta_{\rm D}(s^i)  \delta_{\rm D}(s^j) dV$ in 2D and 3D respectively.
Averaging over statistical distribution in regions above threshold
$\eta$  we obtain the mean density of the integrated critical line curvature 
${\cal C}=\langle dC/dV \rangle$
\begin{eqnarray}
(2D) \quad {\cal C}(\eta_{>}) &=& \frac{1}{R_* \tilde R}
\int_{\eta>x}   d x d^{2}x_{k} d^{3} x_{kl}  d^{4} x_{klm}  d^{5} x_{klmn}\,
{\kappa}(x_{k},x_{kl}, \cdots) 
|\nabla s| \delta_{\rm D}(s)P(x,x_k,x_{kl},\cdots)\,, \label{eq:defC} \\
(3D) \quad {\cal C}(\eta_{>}) &=&  \frac{1}{R_*^2 \tilde R}
\int_{\eta>x}  d x d^{3}x_{k} d^{6} x_{kl}  d^{10} x_{klm}  d^{15} x_{klmn} \,
{\kappa}(x_{k},x_{kl}, \cdots) \left|\nabla s^{i}\times \nabla s^{j} \right|
\delta_{\rm D}( s^{i})\delta_{\rm D}( s^{j}) {P}(x,x_{k},x_{kl}\cdots) \, ,
\label{eq:pdfC}
\end{eqnarray}
where the integration is carried over all the derivatives up to the
fourth order. The required joint probability function is given in
equations~(\ref{eq:fquad.un.trois}) and (\ref{eq:fquad.zero.deux.quatre}).

\begin{figure*}
  \centering
 \includegraphics[angle=0,width=7cm]{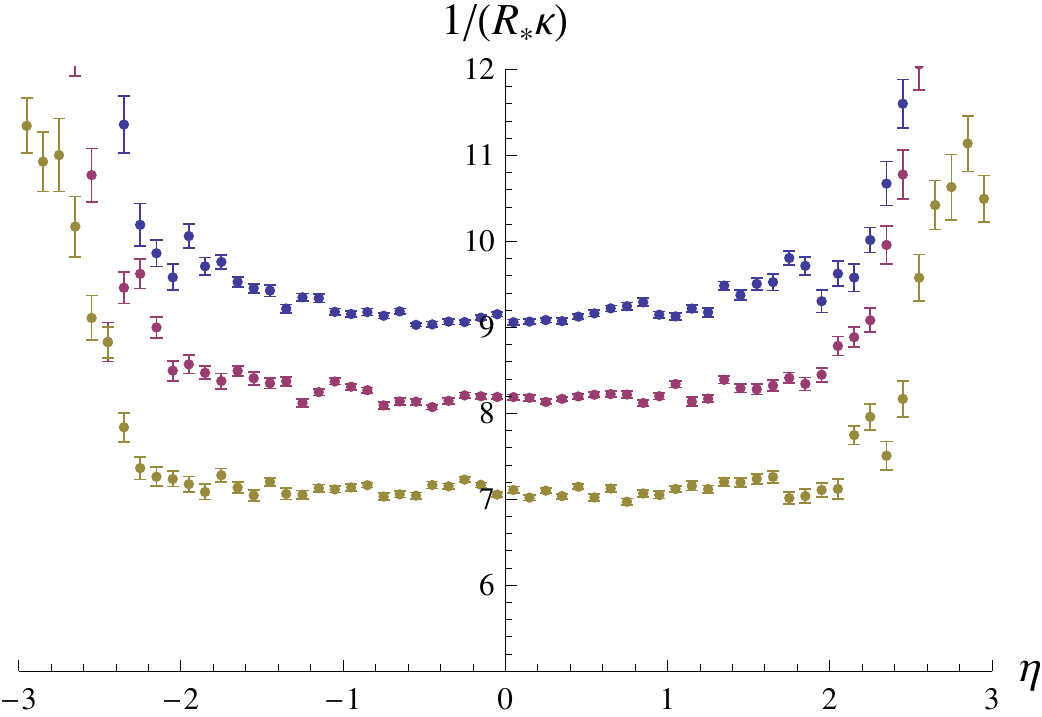}
  \hspace{1cm}
   \includegraphics[angle=0,width=7cm]{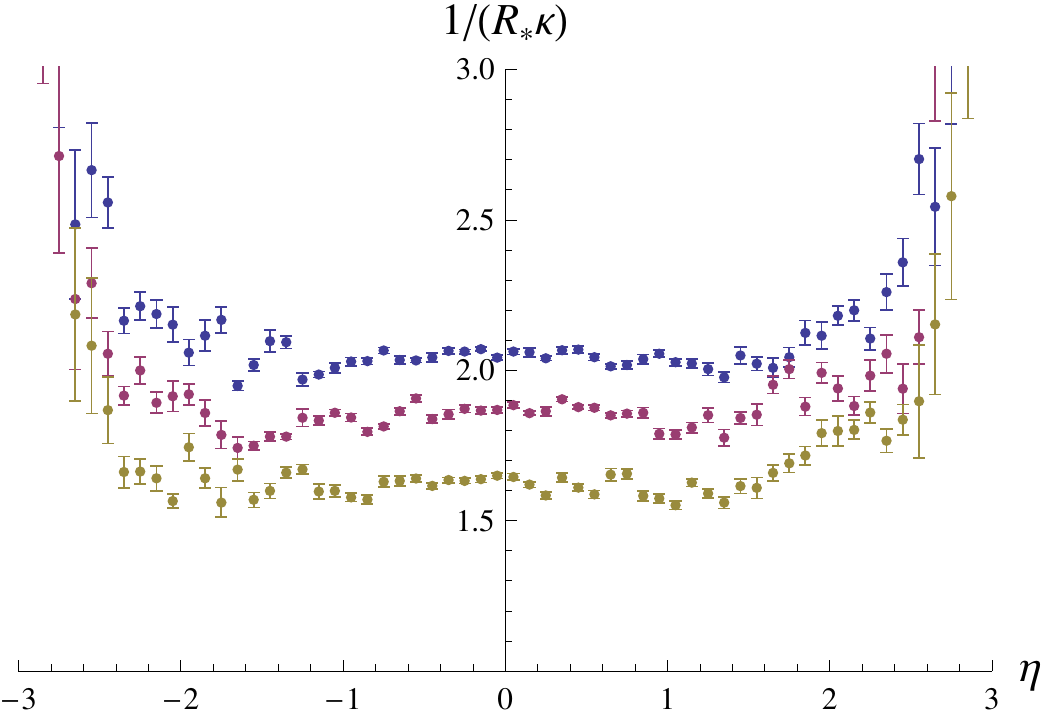}
  \caption{$1/(R_* \langle \kappa \rangle)$, 
the mean curvature {\sl radius} in units of $R^*$ as a function of $\eta$
measured in simulation of 2D ({\sl left}) and 3D ({\sl right}) Gaussian
random fields with scale invariant power-law spectra.  
The top curves correspond to spectra with more power at small scales and
higher $\gamma$ and $\tilde \gamma$ spectral parameters.
}
\label{fig:curv}
\end{figure*}

Let us consider 2D case and estimate the curvature by using 
stiff approximation for the tangent vector $u$ while following
its local variation which involve higher derivatives of the underlying
field. In the Hessian eigenframe,  assuming the skeleton lies along $1$,
if  $\mathbf u$ is approximated by its stiff counterpart we have: 
\begin{equation}
\left(\frac{ | \mathbf{u} \times \dot{\mathbf{u}} |  }{  | \mathbf{u}            |^2      } \right)^{\rm stiff}=|
x_1| \left|(\lambda_{1}+\lambda_{2}  ) x_{112}\right|\,, \label{eq:curvstiff}
\end{equation}
so that  (taking into account the measure in the eigenframe and the 
$\delta_{\rm D}$ function of $S$ in $x_2$): 
\begin{eqnarray}
\frac{\partial {\cal C}^{\rm stiff}}{\partial \eta} &=&
\pi \int d \lambda_{1} d \lambda_{2}d^{4} x_{ijk}
 \left| (\lambda_{1}+\lambda_{2}  ) x_{112}\right| 
P_0(\eta,x_{kl})P_1(x_{ijk}) \nonumber \\
&=& \frac{\sqrt{2-\tilde\gamma^2}}{4 \pi R_* \tilde R}
\left(\sqrt{2(1-\gamma^2)}
\exp\left[-\frac{\gamma^2 \eta^2}{2(1-\gamma^2)}\right]
+\sqrt{\pi} \mathrm{Erf}\left[\frac{\gamma\eta}{\sqrt{2(1-\gamma^2)}}\right]
\gamma\eta \right)
\frac{1}{\sqrt{2 \pi}} \exp\left[-\frac{\eta^2}{2}\right]
\,, \label{eq:defCstiff}
\end{eqnarray}
the last evaluation being done for the primary critical lines.

We have measured the mean curvature of the skeleton lines at the
threshold $\eta$, 
\begin{equation}
\left\langle \kappa \right\rangle \equiv
\frac{\partial {\cal C}/\partial \eta}{\partial {\cal L}/\partial \eta}\, ,
\label{eq:defK}
\end{equation}
in simulations of the Gaussian random fields of different spectra,
using the global skeleton techniques.
Figure~(\ref{fig:curv}) displays the  2D (left panel) and
3D (right panel) results in terms of the curvature radius, 
$R_{\rm curv}=1/\langle \kappa \rangle$.
The measurements show that for the spectra we consider, the averaged curvature
is insensitive to the density threshold for low-to-moderate
threshold values showing a plateau in the interval $-2 < \eta < 2$.
This indicates that in this regime the curvature of the skeleton
does not depend on $\gamma$, but rather on $\tilde \gamma$ and perhaps 
$\hat \gamma$.
It follows that in 3D the critical lines are relatively more wiggly than in 2D.
If we use the lower value of $ R^{\rm 3D}_{\rm curv}$  as a guidance, it seems
the stiff approximation is less accurate in 3D that in 2D, as could be
expected.
The stiff estimate~(\ref{eq:defCstiff}) 
gives the threshold-averaged mean density of the integrated  curvature
${\cal C}^{\rm stiff}$ 
and, using equation~(\ref{eq:Ltot_skel+antiskel_2D}),
the mean curvature radius, $R_\mathrm{curv}^\mathrm{stiff}$, as
\begin{equation}
{\cal C}^{\rm stiff} = 
\frac{\sqrt{1-{\tilde\gamma}^2/2}}{2 \pi R_* \tilde R} \, , \quad
R_\mathrm{curv}^\mathrm{stiff} \equiv \frac{L}{{\cal C}}
=\frac{\sqrt{2}+\pi/2}{\sqrt{1-\tilde\gamma^2/2}}\tilde R 
\approx 2.985 \frac{\tilde\gamma}{\sqrt{1-\tilde\gamma^2/2}} R_* \,.
\end{equation}
This result  captures the qualitative dependence on $\tilde \gamma$
observed in simulations, but is a factor of three smaller in the magnitude
of the curvature radius. This shows that the global skeleton, used in
the numerical measurement, is notably straighter than the local critical lines,
although the dependence of curvature on the spectral parameters seems
similar.

Note in closing that in 2D, (resp. 3D) the knowledge of the differential length,
curvature  (resp. length, curvature and torsion) corresponds to an exhaustive
global statistical description of the critical lines. 

\subsection{Singular points of critical lines}\label{sec:bif}

Let us now ask ourselves the following question:  are there any 
special points along the skeleton?  The obvious ones are the extrema of the
field itself where critical lines intersect. 
Beyond this, one can anticipate two other types of singular points.
The first type corresponds to points where the curvature transverse to
the direction of the critical line vanishes along at least one axis:
typically, in 2D, they mark regions where a crest becomes a trough, or 
vanish into a plateau.
The second type  correspond to  points where the critical lines would split,
even though the field does not go through an extremum: a bifurcation
of the lines occurs along the slope; the occasional skier or mountaineer will
be familiar with a crest line splitting in two,  even though the gradient of
the field has not vanished.
From the point of view of the theory of random fields,
the frequency of such points is an interesting venue: indeed we expect that
steep power spectrum present relatively more bifurcation points as $\tilde R$,
the distance between inflection points (see Sec.~\ref{sec:defsR}), becomes 
much shorter than $R_*$, the distance between extrema.  In an astrophysical
context, the statistical properties of the first type of  points, and in
particular  their clustering properties are of interest for understanding
the geometry of galactic infall, which in turn is believed to play an important
role in defining the morphological properties of galaxies.  The multiplicity of
the maxima ({\sl i.e.} the number of connected skeleton segments) is also
of interest in the context of galaxy formation and feedback.
In more abstract spaces, such as position-time, identifying bifurcations is
important to pin down merging events (see e.g. \cite{hanami} and
Appendix~\ref{sec:ND}). 

\subsubsection{Defining the skeleton singular points}

Formally a singular condition along the skeleton occurs when at some
point the determination of the critical line direction fails.  It means that
at this point the matrix $\nabla_k S^i$ of equation~(\ref{eq:sss}) has more
than one distinct {\it right} null-vector, or, equivalently,
all $M^k$ defined by equation~(\ref{eq:defM}) are zero.
The only case when it happens exactly is at the extremal points of the field
$\nabla \rho = 0 $. There are no other formal 
singularities on the local critical lines, since when $\nabla \rho \ne 0$,
the requirement $M^k=0$ sets $N$ relations between the field gradient,
second and third
derivatives which have vanishing probability to be simultaneously satisfied
along a line in a random field.

\begin{figure*}
  \centering
  \includegraphics[angle=0,width=10.2cm]{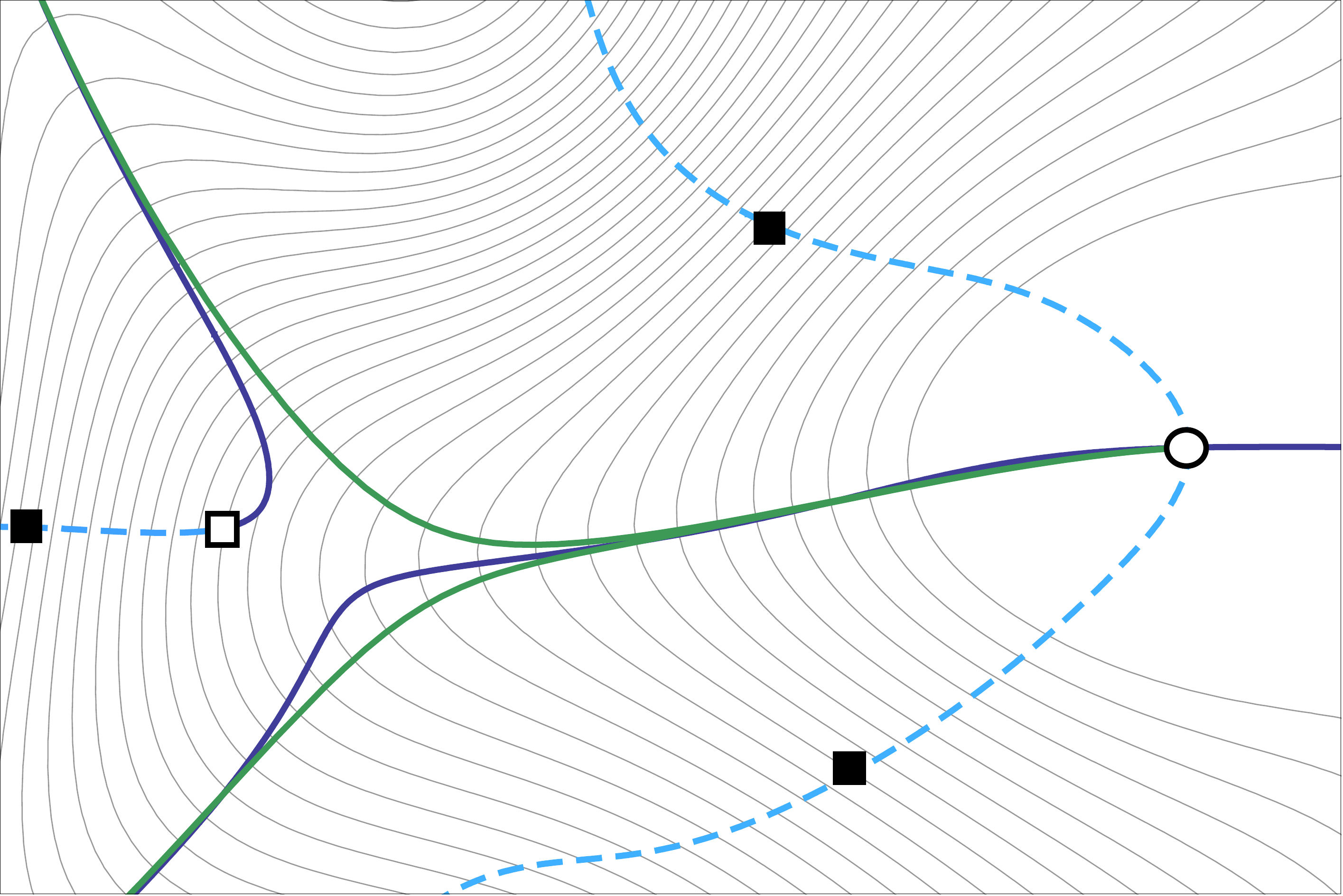}\hfill
  \includegraphics[angle=0,width=6.8cm]{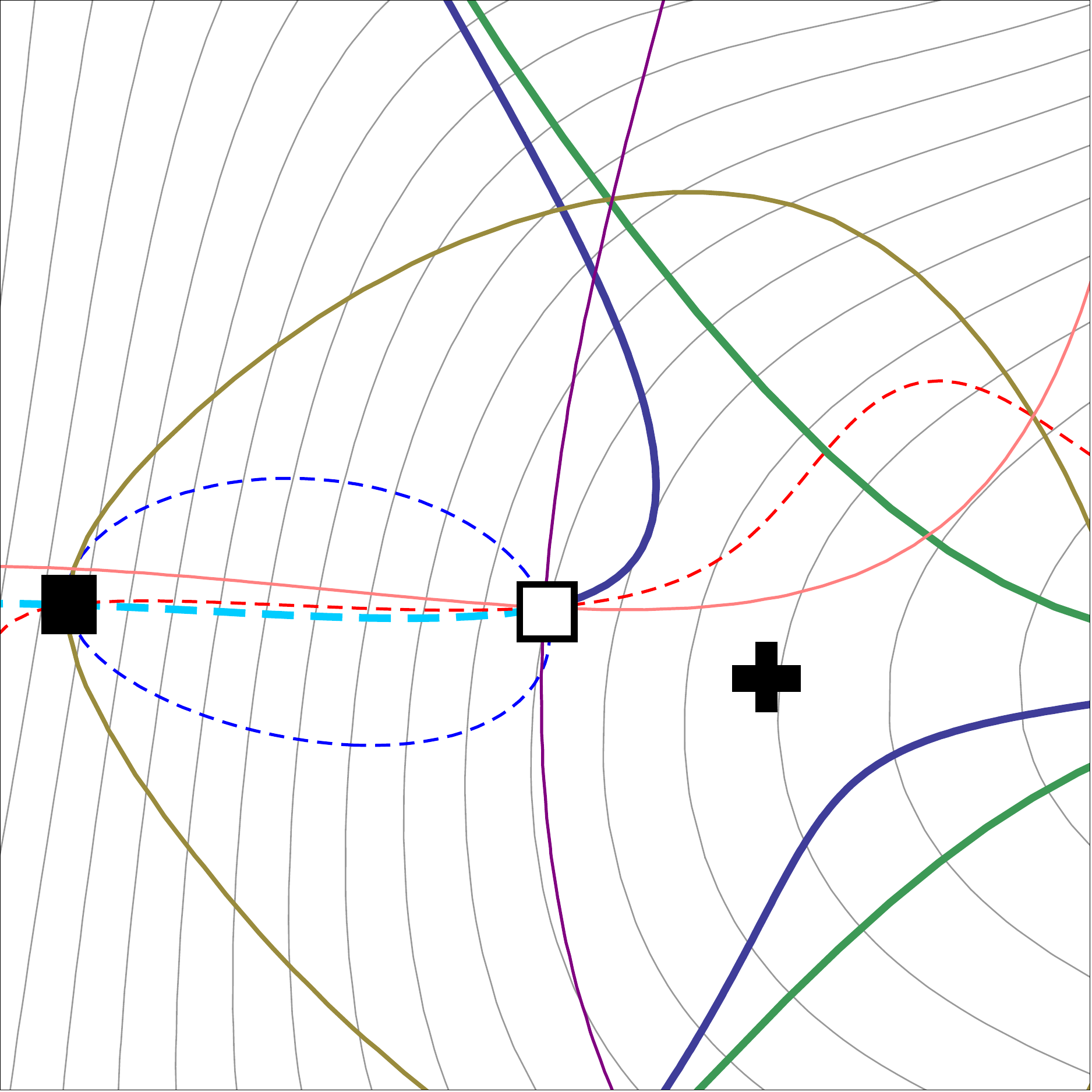}\hfill
  \caption{{\sl Left:} the three types of singular point on the critical
   lines (in solid blue: primary; dashed: secondary; green: gradient
   lines of the global skeleton):
      an extremum ({\sl open circle}), a ``bifurcation'' ({\sl white square}),
      and ``slopping plateaux'' ({\sl black squares}).
      The thin lines are the isocontours of the field.
      {\sl Right:} Detailed view of the bifurcation region:
      The pink and purple lines  mark the conditions $x_{11}=x_{22}$
      and $x_{12}=0$, which when intersect give the point where 
      $\lambda_1=\lambda_2$.
      This point is a singular point of the critical line ({\sl white square}).
      The gold line is the condition $\lambda_2=0$ that
      marks the ``sloping plateau'' on the critical lines.
      The red and blue dashed lines are zero isocontours of two components of
      $\nabla S^{\rm stiff}$. The  $\nabla S^{\rm stiff}=0$ criterium
      pin-points exactly all types of the singular points on the critical lines.
      The black cross marks the position of the point $\nabla S=0$. 
}
\label{fig:bifexample}
\end{figure*}

The failure of the formal definition to identify all the physically interesting
situations primarily reflects the inadequacy of the local skeleton construction,
which only utilizes locally quadratic approximation to the field, to map
the field near the singular points.
\footnote{ Similarly, the bifurcation points for the global fully-connected
skeleton \citep{sousb08} also formally merge with critical points in the
strict sense due to sharp topological theorems \citep{jost}. However they
appear if the skeleton is viewed with a finite resolution.}
Figure~\ref{fig:bifexample} gives a 2D example. In 2D, the critical lines are
zero levels of the scalar $S$-function, while $\nabla S=0$ at the extrema
of $S$ field.  In Figure~\ref{fig:bifexample} the region where a ridge splits
into two is shown. One expect two critical lines cross there, with three
branches following the ridges, and one following the through between two
of the split branches. Instead, the locally defined critical lines are
not allowed two join at the bifurcation point since the formal condition
$\nabla S =0$ is satisfied just {\it off} $S=0$ contour, rather they
artificially reconnect near the bifurcation point into two
non-intersecting segments. 

We conjecture that the critical lines experience a qualitative change in 
behaviour in the vicinity of the points where either the Hessian eigenvalue
of the orthogonal to the gradient direction vanish, or becomes 
equal to the one along the gradient. Namely, if, for definiteness, 
$\nabla \rho$ is taken to be along the first eigen-direction, $\lambda_2=0$, or 
$\lambda_2=\lambda_1$. We call the first case the ``sloping plateau'' as it 
designates the entering of a flat region, and the second, tentatively, 
the ``bifurcation'' as it designates the places of possible reconnection
of critical lines. In particular, at the $\lambda_2=\lambda_1$ points
most of the transitions from primary to secondary behaviour take place.
Remarkably, these special points on the critical lines
are recovered by the formal singular condition $|M^k|=0$ 
if $\nabla_k S^i$ is evaluated in the stiff approximation. As given in
equation~(\ref{eq:Mstiff}), along the ND critical line defined by
$x_2=\ldots=x_N=0$, $|M^{\mathrm{stiff}}|=
x_1^{N-1} \prod_{i>1} \lambda_i \left(\lambda_1-\lambda_i\right) = 0 $
gives rise to three classes of situations: (i) $x_1=0$ corresponding to
extremal points; (ii) one of $\lambda_i=0$ corresponding to
slopping flattened tubes; and (iii) one of $\lambda_i=\lambda_1$,
corresponding to an isotropic bifurcation.

Since it is beyond the scope of this paper to develop the full theory of
these special points, we will focus  here on their number density for isotropic
2D Gaussian random fields, leaving more detailed investigation
to future work.

\subsubsection{Number density of the singular points of the 2D critical lines}

In 2D, the skeleton's singular points correspond to points where 
$S_k \equiv \nabla_k S=\mathbf{0}$.
The number density, $n_{{\rm B}}(\eta)$ of singular points below the
threshold $\eta$ is equal to
\begin{equation}
n_{{\rm B}}(\eta) = \int_{\eta>x}   d x d^{2}x_{k} d^{3} x_{kl}  d^{4} x_{klm}  d^{5} x_{klmn} P(x,x_{k},x_{kl},\cdots)
|{\rm det}\left(\nabla_k \nabla_l s\right)|\delta_{{\rm D}}(s_1)
\delta_{{\rm D}}(s_2)\,.
\end{equation}

The simplest case of the skeleton singular points
$\nabla S=\mathbf{0}$ are, according to
equation~(\ref{eq:defsi}), the extrema of the field itself, $x_1=x_2=0$. 
Indeed when both $x_1$ and $x_2$ vanish
\begin{equation}
|{\rm det}\left(\nabla_k \nabla_l s\right)|
\delta_{{\rm D}}(s_1) \delta_{{\rm D}}(s_2)
=|x_{kl}| \delta_{{\rm D}}(x_{1})\delta_{D}(x_{2})\,,
\label{eq:2Dstiffbif}
\end{equation}
which is exactly the integrand involved in the number density of {\sl extrema}
of the field. The extrema number densities, for
reference, are given in 2D by \citep{1957RSPTA.249..321L}
\begin{eqnarray}
\frac{\partial n_{\mathrm{saddle}}}{\partial \eta} &=& \frac{1}{ {R_*}^2} 
\frac{1}{4 \sqrt{3}}\left[\frac{1}{\sqrt{2 \pi}\sqrt{1-2\gamma^2/3}} 
\exp\left(-\frac{\eta ^2}{2(1-2\gamma^2/3)}\right)\right] \,,\\
\frac{\partial n_{\mathrm{min+max}}}{\partial \eta} &=& 
\frac{\partial n_{\mathrm{saddle}}}{\partial \eta}
+ \frac{1}{4 R_*^2} \gamma^2 (-1 + \eta^2)
\frac{1}{\sqrt{2 \pi}} \exp\left(-\frac{\eta^2}{2}\right)\,.
\end{eqnarray}
The singularity of the extrema from the points of view of the critical line
theory is manifest in the fact that at extrema several critical lines intersect.

The gradient of $S$, evaluated in the stiff approximation,
in the Hessian eigenframe has the components 
\begin{equation}
s_1^\mathrm{stiff} = x_{2} \lambda_1 \left(\lambda_1-\lambda_2\right)\,,
\quad  {\rm and}\quad
s_2^\mathrm{stiff} = x_{1} \lambda_2 \left(\lambda_1-\lambda_2\right)\,.
\label{eq:nB2D-0}
\end{equation}
and involves only second derivatives of the field. Remarkably, within this
approximation, there are new singular points that lie {\sl on} the (local) 
critical lines. The reason is that among two conditions needed for $\nabla S$
to vanish, one is already automatically satisfied by being on a critical line.

To be specific, let us consider the critical line that corresponds to the
$x_2=0$ condition in the Hessian eigenframe. Then $s_1^{\mathrm{stiff}}$
vanishes everywhere along this line. The requirement $s_2^{\mathrm{stiff}}=0$
has a solution at the extremal points, $x_1=0$, but also in
two other cases, namely $\lambda_2=0$ or $\lambda_2=\lambda_1$, that
we conjectured to be of interest.

The first situation, {\it the sloping plateau}
with a flat transverse gradient, only
occurs on secondary critical lines since it implies $\lambda_1> 0$, and
corresponds to 
\begin{equation}
\left|{\rm det}\left(\nabla_{(k} s_{l)}^\mathrm{stiff}\right)\right|
\delta_{{\rm D}}(s_1) \delta_{{\rm D}}(s_2)=
\left|\frac{x_1 x_{112} x_{222}}{(\lambda_1-\lambda_2)}\right|
   \delta_{\rm D}(\lambda_2) \delta_{\rm D}(x_2)\,,
\end{equation}
hence
\begin{eqnarray}
\frac{\partial n_{\rm B }^{\rm F}}{\partial \eta} &=&
\frac{1}{ {\tilde R}^2} \int d \lambda_1 
P_0(\eta, \lambda_1, \lambda_2=0) \times
\int d x_1 d^{4} x_{klm} P_1(x_1,x_2=0,x_{klm})
|x_1 x_{112} x_{222}| ~+~ \left( 1 \to 2, ~ \eta \to -\eta \right)\,,
\nonumber \\
&=& \frac{1}{ {\sqrt{3} \pi^2 \tilde R}^2} 
\left[\frac{1}{\sqrt{2 \pi}\sqrt{1-2\gamma^2/3}} 
\exp\left(-\frac{\eta ^2}{2(1-2\gamma^2/3)}\right)\right]
\left[
\sqrt{1-\tilde\gamma^2} + \frac{1}{4} (2-3\tilde\gamma^2) \mathrm{atan}
\left(\frac{2-3\tilde\gamma^2}{4\sqrt{1-\tilde\gamma^2}}\right)
\right] \,,\nonumber \\
&\equiv& \frac{4}{{\tilde \gamma}^2 \pi^2} {\cal G}_B^F(\tilde \gamma)
\frac{\partial n_{\mathrm{saddle}}}{\partial \eta}\,.
\label{eq:nB2D-3}
\end{eqnarray}

The second situation (isotropic Hessian) corresponds to 
\begin{displaymath}
\left|{\rm det}\left(\nabla_{(k} s_{l)}^\mathrm{stiff}\right)\right|
\delta_{{\rm D}}(s_1) \delta_{{\rm D}}(s_2)= \frac{1}{4}
\left|\frac{x_1 (u_2^2-16(w_1^2+w_2^2))}{(\lambda_1-\lambda_2)}\right|
\delta_{\rm D}(\lambda_1 - \lambda_2) \delta_{\rm D}(x_2)
\,,
\end{displaymath}
therefore
\begin{eqnarray}
\frac{\partial n_{\rm B}^{\rm I}}{\partial \eta} &=&
\frac{1}{ {\tilde R}^2} \int d \lambda_1 
P_0(\eta, \lambda_1, \lambda_2=\lambda_1) \times
\int d x_1 d^{4} x_{klm} P_1(x_1,x_2=0,x_{klm})
\left|x_1 (u_2^2/4-4(w_1^2+w_2^2))\right|
 ~+~ \left( 1 \to 2, ~ \eta \to -\eta \right) \,,\nonumber \\
&=& \frac{1}{ {\pi \tilde R}^2} 
\left[\frac{1}{\sqrt{2 \pi}} \exp\left(-\frac{\eta ^2}{2}\right)\right]
\left[ \frac{2}{\sqrt{2-\tilde\gamma^2}} -\frac{1}{2} (1+ {\tilde \gamma}^2) 
\right]
\equiv \frac{1}{ {\tilde\gamma}^2 \pi R_*^2} {\cal G}_B^I(\tilde \gamma) P(\eta)\,.
\label{eq:nB2D-2}
\end{eqnarray}
Both $ {\cal G}_B^F(\tilde \gamma) $ and $ {\cal G}_B^I(\tilde \gamma) $
are weak functions of $\tilde\gamma$ of order unity. The main $\tilde
\gamma$ dependence $\propto {\tilde\gamma}^{-2}$ reflects $\tilde R$ as the
fundamental scale for the singular points.

We note that the number density of the ``sloping plateaux''
is proportional to the density of the saddle points, hence this type of
singular points is 
predominantly concentrated near mean field values (small $\eta$).
In contrast, the number density of ``bifurcation'' points is proportional just 
to the PDF of the field and, hence, the bifurcation points
are as frequent in the regions
of high field values as in the low ones. This may provide explanation
for the observed insensitivity of the curvature of the skeleton
to the threshold, if we conjecture that most of the curvature accumulates
near the ``bifurcation'' points.
   
\section{Conclusion \& perspectives}\label{sec:conclusion}

The filamentary structure  is a dramatic feature of the observed or simulated
Cosmic Web.  This paper investigated how the set of critical lines of a given
field corresponds to an intermediate representation of the field, which is more
extended than the knowledge of the critical points, but nevertheless much more
compact than the field itself.  It introduced the stiff approximation, which
states that the tangent vector to the critical lines only involves up to the
the second  derivative of the fields.  Within its framework it has been
demonstrated that, for stationary Gaussian random fields, ergodicity allows one
to recast the description of the ND critical lines into a point process, which
only involve the first spectral parameter, $\gamma$, when considering the
differential length as a function of the contrast, and the second spectral
parameter, $\tilde \gamma$, when considering it as a function of the modulus
of the gradient.  The former probability distribution was shown to involve
the average flux of  the Gaussian curvature of the 1D sections. In turn, these
averages can be carried out analytically almost to the last integral in 2D and
3D, and provide simple asymptotics at large and small contrast.  The detailed
contribution of all types of critical lines as a function of thresholding was
described.
The main results of this investigation corresponds to
equations~(\ref{eq:expl_hessian_deriv}) and (\ref{eq:deriv_explicit})
for the differential length of the skeleton and the total set of critical lines
in 2D and equations~(\ref{eq:3Ddiff}) and (\ref{eq:dLdt_3D_skel}) in 3D.
Their generalization to N dimensions is given by equation~(\ref{eq:NDdiff}) in
Appendix~\ref{sec:ND}. 
Table~\ref{tab:lengths} summarizes the average integrated fluxes (i.e length
per unit volume) of the critical lines.
\begin{table}
\begin{tabular}{lclc}
\multicolumn{2}{c}{2D} & \multicolumn{2}{c}{3D} \\
\hline 
Skeleton &$4.21 R_*$ & Skeleton: & $(4.65 R_*)^2$\\
Anti-skeleton & $4.21 R_*$ &Anti-skeleton: & $(4.65 R_*)^2$\\
&&Inter-skeleton: & $(3.38 R_*)^2$\\
\hline
All primary & $2.11 R_*$ & All primary & $(2.36 R_*)^2$\\
\hline
All secondary & $5.54 R_*$ & All secondary & $(3.02 R_*)^2$\\
\hline
Total & $1.55 R_*$ & Total & $(1.86 R_*)^2 $
\end{tabular}
\caption{Inverse average integrated flux
(the characteristic area (3D) or length (2D) per critical line)
of the critical lines of different types.}
\label{tab:lengths}
\end{table}
 For instance in 3D one expect on average one skeleton line crossing a random
$\approx(5 R_*)^2$ surface element.

These findings were illustrated on scale free power spectra with spectral
parameters which are relevant to cosmology\footnote{in other fields, the stiff
approximation might be less well motivated (see Section~\ref{sec:stiff}), 
but the calculations hold.}.
The prediction of the stiff approximation was checked against measurements
for global skeletons \citep{sousb08} on  realizations of these fields in two
and three dimensions and was found to be in good qualitative agreement.
The differential curvature of the corresponding lines was also measured
(section~\ref{sec:curv}) and the corresponding radii were found to be
$ \approx 8 R_*$  and $ \approx 2.5 R_* $ near $\eta=0$ in two and three
dimensions respectively.
Hence an access to both the curvature and the length of the skeleton provides
the means of constraining two shape parameters, $\gamma$ and $\tilde \gamma$.
The stiff approximation is also implemented to compute the differential
curvature in 2D.  Finally  (section~\ref{sec:bif}), the stiff theory of the
singular points of the critical lines was laid out in general,
identifying generically three types of points:  critical points of the
underlying field, bifurcation points and  slopping plateaux.
Again, the stiff approximation provide means of computing the number density
of these points.  Appendix~\ref{sec:cardoso} derived the general joint
probability of the field and its successive derivative in arbitrary dimensions,
which come into play when computing these higher order statistics.

Clearly the formalism developed in this paper will be useful in the context of
the upcoming surveys such as the LSST, or the
SDSS-3 BAO surveys since it  yields access to the shape of the power-spectrum
without artifacts related to varying light to mass ratio. 
For instance, \cite{skellet} first applied the corresponding theory to the
SDSS-DR4 catalogue in order to constraint the global dark matter content of
the universe, since the cosmological parameters are directly a function of
the spectral parameter, $\gamma$. Its application to CMB related full sky data,
such as WMAP or Planck should provide insight into, e.g. the level of
non-Gaussianity in these maps (see SPCNP for a discussion). 
Similarly, upcoming large scale weak lensing surveys could be analyzed in terms
of these tools \citep{ASKI}. 

A natural extension of the theoretical component of this work would be to
investigate the properties of the bifurcation points in anisotropic settings
{and extend beyond the stiff approximation the  preliminary  results of
Section~\ref{sec:bif}}.  This will be the topic of a forthcoming paper.
Another natural venue would be to also investigate the statistical properties
of, {\sl e.g.}  the peak patch walls (surface, curvature) defined as 
$x_{3}=0$ in the eigenframe of the Hessian. Eventually, a global theory of the
critical manifolds beyond the local approximation should also be developed to
provide a framework to study the connectivity of  the critical lines. 

\section*{Acknowledgments}
We thank D.~Aubert and K.~Benabed for comments and D.~Munro for freely
distributing his Yorick programming language and opengl interface (available
at {\em\tt http://yorick.sourceforge.net/}). DP thanks the CNRS (France) for
support through a ``poste rouge'' visiting position during Summer 2007 when
this investigation was originated. CP, TS, SP and CG also thank the hospitality
of the University of Alberta, and the ``Programme National de Cosmologie'' for
funding. Finally, CP and DP thank the Canadian Institute for Theoretical
Astrophysics for hosting the work involved in
finalizing this paper.  This   investigation  carried  within  the framework of
the  Horizon  project, \texttt{www.projet-horizon.fr}.

\bibliographystyle{mn2e}
\bibliography{ms}

\appendix

\section{The stiff skeletons of ND fields}\label{sec:ND}

The emphasis in this paper is on developing the analytical theory of the 
critical lines of a given GRF in two and three dimensions. Yet the critical
lines in higher dimensions are of interest in more abstract spaces such as
space-time or space-smoothing etc. . In 3+1 Dimensions, corresponding to
3D space+time, the 4D critical lines are the dynamical tracks of critical
points in  3D. An alternative view is to think of the 4D skeleton as event
lines of over densities, while the critical points correspond to the position
and time of merging events.  In fact \cite{hanami} explored sloping saddles
(i.e. points in position-smoothing space corresponding  degenerate saddle
points) as a mean of identifying merging events, and argued that the ridges
(the path of the maxima in position-smoothing space as a function of smoothing)
form a 4D skeleton.  Clearly these higher dimensional spaces 
would typically not be strictly isotropic, stationary nor Gaussian. 
As a first step, let us nonetheless investigate these N dimensional lines.

\subsection{Critical Lines in ND}
The local critical lines in $N$ dimensions are defined as
the points where the condition 
\begin{equation}
\vH\cdot \nabla
\rho = \lambda_{i} \nabla \rho \,,\label{eq:defS3N}
\end{equation}
is satisfied. This can be expressed as the condition of the vanishing
of the $N-2$ antisymmetric tensor 
\begin{equation}
\mathbf{S}=S^{i_1,i_2,\ldots,i_{N-2}} \equiv \sum_{klm} \nabla_k \rho {H^k}_l
\epsilon^{i_1, \ldots, i_{N-2},l,m} \nabla_m \rho = 0 
\label{eq:defS_A}
\end{equation}
as defined in equation~(\ref{eq:defS2}). In spaces of dimension $N > 4$
it is more compact to consider the Hodge-dual rank 2 tensor
\begin{equation}
(*\mathbf{S})=(*S)^{ij} = \frac{1}{(N-2)!} \sum_{i_1,\ldots,i_{N-2}}
S^{i_1,i_2,\ldots,i_{N-2}}
\epsilon_{i_1, \ldots, i_{N-2},i,j}  = 0 \quad .
\end{equation}
Local direction of the filament corresponds to the {\it right} null-vector
$ \delta r^k $
of the $N-2+1$-rank tensor  
of the derivatives of ${\mathbf S}$
\begin{equation}
\sum_k \left(\nabla_k S^{i_1,i_2,\ldots,i_{N-2}}\right) \delta r^k = 0 \quad . \label{eq:defgradSijk}
\end{equation}
A non-trivial solution of this set of $C^{2}_N$ homogeneous equations 
generally exists, 
since the existence of the {\it left} null-vector
$ \sum_{i_1} \nabla_{i_1} \rho \left( \nabla_k S^{i_1,i_2,\ldots,i_{N-2}} \right) = 0 $
imposes $C^{2}_{N-1}$ linear relations leaving exactly
$C^{2}_N-C^{2}_{N-1}=N-1$ independent equations to define a line.

The notion of {\it primary} skeleton lines is automatically generalized for
$N$ D as the subset of critical lines obeying
\begin{equation}
\vH\cdot \nabla
\rho = \lambda_{1} \nabla \rho\,, \quad{\rm and} \quad
\lambda_1 + \lambda_2  \le 0 \,,\label{eq:defS2N}
\end{equation}
where $\lambda_{1}$ is the largest and $\lambda_2$ is
the second largest of the sorted eigenvalues.

Let us derive the general expression for the statistical
average of the flux of
the lines arbitrarily defined over the properties of the ND random field 
by $N-1$ equations $S^i = 0, ~i=1 \cdots N-1$, where 
$S^i(x,x_k,x_{kl},\ldots) $ are functions of the field and it's derivatives. 
We can shortcut the procedure of flux evaluation 
by marking each line with one intersection point
with a fiducial surface $\Sigma$, orthogonal to it, and finding the
N-1 number density of the intersection points on the surface $\Sigma=0$.
The average number density of the points defined as the intersection of $n$ 
non degenerate hypersurfaces $\sigma^i,\ldots,\sigma^N$  is given by
\begin{equation}
n=\int d x d^N x_k \cdots P(x,x_k,\cdots) \delta_{\rm D}(\sigma^1)\cdots
\delta_{\rm D}(\sigma^n) |\mathrm{det}(\nabla \sigma^1, \cdots \nabla\sigma^n)|\,.
\end{equation}
Let us  choose $S^1\cdots S^{N-1}$ as $\sigma^2\cdots \sigma^n$
and $\Sigma$ to be $\sigma^N$. 
Expanding the determinant 
$ |\mathrm{det} (\nabla S^1, \cdots, \nabla S^{N-1},\nabla \Sigma)|$
along its last row we obtain
\begin{equation}
n=\int d x d^N x_k \cdots  P(x,x_k,\cdots) 
\delta_{\rm D}(S^{1})\cdots  \delta_{\rm D}(S^{N-1})
\delta_{\rm D}(\Sigma)
\left|\sum_k  M^k\,\, \nabla_k \Sigma \right| \,,
\label{eq:defM}
\end{equation}
where
\begin{equation}
M^k=\, (-1)^{k+1}\mathrm{det} 
\left( \nabla_l S^i \right)^{i=1,\ldots,N-1}_{l=1,\ldots,k-1,k+1,\ldots,N}
=\sum_{l_1,\ldots,l_{N-1}} \epsilon^{k,l_1,\ldots,l_{N-1}}
\nabla_{l_1} S^1 \ldots \nabla_{l_{N-1}} S^{N-1}
\end{equation}
are the corresponding minors.
By design the $\Sigma=0$ surface is to be orthogonal to
the line and therefore its normal $\boldsymbol{\nabla}\Sigma$ and the ``vector''
$\mathbf M\equiv (M^k)$ are parallel, 
\begin{displaymath}
\left|\sum_k  M^k\,\, \nabla_k \Sigma \right| = 
\left| \mathbf{M} \right| \left|\boldsymbol{\nabla}\Sigma\right| \quad .
\end{displaymath}
Without loss of generality, we can consider the intersection point to be
at $\mathbf{r}=0$ and take $\Sigma =\mathbf{e}_{\Sigma} \cdot \mathbf{r}$
where $\mathbf{e}_{\Sigma} $
is the unit vector in the local direction of the filament, hence
$|\boldsymbol{\nabla}\Sigma|=1$.
The average (N-1)D number density of intersection points on $\Sigma$ surface
that gives us the average flux ${\cal L}$ is obtained
by integrating the volume number density over the coordinate along
$\mathbf{e}_\Sigma$, $z=\mathbf{e}_\Sigma \cdot \mathbf{r}$,
with $\delta_D(\Sigma)$ in equation~(\ref{eq:defM}) properly counting 
exactly one intersection per line
\begin{equation}
{\cal L}\equiv \int n \; d (\mathbf{e}_\Sigma \cdot \mathbf{r})
=\int dx d^N x_k \cdots  P(x,x_k,\cdots)
\delta_{\rm D}(S^1)\cdots  \delta_{\rm D}(S^{N-1}) 
\left| \mathbf{M} \right|  \quad .
\label{eq:fluxND}
\end{equation}
To apply this general formula to the critical lines one must choose an
arbitrary subset of $N-1$ linearly independent
$\nabla S^{i_1,i_2,\ldots,i_{N-2}}$ from the set of all $C^{2}_N$ of them. 

Note that one can also think of ${\cal L}$ as the average length of lines
per unit volume, which is the interpretation we focus on in the main text.

\subsection{Stiff critical lines in ND}
In the theory of ND critical lines, the N-1 independent functions 
$S^i$ that define the critical condition (\ref{eq:defS_A}) acquire
the following simple form in the eigenframe of the Hessian of the field
\begin{equation}
s^i= x_a x_i \left(\lambda_a-\lambda_i\right)=0, \quad i \ne a \quad .
\end{equation}
Here $a$ is the index of the Hessian eigenvector that the gradient
is aligned with, as is obvious from the solution $x_i=0, ~ i \ne a $.

In the stiff approximation, the gradients ${s^i}_k \equiv
\nabla_{} s^i $ have just two non-zero components,
${s^i}_a= x_i \lambda_a (\lambda_a-\lambda_i)$
(which vanishes on the critical line) 
and ${s^i}_i = x_a \lambda_i (\lambda_a-\lambda_i) $.
The vector that determines the direction of the critical line becomes
\begin{equation}
M^k=x_a^{N-2} x_k \prod_{i\neq k} \lambda_i
\prod_{i \neq a}  \left(\lambda_a-\lambda_i\right)
\quad .
\label{eq:Sk}
\end{equation}
On the critical line, it has just one non-vanishing component
\begin{equation}
M^a=x_a^{N-1} \prod_{i\neq a} \lambda_i \left(\lambda_a-\lambda_i\right) =
\left | \mathbf{M} \right |\,,
\label{eq:Mstiff}
\end{equation}
which shows that in the stiff approximation we equate the direction of the
line with the gradient of the field. Substituting this expression into
equation~(\ref{eq:fluxND}) and integrating over $\delta_{\rm D}(s^i)=
 \delta_{\rm D}(x_i)/({x_a (\lambda_a-\lambda_i)})$
we obtain a simple expression for the flux of the critical lines
in the stiff approximation
\begin{equation}
{\cal L}
=\int dx d x_{kl} P(x,0,x_{kl})
\left| \prod_{i\neq a} \lambda_i \right | \quad .
\label{eq:fluxNDstiff}
\end{equation}
i.e the flux of critical lines (or the length
per unit volume) is given by the average absolute value of the
Gaussian curvature of the field in the space orthogonal to the skeleton.
 
Let us write the probability of measuring the set $\{\lambda_{i}\}$ as 
\begin{equation}
\prod_{i\le {\rm N}}  d \lambda_i 
 \prod_{i<j} (\lambda_i-\lambda_j)
  \exp\left( -\frac{1}{2}
Q_{\gamma}(\eta,\{\lambda_{i}\})
\right)\,,
\end{equation}
where $Q_{\gamma}$ is a quadratic form in $\lambda_{i}$ and $\eta$ which
functional form is
\begin{equation}
Q_{\gamma}(\eta,\{\lambda_{i}\})=\eta^2+ \frac{\left(\sum_{i}\lambda_{i}+\gamma \eta\right)^2}{ (1-\gamma^2)}+
{\cal Q}_{N}(\{\lambda_{i}\})\,,
 \end{equation}
and  $ \prod_{i<j} (\lambda_i-\lambda_j)$ is the Jacobian of the transformation to the  Hessian eigenframe.
Here ${\cal Q}_{ N}$ involves polynomial combinations of the eigenvalues of the traceless part of the Hessian (see Appendix~\ref{sec:cardoso}):
\begin{equation}
{\cal Q}_{N}(\{x_{ij}\})= \frac{N(N+2)}2\
\sum_{ij}  \overline x_{ij}
  \overline x_{ij}   \,,{\quad} {\rm with}\quad \overline x_{ij} = x_{ij} -\delta_{ij} \frac{1}{N}\sum_{i} x_{ii}\,,
\end{equation} 
  which can be rearranged explicitly in terms of $\lambda$s as:
\begin{equation}
{\cal Q}_{N}(\{\lambda_{i}\})= (N+2)\left[\frac{1}{2}(N-1) \sum_{i} \lambda^{2}_{i}- \sum_{i\neq j} \lambda_{i}\lambda_{j}\right].
\end{equation}
It now follows that the differential length of the ND-critical lines is for the stiff approximation: 
\begin{equation}
\frac{\partial {\cal L}^{\mathrm{ND}}}{\partial \eta}\propto
\left( \frac{1}{R_{*}}\right)^{N-1}\frac{1}{\sqrt{1-\gamma^2}}
\int \cdots \int \prod_{i\le n}  d \lambda_i 
 \prod_{i<j} (\lambda_i-\lambda_j)
\left| \prod_{i>1}   \lambda_i \right|  \exp\left( -\frac{1}{2}
Q_\gamma(\eta,\{\lambda_{i}\})
\right)\,. \label{eq:NDdiff}
\end{equation}
Equation~(\ref{eq:NDdiff}) is the formal generalization of
equations~(\ref{eq:skeleton}) and (\ref{eq:3Dlength}).
For the ND-skeleton, equation~(\ref{eq:NDdiff}) also holds but the integration
region should be restricted to the corresponding condition on the sign of
the eigenvalues.
Since the argument of $Q_\gamma$ is extremal as a function of $\eta$ when
$\gamma \eta \sim \sum_i \lambda_i$, the largest contribution at large
$\gamma \eta$ in the integral should arise when $\lambda_i \propto \gamma \eta$
since near the maximum at high contrast all eigen values are equal \citep{PB}. 
Hence given that $\prod_{i<j} (\lambda_i-\lambda_j)$ is  the measure, the only
remaining contribution in the integrand comes from
$\left| \prod_{i>1}   \lambda_i \right|  \propto (\lambda \eta)^{N-1}$, and
the dominant term at large $\eta$ is given by 
\begin{displaymath}
\frac{\partial {\cal L}^{\mathrm{ND}}}{\partial \eta}\stackrel{\gamma \eta \to \infty}{\sim}
\frac{1}{\sqrt{2 \pi}} \exp \left[-\frac{1}{2} \eta^ 2\right] \left(\frac{\eta}{R_0}\right)^{N-1}\,,
\end{displaymath}
where $R_0=R_*/\gamma $ is defined in equation~(\ref{eq:defR0}).

\begin{figure*}
  \centering
  \includegraphics[angle=0,width=8cm]{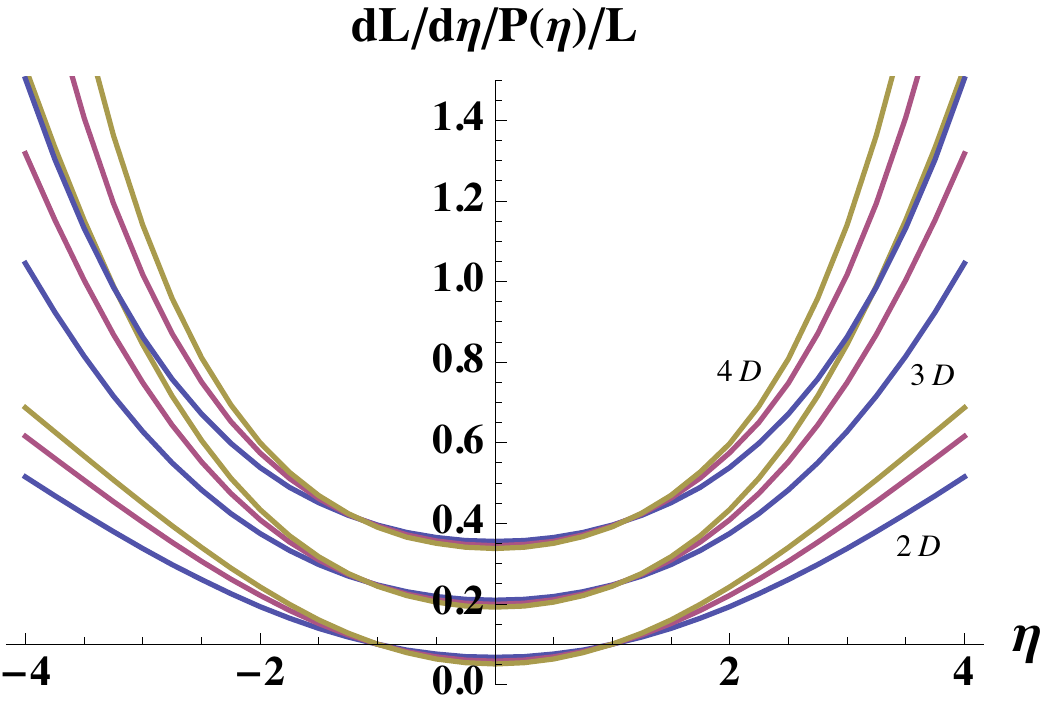}
  \caption{${\partial {\cal L}}/{\partial \eta}/ {P(\eta)/{ L}}$ 
  in 2D, 3D and 4D as labeled for the spectral parameter values
  $\gamma=0.57,0.65,0.70$ ({\sl from bottom to top}).
  These quantities are derived here by direct numerical integration of
  equation~(\ref{eq:NDdiff}).  The different bundles corresponding to different
  dimensions have been shifted down (by $0.3$ for 2D and $0.15$ for 3D) for
  clarity. Note the change in the power of the asymptotic curves.
  }
\label{fig:4Dlength}
\end{figure*}

\section{Secondary critical lines in 2D}\label{app:2Dcrit}
In this Appendix we present a study of asymptotic behaviour of the
lengths statistics of secondary critical lines for 2D Gaussian field. 
Secondary critical lines are the ones that have a gradient of the field aligned
with the Hessian eigenvector that corresponds to the largest by magnitude
eigenvalue, i.e with the direction of maximum curvature of the field.
In 2D, this is the direction of $\lambda_2$ in the skeleton region, 
$|\lambda_1| < |\lambda_2|$, and is the direction of $\lambda_1$ 
in the anti-skeleton region. We shall explicitly consider the first type,
realizing that the second type is a mirror case with $\eta \to - \eta$.
 
Our starting point is the part of  expression~(\ref{eq:expl_hessian_deriv})
that corresponds to the lines where 
the gradient is aligned with the second eigen-direction, in the
region when they are secondary, $\tilde u > 0$
\begin{equation} 
\frac{\partial {\cal L}^\mathrm{sec}}{\partial \eta} = 
\frac{4 \sqrt{2}}{(2 \pi)^{3/2}\sqrt{1-\gamma^2}} \exp \left[-\eta^2/2 \right] 
\int_0^\infty \!\! d \tilde w \tilde w 
\int_{0}^\infty \!\!\!\! d\tilde u
\left| 2 \tilde w - \tilde u \right|  
\exp \left[-\frac{(\tilde u-\gamma \eta)^2}{2 (1-\gamma^2)} - 
 4 \tilde w^2 \right] \quad .
\end{equation}
The absolute value of the transverse to the gradient curvature
$2 \lambda_1=2 \tilde w - \tilde u$ is evaluated differently for 
$\tilde u \le 2 \tilde w$ and $ \tilde u > 2 \tilde w$.
It is convenient to make the 
inner integration to be over $\tilde w$, since it can be carried out
analytically.  The integral splits into two terms
\begin{equation} 
\frac{\partial {\cal L}^{\mathrm{sec}}}{\partial \eta} = 
\frac{4 \sqrt{2}}{(2 \pi)^{3/2}\sqrt{1-\gamma^2}} \exp \left[-\eta^2/2 \right] 
\left( I_1+ I_2 \right)\,,
\end{equation}
where
\begin{eqnarray}
I_1 &=& \int_0^\infty \!\! d \tilde u 
\exp\left[-\frac{(\tilde u-\gamma \eta)^2}{2(1-\gamma^2)} \right]
\int_{\tilde u/2}^\infty \!\! \tilde w (2 \tilde w -\tilde u) 
d \tilde w e^{-4 \tilde w^2}
= \frac{\sqrt{\pi}}{16} \int_0^\infty \!\! d \tilde u 
\exp\left[-\frac{(\tilde u-\gamma \eta)^2}{2(1-\gamma^2)} \right]
\mathrm{Erfc}(\tilde u) \,,
\\
I_2 &=& \int_0^\infty \!\! d \tilde u 
\exp\left[-\frac{(\tilde u-\gamma \eta)^2}{2(1-\gamma^2)} \right]
\int_0^{\tilde u/2} \!\!\!\! \tilde w (\tilde u - 2 \tilde w) 
d \tilde w e^{-4 \tilde w^2}
= \frac{\sqrt{\pi}}{16} \int_0^{\infty} \! d \tilde u 
\exp\left[-\frac{(\tilde u-\gamma \eta)^2}{2(1-\gamma^2)} \right]
\left[ \frac{2}{\sqrt{\pi}} \tilde u - \mathrm{Erf}(\tilde u) \right]\,.
\end{eqnarray}
so that finally
\begin{equation}
\frac{\partial {\cal L}^{\mathrm{sec}}}{\partial \eta} = 
\frac{1}{4 \pi^{3/2}\sqrt{1-\gamma^2}} \exp \left[-\eta^2/2 \right] 
\int_0^\infty \!\! d \tilde u \, 
\exp\left[-\frac{(\tilde u-\gamma \eta)^2}{2(1-\gamma^2)} \right]
\left[\tilde u  - \sqrt{\frac{\pi}{4}} \mathrm{Erf}(\tilde u) 
+ \sqrt{\frac{\pi}{4}} \mathrm{Erfc}(\tilde u) \right]  \quad .
\end{equation}

The integrated length of critical lines $L^{\mathrm{sec}}$ is obtained
by marginalization over all threshold values $\eta$.
Performing this integration first
\begin{equation}
L^{\mathrm{sec}} = \frac{1}{2 \sqrt{2} \pi} 
\int_0^\infty \!\! d \tilde u  \exp\left[-\frac{{\tilde u}^2}{2}\right] 
\left[\tilde u  - \sqrt{\frac{\pi}{4}} \mathrm{Erf}(\tilde u) 
+ \sqrt{\frac{\pi}{4}} \mathrm{Erfc}(\tilde u) \right] 
= \frac{\sqrt{2} - \mathrm{ArcCot}(2 \sqrt{2})}{4 \pi} = 0.08550 
\end{equation}
Thus secondary critical lines are on average almost three times rarer
that the primary ones.

\subsection{Special cases: $\eta \to \infty$}
At high density thresholds the leading asymptotic behaviour 
for $\gamma \eta \gg 1$ is obtained by
using $\frac{1}{\sqrt{2 \pi (1-\gamma^2)}}
\exp\left[-\frac{(\tilde u-\gamma \eta)^2}{2(1-\gamma^2)} \right]
\stackrel{\eta \to \infty}{\to} \delta_{\rm D}( \tilde u-\gamma \eta) $.
Therefore
\begin{equation} 
\frac{\partial {\cal L}^{\mathrm{sec}}}{\partial \eta} \stackrel{\eta \to \infty}{\sim}
\frac{1}{\sqrt{2 \pi}} \exp \left[-\eta^2/2 \right] 
\frac{1}{4} \left(\frac{2}{\sqrt{\pi}}\gamma \eta -1 \right) \quad .
\label{eq:dlt_to_infty}
\end{equation}

\subsection{Special cases: $\eta \to 0$}
\begin{figure}
  \centering
  \includegraphics[angle=0,width=8cm]{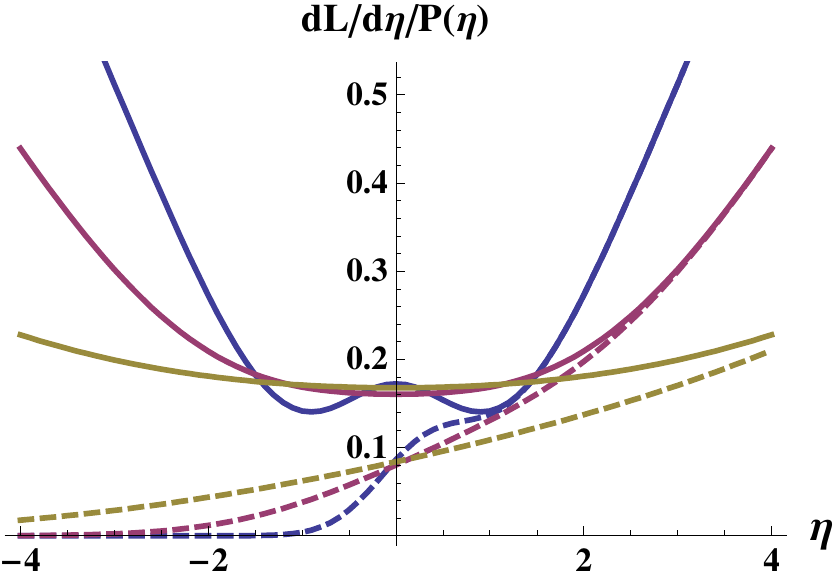}
    \caption{Differential length of the secondary critical lines 
    ${\partial {\cal L}}/{\partial \eta}/\mathrm{PDF}$ in 2D for the complete
    set (solid) and the ones with $\nabla \rho$ aligned with $\lambda_2$ 
    direction in the skeleton $|\lambda_1| \le |\lambda_2| $ region (dashed). 
    Different curves from purple to green correspond to
  the spectral parameter values $\gamma=0.3,0.6,0.95$.}
\label{fig:dLdtsec}
\end{figure}

At small threshold $\eta$ series representation
\begin{equation}
\exp\left[-\frac{(\tilde u-\gamma \eta)^2}{2(1-\gamma^2)} \right] =
\exp\left[-\frac{\tilde u^2}{2(1-\gamma^2)}\right]
\sum_{n=0}^\infty \frac{1}{n! (1-\gamma^2)^{n/2}}
H_{n} \left(\frac{\tilde u}{\sqrt{1-\gamma^2}}\right)
(\gamma \eta)^{n} ~,
\label{eq:Hermite_exp}
\end{equation}
where Hermite polynomials $H_{2n}$ are taken in probabilistic notation, gives 
\begin{eqnarray}
\frac{\partial {\cal L}^{\mathrm{sec}}}{\partial \eta}(\eta \to 0)\!\!\!\! &=&
\!\!\!\!  \frac{1}{\sqrt{2 \pi}} \exp \left[-\eta^2/2 \right] 
\sum_{n=0}^\infty A_{n} (\gamma \eta)^{n} \quad \mathrm{where} \nonumber \\
A_{n} &\equiv & \!\!\!\!\! \frac{1}{2 \sqrt{2} \pi n! (1-\gamma^2)^{n/2}} 
\int_0^{\infty}\!\!\! d \bar u \exp\left[-\bar u^2/2\right]
H_{n}(\bar u ) \left(\sqrt{1-\gamma^2} \bar u 
- \sqrt{\frac{\pi}{4}} \mathrm{Erf}\left[\sqrt{1-\gamma^2} \bar u\right] 
+ \sqrt{\frac{\pi}{4}}
\mathrm{Erfc}\left[\sqrt{1-\gamma^2} \bar u\right] \right) \quad .
\label{eq:tto0}
\end{eqnarray}
The first three coefficients are
\begin{eqnarray}
A_{0} & =& \frac{\sqrt{2 (1-\gamma^2)} + 
\mathrm{acot}[\sqrt{2(1-\gamma^2)}]-
\mathrm{atan}[\sqrt{2(1-\gamma^2)}]}{4 \pi} ~,\quad \nonumber \\
A_{1} &=& \frac{1}{4 \sqrt{\pi}}
\left(1+\frac{1}{\sqrt{2 (1-\gamma^2)}}- \frac{2}{\sqrt{(3-2 \gamma^2)}}
\right)~, \quad \nonumber \\
A_{2}  &=& \frac{\sqrt{2}}{8 \pi} \frac{1-2\gamma^2}{3-2 \gamma^2}
(1-\gamma^2)^{-\frac{1}{2}} ~,\quad
\end{eqnarray}
If we add all secondary critical lines, the odd power terms of the 
expansion~(\ref{eq:Hermite_exp}) cancel, while the even double recovering 
symmetrical behaviour of the differential length with the threshold.
In Figure~\ref{fig:dLdtsec} this behaviour is illustrated.

Under our definition of the secondary critical lines, for $\gamma > 1/\sqrt{2}$
there is an excess of  critical lines near zero threshold.
The curvature at $\eta=0$ is positive and diverges in the limit $\gamma \to 1$
when our series expansion formally fails. This divergence in the second
derivative of the differential length is exactly opposite the one the primary 
lines demonstrate in this limit. We should emphasize, that near $\eta=0$ the 
behaviour of critical lines of individual type depend significantly on 
how exactly they are defined. 

\section{Asymptotic behaviour of critical lines in 3D}\label{sec:3D_crit}

\begin{figure*}
  \centering
  \subfigure{\includegraphics[angle=0,width=5cm]{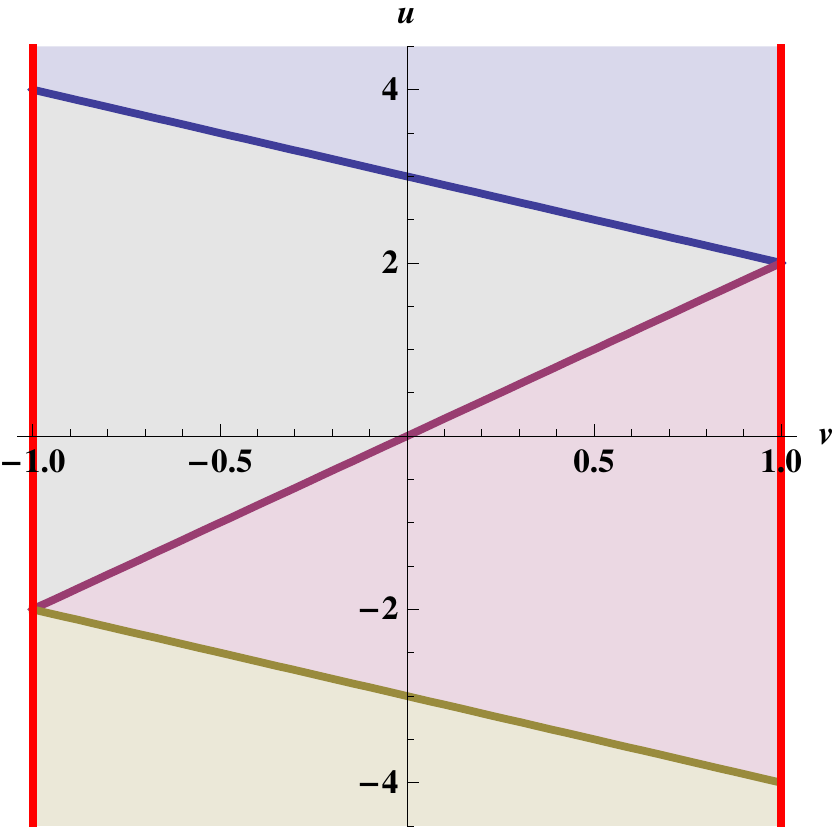}}
  \subfigure{\includegraphics[angle=0,width=5cm]{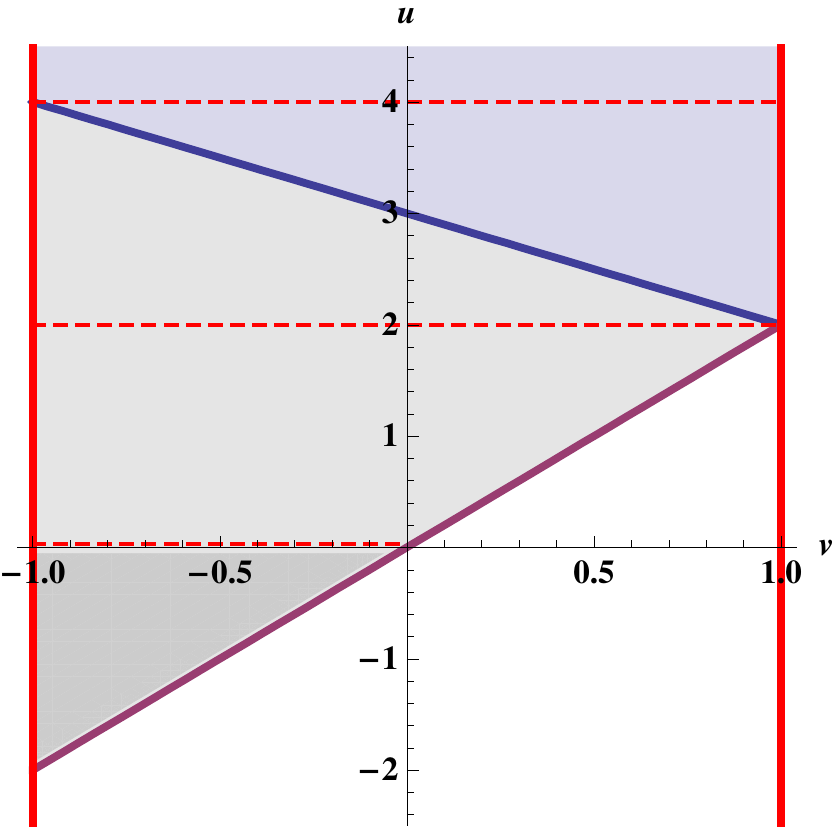}}
  \subfigure{\includegraphics[angle=0,width=5cm]{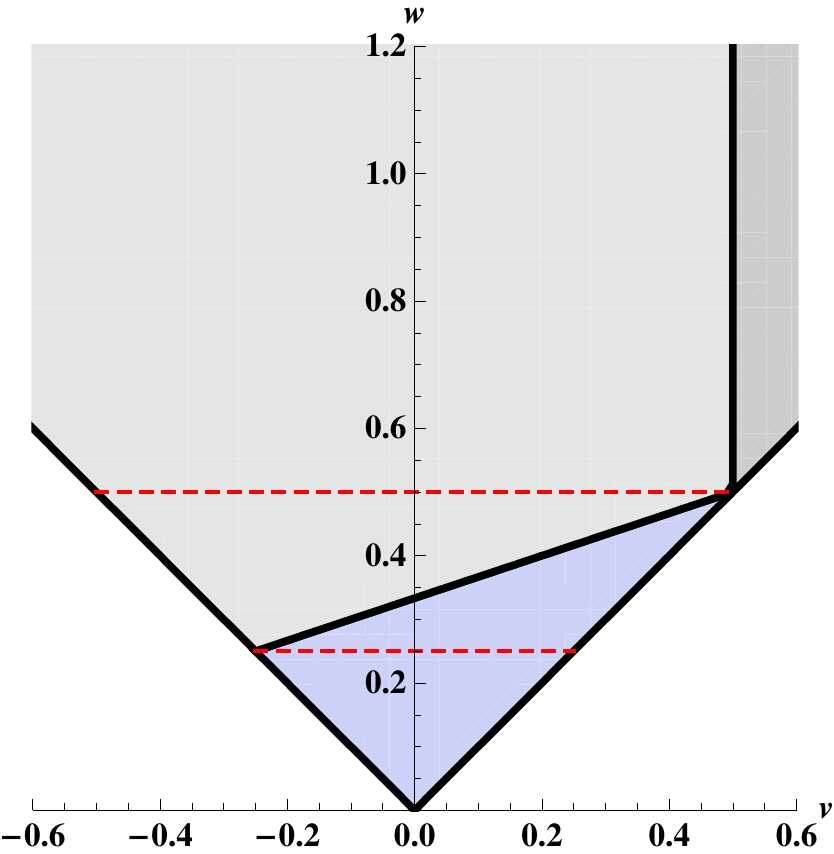}}
  \caption{{\sl Left}: Integration zones in $\tilde v, \tilde u$ plane based on the
  signs of the eigenvalues. Variables are given in units of $\tilde w$.
 Here  $\tilde v$ varies from $-\tilde w$ to $+\tilde w$, while $\tilde u$ is 
  unrestricted. Three inclined lines are (from top to bottom) 
  a) $\lambda_1 = 0 \Rightarrow \tilde u = -\tilde v + 3 \tilde w $,
  b) $\lambda_2 = 0 \Rightarrow \tilde u = 2 \tilde v $ and
  c)
  $\lambda_3 = 0 \Rightarrow \tilde u = -\tilde v - 3 \tilde w $.
  In the upper sector I (that stretches to infinity in $\tilde u$) 
  $( 0 > \lambda_1 \ge \lambda_2 \ge \lambda_3 )$, next is the zone II
  $(\lambda_1 \ge 0, 0 > \lambda_2 \ge \lambda_3 ) $, then III
  $(\lambda_1 \ge \lambda_2 \ge 0, 0 > \lambda_3 ) $, and, finally 
  extending to minus infinity in $\tilde u$ is the sector IV
  where $(\lambda_1 \ge \lambda_2 \ge \lambda_3 \ge 0)$.
  {\sl Centre}: two zones of integration after variable change leading to 
  equation~(\ref{eq:wvu_integral}). Horizontal dashed lines mark the further
  subdivision of the integration space if the order of integration is
  changed according to equation~(\ref{eq:uwv_integral}).
  {\sl Right}: Integration zones in the $(\tilde v , \tilde w )$ plane
  after $\tilde u$ has been mapped to the $[0-\infty]$ interval.
  Variables are given in units of $\tilde u$.
  The lower triangular zone corresponds to semi-open upper band in 
  $\tilde v-\tilde u$ of the centre panel. In this
  region, the integrand is given by  the terms I+III of
  equation~(\ref{eq:wvu_integral}).  Vice-versa, the open upper band in
  the $(\tilde v , \tilde w)$ plane corresponds to the II+IV integration
  over the lower triangular zone in  the $\tilde v-\tilde u$ space.
  The right-most sector of this region, however, corresponds to negative 
  $\tilde u$, so the integrand in this sector 
  has coordinate change $\tilde u \to -\tilde u, \tilde v \to -\tilde v$
  The dashed lines show the subdivided integrals given in 
  equation~(\ref{eq:uwv_integral}), which corresponds to subdivisions in
  the centre panel.
  }
\label{fig:uv_plane}
\end{figure*}

There are four regions with the different signs of sorted eigenvalues in
3D:
I~---~$( 0 > \lambda_1 \ge \lambda_2 \ge \lambda_3 )$,
II~---~$(\lambda_1 \ge 0, 0 > \lambda_2 \ge \lambda_3 ) $,
III~---~$(\lambda_1 \ge \lambda_2 \ge 0, 0 > \lambda_3 ) $ and
IV~---~$(\lambda_1 \ge \lambda_2 \ge \lambda_3 \ge 0)$. 
Since $\tilde w$ is non-negative, the correspondent zones of integration for
equations~(\ref{eq:3Ddiff}) and (\ref{eq:3Dlength}) 
are easy to visualize 
in $(\tilde v, \tilde u)$ plane (see Figure~\ref{fig:uv_plane}).
The integration limits and the integrand acquire the following form
\begin{equation}
\left.
\begin{array}{rl}
I: & \displaystyle \int_0^\infty d \tilde w \int_{-\tilde w}^{\tilde w} d \tilde v
\int_{-\tilde v + 3 \tilde w}^\infty d \tilde u  \;
\left[\frac{1}{3} \tilde u^2 - \frac{1}{3} \tilde v ^2 -\tilde w^2 \right] \\
II: &  \displaystyle \int_0^\infty d \tilde w \int_{-\tilde w}^{\tilde w} d\tilde v
\int_{2 \tilde v}^{-\tilde v + 3 \tilde w} d \tilde u
\left[ \tilde w^2 - \frac{1}{9} ( \tilde u + \tilde v)^2 + \frac{2}{3} \tilde w
( \tilde u - 2 \tilde v)\right]\\
III: & \displaystyle  \int_0^\infty d \tilde w \int_{-\tilde w}^{\tilde w} d \tilde v
\int_{-\tilde v - 3 \tilde w}^{2 \tilde v} d \tilde u 
\left[ \tilde w^2 - \frac{1}{9} ( \tilde u + \tilde v)^2 - \frac{2}{3} \tilde w
( \tilde u - 2 \tilde v)\right]\\
IV: & \displaystyle  \int_0^\infty d \tilde w \int_{-\tilde w}^{\tilde w} d \tilde v
\int_{-\infty}^{-\tilde v - 3 \tilde w} d \tilde u \;
\left[\frac{1}{3} \tilde u^2 - \frac{1}{3} \tilde v ^2 -\tilde w^2 \right]
\end{array}
\right\} \times \tilde w \left(\tilde w ^2 - \tilde v ^2\right)
\exp \left[-\frac{(\tilde u-\gamma \eta)^2}{2 (1-\gamma^2)} - 
 \frac{15}{2} \tilde w^2  - \frac{5}{2} \tilde v^2  \right] \; .
\end{equation}
Changing variables $\tilde u \to - \tilde u, \tilde v \to - \tilde v$
in III and IV one can combine the last two cases with the first two
\begin{equation}
\left.
\begin{array}{rl}
I+IV: & \displaystyle  \int_0^\infty d \tilde w \int_{-\tilde w}^{\tilde w} d \tilde v
\int_{3 \tilde w -\tilde v}^\infty d \tilde u  \;
\left[\frac{1}{3} \tilde u^2 - \frac{1}{3} \tilde v ^2 -\tilde w^2 \right] \\
II+III: &  \displaystyle  \int_0^\infty d \tilde w \int_{-\tilde w}^{\tilde w} d\tilde v
\int_{2 \tilde v}^{3 \tilde w -\tilde v} d \tilde u
\left[ \tilde w^2 - \frac{1}{9} ( \tilde u + \tilde v)^2 + \frac{2}{3} \tilde w
( \tilde u - 2 \tilde v)\right]
\end{array}
\right\} \times \tilde w \left(\tilde w ^2 - \tilde v ^2\right)
\exp \left[-\frac{15}{2} \tilde w^2  - \frac{5}{2} \tilde v^2  \right] 
\Phi_\gamma(\tilde u, \eta) \; .
\label{eq:wvu_integral}
\end{equation}
where \[\Phi_\gamma(\tilde u, \eta) = \exp \left[-\frac{(\tilde u-\gamma \eta)^2}{2 (1-\gamma^2)}\right]+
 \exp \left[-\frac{(\tilde u+\gamma \eta)^2}{2 (1-\gamma^2)} \right]\,.\]

Direct evaluation of the integrated length gives
\begin{equation}
L = 0.289627  ~ (\times R_*^{-2}) \quad.
\end{equation}

To study high threshold regime
it is advantageous to make the $\tilde u$ integration the outmost one,
since it depends on the variable threshold
\begin{eqnarray}
I+IV: && \int_0^\infty \!\! d \tilde w 
\int_{-\tilde w}^{\tilde w} \!\! d \tilde v
\int_{3 \tilde w -\tilde v}^\infty \!\!\!\!\!\!\! d \tilde u  \to
\int_0^\infty \!\! d\tilde u \int_0^{\tilde u/4} \!\! d\tilde w 
\int_{-\tilde w}^{\tilde w} d \tilde v  + 
\int_0^\infty \!\! d\tilde u \int_{\tilde u/4}^{\tilde u /2} d \tilde w  
\int_{3\tilde w -\tilde u}^{\tilde w} d \tilde v  \label{eq:I+IV}  \\
II+III: && \int_0^\infty \!\!  d \tilde w  
\int_{-\tilde w}^{\tilde w} \!\! d \tilde v
\int_{2 \tilde v}^{3 \tilde w -\tilde v} \!\!\!\!\!\!\!\!\! d \tilde u  \to
\int_0^\infty \!\! d\tilde u \int_{\tilde u/4}^{\tilde u /2} \!\!\!\! d \tilde w  
\int_{-\tilde w}^{3\tilde w -\tilde u} \!\!\!\!\!\!\! d \tilde v  + 
\int_0^{\infty} \!\!\! d \tilde u  \int_{\tilde u/2}^\infty \!\! d\tilde w
\int_{-\tilde w}^{\tilde u/2} \!\!\! d \tilde v +
\int_0^{\infty} \!\!\! d \tilde u  \int_{\tilde u/2}^\infty \!\! d\tilde w
\int_{\tilde u/2}^{\tilde w} \!\!\! d \tilde v
(\tilde u,\tilde v \to -\tilde u, -\tilde v) \label{eq:II+III} \quad . \nonumber
\label{eq:uwv_integral}
\end{eqnarray}
The parenthesis in the last term indicate the substitution that  must be performed in the integrand.
Right panel in Figure~\ref{fig:uv_plane} illustrates the integration zones
now in the $(\tilde v , \tilde w )$ plane.

Although one can perform the $\tilde v$ integral analytically and reduce
the problem to two-dimensional integration,
the resulting expression is too cumbersome. We can obtain useful limits 
already from unreduced formulae.
In particular, at high density threshold, $\gamma \eta \to \infty$,
only the first integral in the term~(\ref{eq:I+IV}), which contains
$\tilde w,\tilde v \sim 0$ neighbourhood, is not exponentially small.
Moreover, in the leading order the upper limit of the integral
over $\tilde w$ can be set to infinity. 
\begin{eqnarray}
\frac{\partial {\cal L}}{\partial \eta}
&\stackrel{\gamma \eta \to \infty}{\sim}&
\frac{3^2 5^{5/2}} {4 \pi^2 \sqrt{2 \pi (1-\gamma^2)}} 
\exp \left[-\frac{1}{2} \eta^ 2\right] 
\int_0^\infty \!\!\!\!\! d \tilde u \int_0^\infty \!\!\!\!\!\!\!\!\ d \tilde w
\int_{-\tilde w}^{\tilde w} \!\!\!\!\! d \tilde v
\tilde w (\tilde w^2 - \tilde v^2) 
\left[\tilde u^2 - \tilde v ^2 -3 \tilde w^2 \right]
 \exp \left[-\frac{(\tilde u-\gamma \eta)^2}{2 (1-\gamma^2)} - 
 \frac{15}{2} \tilde w^2  - \frac{5}{2} \tilde v^2  \right]
\nonumber \\
&\stackrel{\gamma \eta \to \infty}{\sim}&
\frac{1}{\sqrt{2 \pi}} \exp \left[-\frac{1}{2} \eta^ 2\right] 
\frac{(\gamma \eta)^2-\gamma^2}{2 \pi}\,.
\end{eqnarray}


\section{Joint distribution of the field and its derivatives for a GRF}
\label{sec:cardoso}

\newcommand\smallheader[1]{\medskip\par\noindent\underline{#1}}

The joint point distribution functions that are needed for the study
of the critical lines in this paper are ${P}_{0}(x,x_{kl})$ and
${P_{1}}(x_{i},x_{ijk})$, taking into account that for Gaussian random
field there is no cross-correlation between odd order derivatives and
the field itself or even order derivatives.  When considering the
curvature of the critical lines, fourth order derivatives, and, thus,
more general ${P}_{0}(x,x_{kl},x_{klmn})$ have to be considered.  Some
well known results in 2D and 3D are first summarized in
section~\ref{sec:jdfprevres}.  More general results can be obtained by
resorting to a general framework which is sketched in
section~\ref{sec:jdftheo} and applied in section~\ref{sec:jdfappli}
for the various cases of interest.

\subsection{ Lower order joint distributions
}\label{sec:jdfprevres}

\textbf{Distribution of the Gaussian field and its second derivative
in 3D.}
The full expression for ${P}_{0}(x,x_{kl})$ for the Gaussian field 
is given in \cite{1986ApJ...304...15B}. Introducing the variables
\begin{equation}
u  \equiv - \Delta x = -(x_{11}+x_{22}+x_{33})\,,\quad
w  \equiv\frac{1}{2}  (x_{11}-x_{33})\,,  \quad
v  \equiv   \frac{1}{2}(2 x_{22}-x_{11} - x_{33})\,, 
\end{equation}
in place of diagonal elements of the Hessian $(x_{11},x_{22},x_{33})$
one finds that $u,v,w,x_{12} ,x_{13},x_{23}$ are
uncorrelated. Importantly, the field, $x$ is only correlated with
$u=\Delta x$ and
\begin{equation}
\langle x u \rangle= \gamma, \quad \langle x v \rangle =0, \quad 
\langle x w \rangle=0, \quad \langle x x_{kl} \rangle=0, \ k\neq l,
\end{equation}
where $\gamma$ is the same quantity as in equation~(\ref{eq:gammadef}).
The full expression of ${P}_{0}(x,x_{kl})$ is then
\begin{displaymath}
  P_{0}(x,x_{kl}) dx d^6 x_{kl} =
  \frac{5^{1/2}15^2}{(2\pi)^{7/2}({1-\gamma^2})^{1/2}} 
  \exp\left(-\frac12 \left[ Q_0(x,u) + Q_2(v,w,x_{12},x_{13},x_{23}) \right]\right)
  dx\; du\;  dv\; dw\; dx_{12}\;  dx_{13}\;  dx_{23}\,,
\end{displaymath}
with the quadratic forms $Q_0$ and $Q_2$ given by 
\begin{equation}\label{eq:quadformdimadd}
  Q_0 = x^2 + \frac{(u -\gamma x)^2}{(1-\gamma^2)}
  \qquad
  Q_2 = 5 v^{2} + 15  (w^2 + x_{12}^{2}+x_{13}^{2}+x_{23}^{2}) .
\end{equation}
It depends only one a single correlation parameter: $\gamma$. \\

\noindent\textbf{First and third derivatives of the Gaussian field in 3D.}
A similar procedure can be performed for the joint probability of the
first and third derivatives of the fields, ${P_{1}}(x_{i},x_{ijk})$ by
defining the following nine parameters (see also \citep{hanami}):
\begin{equation}
u_i \equiv \nabla_{i} u,
\quad v_i \equiv \frac{1}{2} \epsilon^{ijk} \nabla_{i }
\left( \nabla_{j} \nabla_{j}- \nabla_{k} \nabla_{k}\right)x \,,\,\,\,
{\rm with}\,\,\, j<k\,,\quad {\rm and } \quad
w_i \equiv \sqrt\frac{5}{12}\nabla_{i}
\left( \nabla_{i} \nabla_{i}  -\frac{3}{5}\Delta \right)x\,,
\end{equation}
and replacing the variables $(x_{i11},x_{i22},x_{i33})$ with $(u_i,v_i,w_j)$. 
In that case, the only cross-correlations in the vector 
$(x_1,x_2,x_3,u_1,v_1,w_1,u_2,v_2,w_2,u_3,v_3,w_3,x_{123})$ which 
do not vanish are between the same components of the gradient and
the gradient of the Laplacian of the field:
\begin{eqnarray}
\langle x_i u_i \rangle & = & {\tilde \gamma}/3,\quad i=1,2,3,
\end{eqnarray}
where $\tilde \gamma$ is the same quantity as in equation~(\ref{eq:gammadef}).
This allows us to write:
\begin{equation}
{P_{1}}(x_{i},x_{ijk}) d^3x_i\, d^{10} x_{ijk} =
\frac{105^{7/2} 3^{3}}{(2\pi)^{13/2}(1-{\tilde \gamma}^2)^{3/2}} 
\  \exp\left( -\frac12 \left (Q_1  +Q_3  \right)\right)
d^3x_i\, d^3u_i \, d^{3}w_i \, d^{3}v_i\,dx_{123} 
. \label{eq:defp13D}
\end{equation}
with the quadratic forms:
\begin{equation}
Q_1 = 3 \sum_i \left( \frac{(u_i -{\tilde \gamma} x_i)^2}{(1-{\tilde \gamma}^2)} +  x_i^2 \right) \, ,
\qquad
Q_3 = 105 \left( x_{123}^{2}+\sum_{i=1}^{3}(v_{i}^{2}+w_{i}^{2})\right)  \,
. \label{eq:defp1Q3D}
\end{equation}

\noindent\textbf{The Gaussian field and its second derivative in 2D.}
Introducing the variables
\begin{equation}
u  \equiv - \Delta x = -(x_{11}+x_{22})\,,\quad
w  \equiv\frac{1}{2}  (x_{11}-x_{22})\,, 
\end{equation}
one finds again that $u,w, x_{12}$ are
uncorrelated. 
The expression for ${P}_{0}(x,x_{kl})$ is then
\begin{displaymath}
  P_{0}(x,x_{kl}) dx d^3 x_{kl}
  =
  \frac{8}{(2\pi)^2 ({1-\gamma^2})^{1/2}} 
  \exp\left( -\frac12 \left[ Q_0(x,u)  + Q_2(w, x_{12}) \right]\right)
  dx\; du\;   dw\; dx_{12}\,,
\end{displaymath}
where the quadratic forms $Q_0$ and $Q_2$  are
\begin{equation}\label{eq:defpztdt}
  Q_0 = x^2 + \frac{(u -\gamma x)^2}{(1-\gamma^2)}\,,
  \qquad
  Q_2 =  8 (w^2 + x_{12}^{2}) .
\end{equation}\\

\noindent\textbf{First and third derivatives of the Gaussian field in 2D.}
Defining the following 4 uncorrelated parameters:
\begin{equation}
u_i \equiv \nabla_{i} u, \quad
w_i \equiv \nabla_{i}\left(\nabla_{i}\nabla_{i} -\frac{3}{4}\Delta \right)x\,,
\end{equation}
yields
\begin{equation}
{P_{1}}(x_{i},x_{ijk}) d^2 x_i\, d^{4} x_{ijk} =
\frac{128}{(2\pi)^{3}(1-{\tilde \gamma}^2)} 
\exp\left( -\frac12 \left (Q_1  +Q_3  \right)\right)
d^2 x_i \, d^2 u_i d^2 w_i\
\,. \label{eq:defp12D}
\end{equation}
with the quadratic forms:
\begin{equation}
  Q_1 = 2 \sum_{i=1}^{2} \left(
    \frac{(u_i -{\tilde \gamma} x_i)^2}{(1-{\tilde \gamma}^2)} +  x_i^2 \right) \, ,
  \qquad
  Q_3 = 32 \sum_{i=1}^{2} w_{i}^{2}
  \, . \label{eq:defp1Q2D}
\end{equation}
It is the purpose of the next section to elucidate the nature of these
quadratic forms and to show how similar expressions can be obtained
for any combination of derivatives in a space of any dimension.

\subsection{Theory}\label{sec:jdftheo}

To proceed further, a more systematic way of computing the
correlations between the field derivatives is needed.  This can be
provided by the harmonic decomposition of symmetric tensors (such as
derivative tensors).  The main results are outlined hereafter, the
reader being referred to~\cite{2007.JFC.CorrDerIso} for a detailed
exposition.

\smallheader{Harmonic decomposition of symmetric tensors.} 
The harmonic decomposition of symmetric tensors amounts to projection
onto the irreducible representations of $\mathrm{SO}(n)$.  It is obtained in
close form as follows.
A symmetric tensor $T$ of rank $n$ is associated with a set
$\{T\spi{\ell} \mid 0\leq\ell\leq n, \ n-\ell \text{ even} \}$ of
``harmonic components'' where each $T\spi{\ell}$ is a symmetric
trace-free tensor of rank $\ell$.  
Index $\ell$ can be understood as an \emph{angular frequency}.  We
refer to it as the ``frequency'' of the component.
The harmonic component at frequency $\ell=n-2k$ of a rank $n$ tensor
is obtained as
\begin{displaymath}
  T\spi{n-2k} =  \overline{\tra^k T} \,,
\end{displaymath}
where $(\tra^k\cdot\quad)$ means applying $k$ times the trace operator
(contraction over any pair of indices) and where $\overline{T}$
denotes the traceless part of tensor $T$.
In indexed notations, the first (ranks $0,\ldots, 5$) de-traced tensors
on $R^3$ are given by %
$\overline t=t$, %
$\overline t_i = t_i$, %
$\overline t_{ij} =t_{ij} - \frac13 t_{aa} \delta_{ij}$,
\begin{equation}\label{eq:detrace.trois.quatre}
  \overline t_{ijk} = t_{ijk} - \frac35 \, t_{aa(j} \delta_{kl)},
  \ \ \
  \overline t_{ijkl} = t_{ijkl} - \frac67\, t_{aa(ij} \delta_{kl)} + \frac3{35}\, t_{aabb} \delta_{(ij} \delta_{kl)},
  \ \ \
  \overline t_{ijklm} = t_{ijklm} 
  - \frac{10}9\, t_{aa(ijk} \delta_{lm)}
  + \frac5{21}\, t_{aabb(i} \delta_{jk} \delta_{lm)}\,,
\end{equation}
with an implicit summation over repeated indices and symmetrization
between parenthesized indices (for instance: 
$ t_{aa(j} \delta_{kl)} =  [t_{aaj} \delta_{kl} + t_{aak}
\delta_{lj} + t_{aal} \delta_{jk}]/3$ and so on).

\smallheader{Invariant statistics.}
Let $\mathcal{T}= \left\{ T_0, T_1,\ldots\right\}$ be a set of
symmetric tensors which are jointly isotropically distributed.
A consequence of isotropy is frequency decoupling: $T_a\spi{\ell}$ is
uncorrelated with $T_b\spi{\ell'}$ if $\ell\neq\ell'$.
Further, at any frequency $\ell$, the scalar product $ \pscal{
  T_a\spi{\ell}}{ T_b \spi{\ell}}$ is invariant under rotations.
It is convenient to arrange these products at frequency $\ell$ into a
$m_\ell\times m_\ell$ Gram matrix $\widehat R_\ell$ where $m_\ell$
denotes the number of tensors in $\mathcal{T}$ having an harmonic
component at frequency $\ell$ (this occurs whenever
$\mathrm{rank}(T)-\ell$ is a non-negative even integer):
\begin{displaymath}
  [ \widehat R_\ell  ] _{ab} 
  =
  \pscal{ T_a\spi{\ell}}{ T_{b} \spi{\ell}}\,,
\end{displaymath}
where indices $a$ and $b$ run only over the $m_\ell$ relevant values
(the specific ordering does not matter).  
A further consequence of isotropy is that, in the Gaussian case, these
matrices form a set of sufficient statistics: the joint distribution
of $\mathcal{T}$ can be expressed as a function of those matrices and
nothing else, as seen next.

\smallheader{Spectral matrices.}
The `spectral matrix' $R_\ell$ at frequency $\ell$ is defined as the
expected value of $\widehat R_\ell$, that is, $ R_\ell = \E \bigl(
\widehat R_\ell \bigr)$.  
For a set $\mathcal{T}$ of symmetric random tensors with a
rotationally invariant joint distribution, one finds
\begin{displaymath}
  \mathcal{T}\adj \cov(\mathcal{T})\inv \mathcal{T}
  = \sum_\ell w_\ell 
  \tra\bigl(   \widehat R_\ell   R_\ell\inv \bigr)\,,
\end{displaymath}
where $w_\ell$ is a positive scalar, which is equal to $2\ell+1$ for
tensors in $\Rset^3$.

\smallheader{Spectral matrices for a GRF.}
Now, we consider the case when in $\mathcal{T}= \left\{ T_0, \ldots,
  T_Q \right\}$, the $q$-th tensor $T_q$ is the $q$-th derivative at a
given point: $ t_{i_1\cdots i_n} = {\partial^n \rho}/ {\partial
  r_{i_1} \cdots \partial r_{i_n} }$ of a stationary random field
$\rho$ with spectrum $P(\nu)$.  Then $\mathcal{T}$ is a set of
isotropically distributed symmetric tensors and each spectral matrix
$R_\ell$ can be expressed as a function of the spectrum.
Indeed, if $\ell-q$ and $\ell-q'$ are non negative even integers,
matrix $R_\ell$ has an entry $[ R_\ell ] _{qq'} $ related to the
derivatives of orders $q$ and $q'$ given by
\begin{displaymath}
  [  R_\ell  ] _{qq'} 
  =
  (-1)^{\frac{q-q'}2}\, g_\ell\,  \sigma_{\frac{q+q'}2}^2\,,
\end{displaymath}
with the spectral moments $\sigma_p^2$ defined at
eq.~(\ref{eq:sigmadef}).
The geometric factor $g_\ell$ is the squared ratio $g_\ell =
(\|\overline{\xi^\ell}\|/\|\xi^\ell\|)^2$ by which the norm of the
$\ell$-th tensor product $\xi^\ell$ of any vector $\xi$ is decreased
upon detracing.  It is equal to $g_\ell={\ell!}/{(2\ell-1)!!}$ in
dimension $D=3$.
We do not provide explicit expressions for $w_l$ and $g_\ell$ in
arbitrary dimension since only their ratio $w_\ell/g_\ell$ is needed
and turns out to have a simpler expression than either $w_\ell$ or
$g_\ell$:
\begin{equation}\label{eq:wlovergl}
  \frac{w_\ell}{g_\ell} = \frac{(2\ell+D-2)!!}{\ell! \ (D-2)!!}\,.
\end{equation}
Some precomputed values are listed in Table~\ref{tab:quadcoef}.
\begin{table}
  \centering
  \begin{tabular}{|c|cccccc|}\hline
    & $\ell=0$  & $\ell=1$  & $\ell=2$  & $\ell=3$  & $\ell=4$  & $\ell=5$ \\ \hline 
    D=2 &  1    &  2    &  4    &  8    &  16    &  32   \\
    D=3 &  1    &  3    & 15/2   & 35/2   & 315/8   & 693/8  \\
    D=4 &  1    &  4    &  12    &  32    &  80    &  192   \\
    D=5 &  1    &  5    & 35/2   & 105/2   & 1155/8   & 3003/8  \\
    \hline
  \end{tabular}
  \caption{Values of ${w_\ell}/{g_\ell} ={(2\ell+D-2)!!}/({\ell! \ (D-2)!!})$ in dimensions
    $D=2,3,4,5$ for $0\leq\ell\leq 5$.}
  \label{tab:quadcoef}
\end{table}

\smallheader{Summary and rescaled forms.}
We collect all previous results into a normalized form.  Using the
normalized spectral shape parameters of def.~(\ref{eq:gammadef}) and
normalized derivative tensors $X_n$ defined as:
\begin{displaymath}
  X_n =\frac1{\sigma_n}\nabla^n \rho\,,
  \qquad
  \text{\textit{i.e.}}
  \qquad 
  x_{i_1\cdots i_n}
  =
  \sigma_n^{-1} 
  \frac
  {\partial^n \rho}
  {\partial r_{i_1} \cdots \partial r_{i_n} }\,,
\end{displaymath}
one finds that 
\begin{equation}\label{eq:laveritevraie}
  \mathcal{X}\adj \cov(\mathcal{X})\inv \mathcal{X}
  = \sum_\ell \,
  \frac{(2\ell+D-2)!!}{\ell! \ (D-2)!!}
  \,
  \tra\bigl(   \widehat \Gamma_\ell   \Gamma_\ell\inv \bigr)\,,
  \quad\text{with}\quad
  [\widehat \Gamma_\ell ]_{pq} = \pscal {X_p\spi{\ell}}  {X_{q}\spi{\ell}} 
  \quad\text{and}\quad
  [\Gamma_\ell ]_{pq} =  (-1)^{\frac{p-q}2}\, \gamma_{p,q}\,.
\end{equation}
Note that the diagonal entries of $\Gamma_\ell$ are always equal to~$1$.

\smallheader{Special cases and smaller statistics.}
Our approach compresses a set of derivative tensors into a set
$\widehat\Gamma_\ell$ of symmetric matrices of size $m_\ell\times
m_\ell$, yielding $\sum_\ell m_\ell(m_\ell+1)/2$ invariant scalars.
There are two special cases where even smaller invariant sufficient
statistics can be found.

First, at angular frequency $\ell=0$, the detraced tensors are just
scalars so that, for $\ell=0$, one has 
$ [ \widehat \Gamma_0 ] _{pq} = 
\pscal{ X_p\spi{0}}{ X_{q} \spi{0}} = 
X_p\spi{0} X_{q} \spi{0} $.  
Therefore $\widehat\Gamma_0 $ actually is a rank-one matrix:
$\widehat\Gamma_0=vv\adj$ where the entries of vector $v$ are
$v_p=X_p\spi{0}$. 
Hence, at the null frequency, we can further compress the
$m_0(m_0+1)/2$ statistics (the non-redundant entries of
$\widehat\Gamma_0$) into $m_0$ scalars (the entries of $v$).  
Of course, the $\ell=0$ term in the quadratic form also reads:
\begin{equation}\label{eq:specialcasenullfreq}
  \tra\bigl(   \widehat \Gamma_0   \Gamma_0\inv \bigr)  
  = 
  v\adj  \Gamma_0\inv v.
\end{equation}
Second, there are several cases of interest where $m_\ell=2$.  This
happens for instance   
at $\ell=0$ with derivative orders 0 and 2, 
at $\ell=1$ when considering derivatives of order 1 and 3,  
at $\ell=2$ with derivatives of orders 0,2 and 4, etc.
Then, for such an $\ell$,
\begin{displaymath}
  \tra\bigl( \widehat \Gamma_\ell   \Gamma_\ell\inv  \bigr)  
  =
  \tra\Bigl( 
  \bimat {\pscal{a}{a}}  {\pscal{a}{b}}  {\pscal{b}{a}}  {\pscal{b}{b}} 
  \bimat{1}{-\gamma}{-\gamma}{1}\inv  
  \Bigr)  
\end{displaymath}
where $a$ and $b$ are rank-$\ell$ tensors and $\gamma$ is a scalar.
Simple algebra yields
\begin{equation}\label{eq:specialcasebidim}
  \tra\bigl( \widehat \Gamma_\ell   \Gamma_\ell\inv  \bigr)  
  =
  \|a\|^2 + \frac{\|b+\gamma a\|^2}{1-\gamma^2}\,,
\end{equation}
that is, a form ubiquitous in this paper.  However, an equivalent,
more regular form is 
\begin{displaymath}
  \tra\bigl( \widehat \Gamma_\ell   \Gamma_\ell\inv  \bigr)  
  =
  \frac1{1-\gamma^2} \left( \|a\|^2 + \|b\|^2\right) 
  +
  \frac{2\gamma}{1-\gamma^2}\ \pscal a b \,,
\end{displaymath}
which has the benefit of stressing that, at such $\ell$, a sufficient
statistic is only made of \emph{two} invariant scalars, namely
$\|a\|^2 + \|b\|^2$ and $\pscal a b$.  
In the limit of weak correlation $\gamma\rightarrow0$, one has, of
course, $\tra\bigl( \widehat \Gamma_\ell \Gamma_\ell\inv \bigr) =
\|a\|^2 + \|b\|^2$.  An even more symmetric form, which stresses the
decorrelation between $a+b$ and $a-b$ is
\begin{displaymath}
  \tra\bigl( \widehat \Gamma_\ell   \Gamma_\ell\inv  \bigr)  
  =
  \frac{\|a+b\|^2}{2(1-\gamma)} 
  +
  \frac{\|a-b\|^2}{2(1+\gamma)} .
\end{displaymath}

\subsection{Some applications}\label{sec:jdfappli}

We now work out these expressions in some cases of interest.
\smallheader{Derivative of orders 0+2 in 3D.}
%
The case $\mathcal{X}= \left\{ X_0, X_2 \right\}$ is the simplest
non-trivial case.  The theory sketched at sec.~\ref{sec:jdftheo}
applies straightforwardly.  In the notations of
section~\ref{sec:jdftheo}], we are concerned with frequencies $\ell=0$
and $\ell=2$ for which, in 3D, $w_0/g_0=1$, $w_2/g_2=15/2$ (see
table~\ref{tab:quadcoef}).
The quadratic form~(\ref{eq:laveritevraie}) then reduces to
$\tra\bigl(\widehat \Gamma_0 \Gamma_0\inv \bigr)\, + \frac{15}2
\tra\bigl( \widehat \Gamma_2 \Gamma_2\inv \bigr)$.
For $\ell=0$, we have here $m_0=2$ and we can use the specific
form~(\ref{eq:specialcasebidim}) to work out $\tra\bigl(\widehat
\Gamma_0 \Gamma_0\inv \bigr)$ with $[ a, b ] = [ X_0\spi{0},
X_2\spi{0}] = [ x , x_{aa}]$, that is the (normalized) field and the
trace of its Hessian.
For $\ell=2$, we have here $m_2=1$: we need only scalars.  Following
expressions (\ref{eq:laveritevraie}) again, we have $\widehat\Gamma_2
= \|X_2\spi{2}\|^2 = \|\bar X_2\|^2= \bar x_{ab}\bar x_{ab}$ and
$\Gamma_2=(-)^{(2-2)/2} \gamma_{2,2} = 1$.
In summary:
\begin{equation}\label{eq:fquad.zero.deux}
  Q_0+Q_2
  =
  \tra\bigl(\widehat \Gamma_0 \Gamma_0\inv \bigr)\, 
  + \frac{15}2 \tra\bigl( \widehat \Gamma_2 \Gamma_2\inv \bigr)
  =
  x^2 
  +
  \frac
  {(x_{aa} + \gamma x)^2}
  {1-\gamma^2}
  \
  +
  \frac{15}2\,
  \overline x_{ab}
  \overline x_{ab}\,,
\end{equation}
This is, of course, identical to equation (\ref{eq:quadformdimadd}) using the
local definitions there.  It also shows that the complicated
expression for $Q_2$ in (\ref{eq:quadformdimadd}) is nothing but the
the squared Euclidean norm of the detraced Hessian (with a $15/2$ prefactor).

\smallheader{Result for orders 1+3 in 3D.}
We take $\mathcal{X}= \left\{ X_1, X_3 \right\}$, that is, the first
and third order derivatives of the field.  The rescaled harmonic components are
\begin{displaymath}
  \left[  \sigma_1^{-1} X_1\spi{1} \right]_i = x_i\,,
  \qquad
  \left[  \sigma_3^{-1} X_3\spi{1} \right]_i = x_{iaa}\,,
  \qquad
  \left[  \sigma_3^{-1} X_3\spi{3} \right]_{ijk} = x_{ijk} - \frac35 x_{aa(i} \delta_{jk)} = \overline x_{ijk} .
\end{displaymath}
We need frequencies $\ell=1$ and $\ell=3$ for which, in 3D,
$w_1/g_1=3$, $w_3/g_3=35/2$ (see table~\ref{tab:quadcoef}).  For
frequency $\ell=1$, we have $m_\ell=2$; matrix $\Gamma_1$ is $2\times
2$ with entries given by equation~(\ref{eq:laveritevraie}), that is, diagonal
entries equal to $1$ (as always) and off-diagonal entries given by
$(-1)^{(1-3)/2}\gamma_{1,3}=-\tilde\gamma$.  
Since $\Gamma_1$ is $2\times 2$, we can still use equation
(\ref{eq:specialcasebidim}) and finally obtain
In summary:
\begin{align}\label{eq:fquad.un.trois}
  \frac{w_1}{g_1}  \tra\bigl(   \widehat \Gamma_1   \Gamma_1\inv \bigr)\, +
  \frac{w_3}{g_3}  \tra\bigl(   \widehat \Gamma_3   \Gamma_3\inv \bigr)\,
  &=
  3\,
  \tra
  \biggl\{ 
  \bimat{1}{-\tilde\gamma}{-\tilde\gamma}{1} \inv
  \bimat{x_ix_i}{x_ix_{iaa}}{x_ix_{ibb}}{ x_{icc} x_{idd}}
  \biggr\}
  +
  \frac{35}2\
  \overline x_{ijk} 
  \overline x_{ijk} \\
  &=
  3\,
  \biggl(
  x_{i}
  x_{i}
  +\frac
  {
    (x_{iaa}-\tilde\gamma x_i)
    (x_{ibb}-\tilde\gamma x_i)
  }
  {1-\tilde\gamma^2}
  \biggr)
  +
  \frac{35}2\
  \overline x_{ijk} 
  \overline x_{ijk} \,.
\end{align}
This is consistent with equation~(\ref{eq:defp13D}) and reveals the
meaning of $x_{123}^{2} + \sum_{i=1}^{3} (v_{i}^{2}+w_{i}^{2})$ as
equal to $\frac16 \overline x_{ijk} \overline x_{ijk}$ \textit{i.e.}
the squared norm of the detraced third derivative tensor (with a
prefactor $1/6$).

\medskip The results for other combinations of derivatives can be
derived in the same way.  A few results are listed below without going
into much detail.

\smallheader{Result for orders 0+2+4 in 3D.}
We consider $\mathcal{X}= \left\{ X_0, X_2 , X_4\right\}$.
Hoping to improve clarity, we denote $y_{ij} = [\sigma_4\inv
X_4\spi{2}]_{ij}$, that is, the de-traced contraction of the 4th-order
derivative tensor.  Explicitly, in 3D:
\begin{displaymath}
  y_{ij} = x_{ijaa} - \frac13 x_{aabb} \delta_{ij}\,,
\end{displaymath}
With this notation and recalling that $\overline x_{ijkl}$ denotes the
traceless part of $x_{ijkl}$ (the rescaled 4th-order derivative
tensor) computed according to the
prescription~(\ref{eq:detrace.trois.quatre}), the quadratic form is
\begin{equation}\label{eq:fquad.zero.deux.quatre}
  \begin{bmatrix}    x \\ x_{aa} \\ x_{aabb}  \end{bmatrix} \adj
  \begin{bmatrix}   
    1 &-\gamma & \breve\gamma \\
    -\gamma & 1 & -\hat \gamma \\
    \breve\gamma & -\hat\gamma & 1
  \end{bmatrix}\inv
  \begin{bmatrix}    x \\ x_{aa} \\ x_{aabb}  \end{bmatrix}
  + 
  \frac{15}2
  \tra
  \biggl\{ 
  \bimat{1}{-\hat\gamma}{-\hat\gamma}{1} \inv
  \bimat{\bar x_{ij}\bar x_{ij}}{\bar x_{ij} y_{ij}}{\bar x_{ij} y_{ij}}{y_{ij} y_{ij}}
  \biggr\}
  + 
  \frac{315}8 \  \overline x_{ijkl}  \overline x_{ijkl} \,, 
\end{equation}
where yet another spectral shape parameter has to be defined:
\begin{displaymath}
  \breve\gamma
  = \frac{\sigma_2^2}{\sigma_0\sigma_4}
  = \frac{\tilde R \hat R }{R_0 R_\star} = \frac{\gamma {\tilde \gamma}^2}{{\hat \gamma}}\,,
\end{displaymath}
Needless to say that expression~(\ref{eq:fquad.zero.deux}) obtained
for $\mathcal{X}= \left\{ X_0, X_2 \right\}$ is recovered by cutting
the irrelevant terms from equation (\ref{eq:fquad.zero.deux.quatre}).

\smallheader{Result for orders 1+3+5 in 3D.}
To simplify the notations, we introduce local definitions for the
derivative tensors and their contractions:
\begin{displaymath}
  y_a=x_{abb} \,,
  \qquad
  z_a = x_{abbcc} \,,
  \qquad
  t_{abc} = x_{abcdd} \,,
\end{displaymath}
and, proceeding as above, we obtain the quadratic form:
\begin{equation}\label{eq:fquad.un.trois.cinq}
  3\
  \tra
  \Biggl\{ 
  \begin{bmatrix}   
    1 &-\gamma_{1,3} & \gamma_{1,5} \\
    -\gamma_{1,3} & 1 & -\gamma_{3,5} \\
    \gamma_{1,5} & -\gamma_{3,5} & 1
  \end{bmatrix}\inv
  \begin{bmatrix}   
    x_a x_a & x_a y_a  & x_a z_a \\
    y_a x_a & y_a y_a  & y_a z_a \\
    z_a x_a & z_a y_a  & z_a z_a 
  \end{bmatrix}
  \Biggr\}
  +
  \frac{35}2
  \tra
  \biggl\{ 
  \bimat{1}{-\gamma_{3,5}}{-\gamma_{3,5}}{1} \inv
  \bimat
  {\bar x_{ijk} \bar x_{ijk}} {\bar x_{ijk} \bar t_{ijk}}
  {\bar t_{ijk} \bar x_{ijk}} {\bar t_{ijk} \bar t_{ijk}}
  \biggr\}
  + 
  \frac{693}8 \  \overline x_{ijklm}  \overline x_{ijklm} 
\end{equation}

\smallheader{The 2D case.}
The theory applies to isotropic fields in any dimension.  We have
already provided expressions for the spectral
moments~(\ref{eq:sigmadef}) and  the  coefficients
$w_\ell/g_\ell$ of equation~(\ref{eq:wlovergl}).  It remains to find detracing
coefficients.
In the 2D case, the first (ranks $0,\ldots, 5$) de-traced tensors on
$R^2$ are given by %
$\overline y=y$, %
$\overline y_i = y_i$, %
$\overline y_{ij} =y_{ij} - \frac12 y_{aa} \delta_{ij}$,
\begin{equation}\label{eq:detrace.trois.quatre.Ddeux}
  \overline y_{ijk} = y_{ijk} - \frac34 \, y_{aa(i} \delta_{jk)},
  \ \ \
  \overline y_{ijkl} = y_{ijkl} -\, y_{aa(ij} \delta_{kl)} + \frac18\, y_{aabb} \delta_{(ij} \delta_{kl)},
  \ \ \
  \overline y_{ijklm} = y_{ijklm} 
  - \frac54\, y_{aa(ijk} \delta_{lm)}
  + \frac5{16}\, y_{aabb(i} \delta_{jk} \delta_{lm)}\,.
\end{equation}
For the correlation between the field and its Hessian, we proceed as
above in 3D with $w_0/g_0=1$ and $w_2/g_2=4$  given in
table~\ref{tab:quadcoef}.  Therefore the quadratic form is
\begin{displaymath}
  Q_0+Q_2
  =
  \tra\bigl(\widehat \Gamma_0 \Gamma_0\inv \bigr)\, 
  + 4  \tra\bigl( \widehat \Gamma_2 \Gamma_2\inv \bigr)
  =
  x^2 
  +
  \frac
  {(x_{aa} + \gamma x)^2}
  {1-\gamma^2}
  \
  +
  4\
  \overline x_{ab}
  \overline x_{ab} \,,
\end{displaymath}
in agreement with equation~(\ref{eq:defpztdt}).
For the case of first and third order derivatives, we read $w_1/g_1=2$
and $w_3/g_3=8$ from table~\ref{tab:quadcoef} so that, similar to
equation~(\ref{eq:fquad.un.trois}), one finds
\begin{align}
  \frac{w_1}{g_1}  \tra\bigl(   \widehat \Gamma_1   \Gamma_1\inv \bigr)\, +
  \frac{w_3}{g_3}  \tra\bigl(   \widehat \Gamma_3   \Gamma_3\inv \bigr)\,
  &=
  2\,
  \tra
  \biggl\{ 
  \bimat{1}{-\tilde\gamma}{-\tilde\gamma}{1} \inv
  \bimat{x_ix_i}{x_ix_{iaa}}{x_ix_{ibb}}{ x_{icc} x_{idd}}
  \biggr\}
  +
  8\
  \overline x_{ijk} 
  \overline x_{ijk}\,, \\
  &=
  2\,
  \biggl(
  x_{i}
  x_{i}
  +\frac
  {
    (x_{iaa}-\tilde\gamma x_i)
    (x_{ibb}-\tilde\gamma x_i)
  }
  {1-\tilde\gamma^2}
  \biggr)
  +
  8\
  \overline x_{ijk} 
  \overline x_{ijk} \,,
\end{align}
with $\overline x_{ijk} = x_{ijk} - \frac34 \, x_{aa(i} \delta_{jk)}$
so that $8\overline x_{ijk} \overline x_{ijk}$ can be checked to
equal $Q_3$ in equation~(\ref{eq:defp1Q2D}).

\smallheader{The $d$-dimensional case.}  
We outline some results in the $d$-dimensional case.
%
%
The de-tracing formulae can be extended to the $d$-dimensional case
but, in this paper, we will content ourselves with the correlations
between the field and its Hessian: $\mathcal{X} =\left\{ X_0 , X_2
\right\}$.  Therefore, we need only $\ell=0$ and $\ell=2$ so that
de-tracing remains trivial: the normalized de-traced Hessian given by
$ \bar x _{ij} = x_{ij} - \frac 1d\, \delta_{ij}\, x_{aa}$.
Hence for the correlation between the field and its Hessian, we obtain
the quadratic form
\begin{equation}
  \bivec{x}{x_{aa}}\adj
  \bimat{1}{-\gamma}{-\gamma}{1}\inv
  \bivec{x}{x_{aa}}
  \ +\ 
  \frac{d(d+2)}2\
  \overline x_{ab}
  \overline x_{ab}\,,
\end{equation}
which is a straightforward extension of the 3D case of
equation~(\ref{eq:fquad.zero.deux}).  Just recall that $\gamma$ is
now defined in terms of the spectral moments~(\ref{eq:gammadef}) and
that de-tracing the Hessian requires a factor $1/d$ instead  of $1/3$.

\end{document}